\documentclass[review,number,sort&compress]{elsarticle}

\usepackage{amssymb}
\usepackage{amsmath}
\usepackage{multirow}
\usepackage{booktabs}
\usepackage{lscape,graphicx}
\usepackage{threeparttable}
\usepackage{footnote}
\usepackage[a4paper]{geometry}
\usepackage{footnote}
\usepackage[]{natbib}

\biboptions{comma,square}

\journal{Nuclear Instruments and Methos B}

\begin{document}

\begin{frontmatter}



\title{The in-gas-jet laser ion source: resonance ionization spectroscopy of radioactive atoms in supersonic gas jets}


\author[]{Yu. Kudryavtsev \corref{cor1}}
\ead{Yuri.Kudryavtsev@fys.kuleuven.be, Tel.:+32 16327271, Fax:+32 16327985}
\author{R. Ferrer}
\author{M. Huyse}
\author{P. Van den Bergh}
\author{and \, P. Van Duppen}
\address{\begin{center}
    Instituut voor Kern- en Stralingsfysica, KU Leuven, \\
    Celestijnenlaan 200D, B-3001 Leuven, Belgium
    \end{center}}
\cortext[cor1]{Corresponding author}

\begin{abstract}
New approaches to perform efficient and selective step-wise Resonance Ionization Spectroscopy (RIS) of radioactive atoms in different types of supersonic gas jets are proposed. This novel application results in a major expansion of the In-Gas Laser Ionization and Spectroscopy (IGLIS) method developed at KU Leuven. Implementation of resonance ionization in the supersonic gas jet allows to increase the spectral resolution by one order of magnitude in comparison with the currently performed in-gas-cell ionization spectroscopy. Properties of supersonic beams, obtained from the de Laval-, the spike-, and the free jet nozzles that are important for the reduction of the spectral line broadening mechanisms in cold and low density environments are discussed. Requirements for the laser radiation and for the vacuum pumping system are also examined. Finally, first results of high-resolution spectroscopy in the supersonic free jet are presented for the 327.4 nm 3d$^{10}$4s $^2$S$_{1/2}$\,$\rightarrow$\,3d$^{10}$4p $^2$P$_{1/2}$ transition in the stable $^{63}$Cu isotope using an amplified single mode laser radiation.
\end{abstract}


\end{frontmatter}


\section{Introduction}
\label{}
The method of laser Resonance Ionization Spectroscopy (RIS) \cite{Let1987, Hur1988} was developed at the end of the last century. Nowadays resonance photoionization with pulsed lasers is widely used or planned, in particular, for the production of pure beams of short-lived isotopes at several on-line Radioactive Ion Beam (RIB) facilities \cite{Alk1989, Bar2012, Ver1994, Kud1996, Coc2009, Mis1993, Fed2000, Bac1998, Bac2000, Las2005, Moo2005, Moo2010,Liu2006, Jeo2010, Son2009a, Lec2010}. In addition to the production of element-pure or isomeric-pure beams, RIS can be used to obtain nuclear-model independent information on the properties of nuclear ground and long-living excited states such as nuclear spins, nuclear magnetic dipole moments, quadrupole moments, and changes in the mean-square charge radii from atomic spectra \cite{Ott1989, Bil1995, Neu2002, Klu2003, Neu2006, Che2010, Klu2010}. Numerous two- and three-step ionization schemes have been proposed to produce and investigate radioactive nuclei using resonance ionization techniques \cite{Fed2012}. Since the radioactive isotopes of interest are created in nuclear reactions in very small quantities, together with a huge background of contaminating nuclei, the RIB production and detection methods have to be sensitive, efficient, selective, and fast.
\begin{figure}
\begin{center}
\includegraphics[scale=0.55]{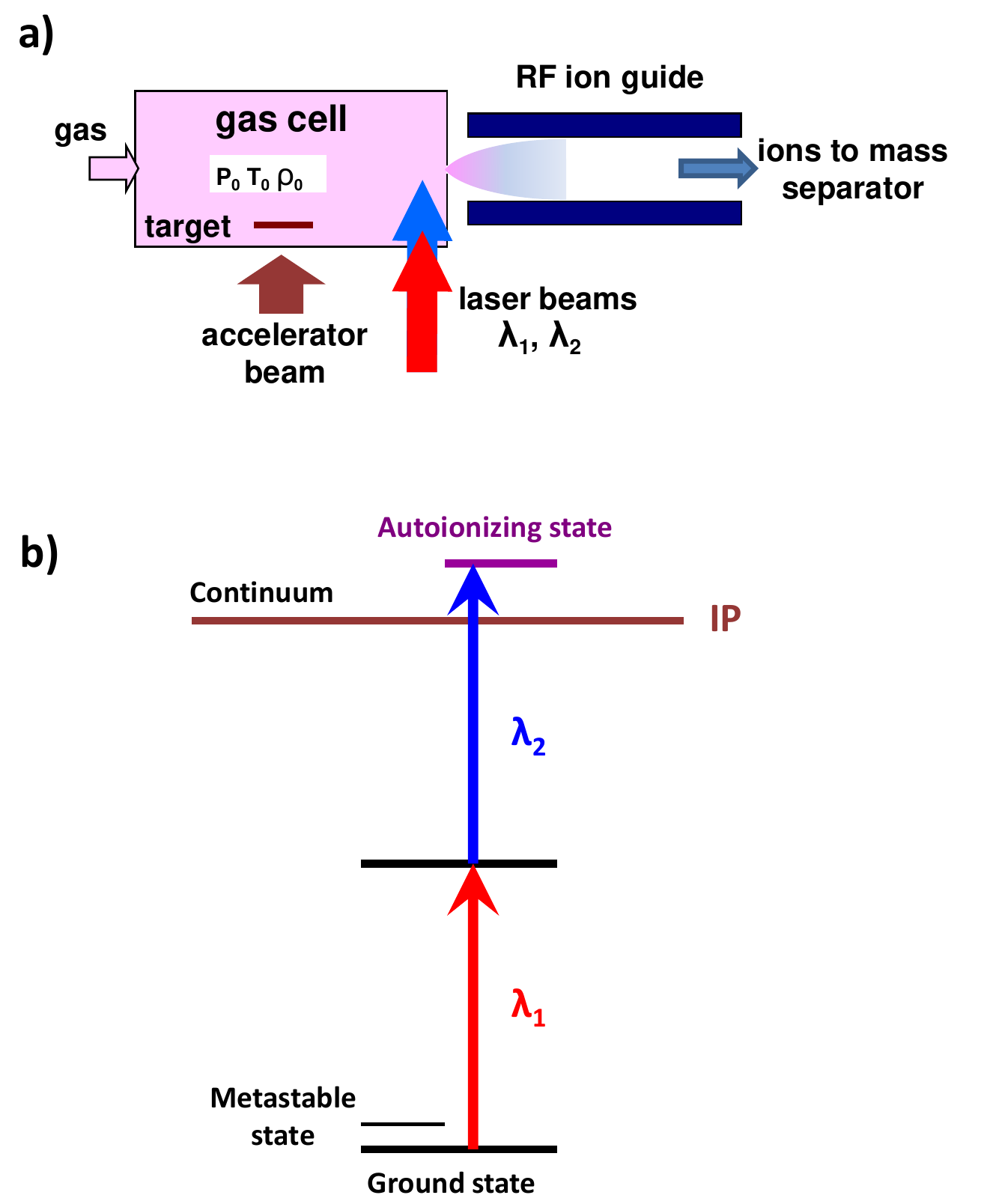} \caption{a) Layout of the in-gas-cell laser ion source for the production of radioactive ion beams, b) two step laser resonance ionization scheme. $P_0$, $T_0$, and $\rho_0$ represent the stagnation conditions of pressure, temperature, and density in the gas cell, respectively.}\label{Fig:1}
\end{center}
\end{figure}
Currently, collinear Resonance Ionization Spectroscopy (CRIS) in an accelerated atomic beam ($>10$ keV) \cite{Kud1982, Sch1991, Wen2000, Fla2008, Pro2012} is one of the most sensitive detection methods. Owing to the reduction of the Doppler width by the electrostatic acceleration \cite{Kau1976}, a high spectral resolution and a high selectivity can be provided for the detection of short-lived- and low-abundant long-lived radioactive isotopes \cite{Kud1992, Mon1993, Wen1997}.

Another sensitive atomic spectroscopy and selective production method is in-source laser resonance ionization spectroscopy. This method has been implemented in two distinctly different ways at Isotope Separator On Line (ISOL) systems to produce RIBs and to perform laser spectroscopy measurements. The two approaches  are based on resonance ionization either in a hot cavity \cite{Alk1992, Fed2003, Kos2003}  or in a buffer gas cell \cite{Ver1994, Kud1996, Fac2004, Coc2009}. The spectral resolution in a hot cavity is limited by Doppler broadening as the temperature has to be above 2000 K in order to keep the reaction products volatile and in their atomic form. In a gas cell, additionally to the room temperature Doppler broadening, the spectral resolution is limited by collision broadening with the buffer gas atoms.  In spite of the limited spectral resolution in comparison with the collinear photo ionization spectroscopy, the in-source technique is very sensitive; results have been obtained with beams of less than 1 atom per second \cite{DeW2007, Coc2011}, and it can be applied for the study of isotopes with a large hyperfine splitting \cite{Lau1992, Coc2010, Sto2008}.

The In-Gas Laser Ionization and Spectroscopy (IGLIS) technique, developed at KU Leuven since the late 1980s \cite{Qam1992, VDu1992, Ver1994, Kud1996}, is used at the Leuven Isotope Separator On Line (LISOL) facility \cite{Kud2003} to produce short-lived radioactive beams in different regions of the chart of nuclides using light and heavy-ion induced fusion or fission reactions. The basic principle of the IGLIS method can be summarized as follows. Nuclear-reaction products are thermalized and neutralized in the high-pressure noble gas in their ground- and possibly in low-lying metastable atomic states (see Fig. \ref{Fig:1}a). They are subsequently transported by the gas flow towards the exit orifice. Shortly before leaving the gas cell, the atoms undergo element-selective two-step resonance laser ionization (Fig. \ref{Fig:1}b). Outside the gas cell the laser-produced ions are captured by the Radio Frequency (RF) field of a SextuPole Ion Guide (SPIG) for further transport towards the mass separator \cite{VdB1997}. The use of a repelling voltage to suppress unwanted ions and laser ionize the nuclei of interest in an RF trap, the so-called Laser Ion Source Trap (LIST) technique, was first proposed for the hot cavity approach \cite{Bla2003} and recently successfully applied in on-line conditions \cite{Fin2012}.  The coupling of the LIST method to the gas cell approach was suggested at Jyv\"{a}skyl\"{a} \cite{Moo2006}. In order to improve the spectral resolution and the selectivity, the possibility of laser ionization in the free gas jet has been investigated at LISOL with 200 Hz \cite{Son2009} and 10 kHz \cite{Fer2012} pulse repetition rate lasers. In these experiments, the ionizing laser beams passed through the gas cell and the exit orifice to reach the expanding free gas jet. By applying a positive potential to the SPIG rods relative to the gas cell the ions created in the gas jet could be separated from those created in the gas cell. Compared to in-gas cell ionization an improved spectral resolution down to 2.6 GHz was achieved for in-gas jet ionization owing to  the low pressure and low temperature environment of the supersonic gas jet, however, major developments were required in order to improve the efficiency, selectivity, and spectral resolution to be able to perform spectroscopic studies of the nuclei.
The supersonic gas jet can be a natural part of the target-ion-source system for on-line mass separators. In this paper we propose new approaches for high-resolution, efficient, and selective step-wise laser resonance ionization of radioactive atoms using different types of supersonic jets. The spectral resolution that can be reached in the supersonic gas jet is calculated and found to be far superior to that in the gas cell. The requirements for the laser radiation and for the vacuum pumping system are also discussed. Finally, first off-line results of two-step high resolution laser resonance ionization spectroscopy in the supersonic free jet that show the feasibility of this method are presented.

\section{Brief introduction to the different jet-formation schemes}
\begin{figure}
\begin{center}
\includegraphics[scale=0.50]{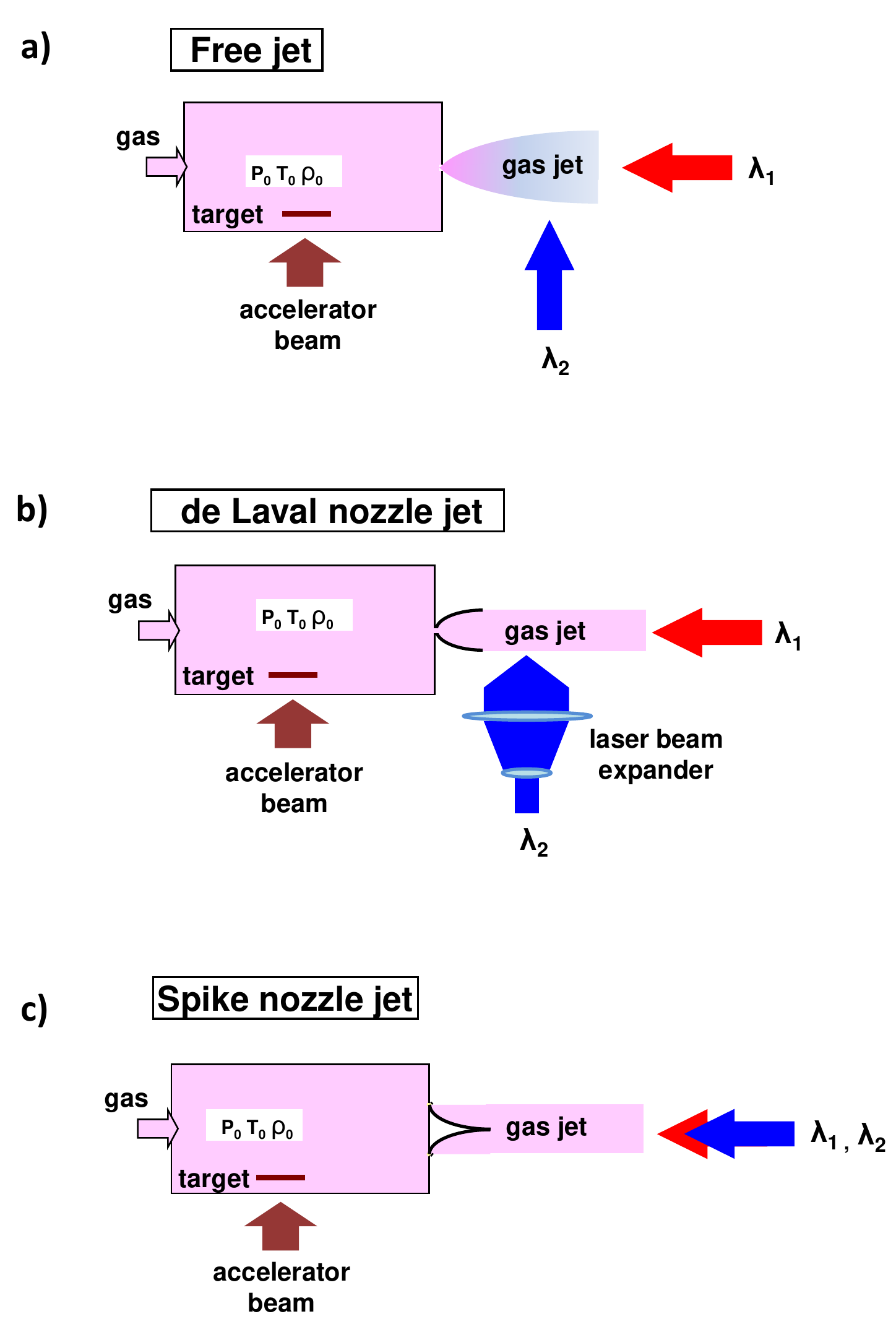} \caption{Proposed setup layouts for the production and spectroscopy of radioactive isotopes in a) a free jet, and in jets produced by b) a de Laval nozzle and c) a spike nozzle. The production target is here located in the gas cell and the primary beam from the accelerator is therefore entering the cell.}\label{Fig:2}
\end{center}
\end{figure}

Three different approaches for gas-jet formation suitable for resonance ionization are considered in this article and can be found schematically illustrated in Fig. \ref{Fig:2}.
\begin{itemize}
\item The supersonic free-jet expansion technique, Fig. \ref{Fig:2}a, proposed in \cite{Kan1951} is used nowadays in different fields of research involving cold atoms and molecules. Effective translation, rotation, and vibration cooling of the molecules seeded into the jet allow to perform fluorescence spectroscopy of complex molecules with very high resolution \cite{Sma1977, Lev1980}. Sub-Doppler resolution was achieved with very well collimated beams, where owing to the strong collimation only a small part of all atoms coming out the gas cell were used. The most important advantage of this low-temperature molecular spectroscopy consists in the simplification of the spectra due to the compression of the population distribution in low-lying vibration and rotational levels. The collimated supersonic free-jet beams in a crossed-beam geometry prepared in well-defined states are used to study chemical reaction dynamics \cite{Kop2006}. Studies of gas-jet formation have been performed in preparation of the gas cell-based LIST project \cite{Rep2011, Rep2012}.
\item Unlike the free-jet nozzle, the de Laval nozzle, Fig. \ref{Fig:2}b, has been used to produce a homogeneous flow of cold molecules to investigate ion-molecular and neutral-neutral reactions at very low temperatures \cite{Row1984, Row1987, Sim1994, Atk1995, Lee2000}. Continuous-flow and pulsed-supersonic-expansion apparatus  have been developed. Carefully designed nozzles allow to obtain a homogenous flow and a low temperature zone of up to 30 cm in length \cite{Dup1985}. A pulsed de Laval nozzle has been used to study neutral-neutral reactions initiated by laser photolysis and consecutive laser photoionization of the product species inside a time-of-flight mass spectrometer \cite{Lee2000}. The uniform, supersonic, and axisymmetric beams have been combined with a high-resolution Fourier transform spectrometer to perform infra-red molecular absorption spectroscopy \cite{Ben2000}. The properties of high-Mach-number jets have been studied to better understand the behavior of astrophysical jets simulated in an Earth Laboratory \cite{Tor2011}.
\item The third type of nozzle that will be considered in this article is an axisymmetric spike nozzle, Fig. \ref{Fig:2}c. This nozzle also allows to produce a quasi-parallel supersonic atomic beam with high Mach numbers. Applications of this type of nozzle though is currently mainly limited to the rocket design \cite{Lem2009, Bes2002}. 

\end{itemize}

\section{In-gas-jet laser ion source. Principle of operation}
\begin{figure}
\begin{center}
\includegraphics[scale=0.6]{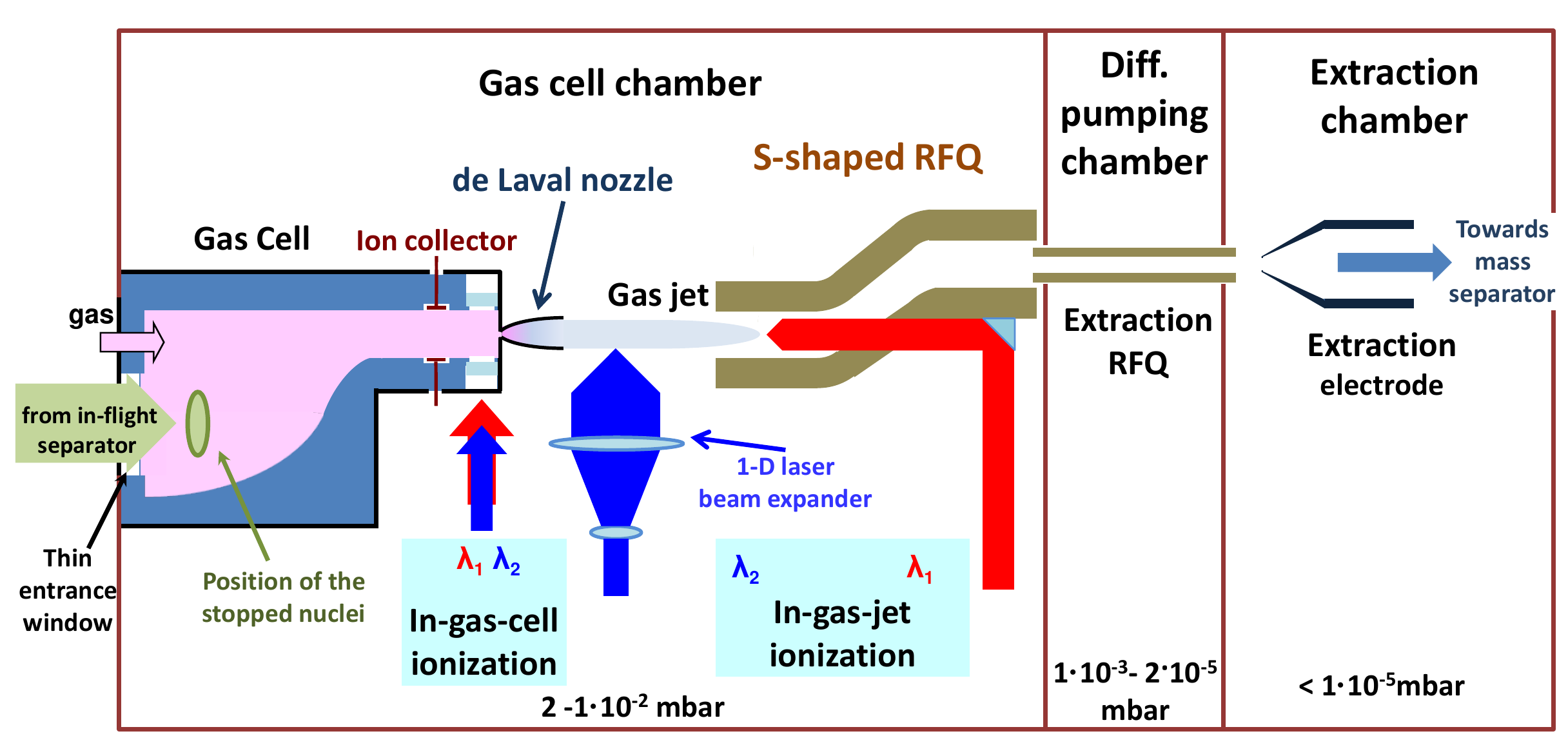} \caption{Generic layout of an In Gas Laser Ionization and Spectroscopy (IGLIS) setup coupled on-line to an in-flight mass separator.}\label{Fig:3}
\end{center}
\end{figure}
Gas catchers are used in radioactive ion beam research as an alternative to solid or liquid catchers owing to the physicochemical limitations of the latter. Nuclear reaction products recoiling out of a target located in the gas cell or coming from an in-flight mass separator  can effectively be stopped in a high pressure noble gas. The buffer gas pressure needed to stop the nuclear reaction products is defined by the energy of the ions and their atomic mass number. Argon at a pressure of 100-500 mbar is the most suitable gas because of its large stopping power. Helium can also be used for nuclear reactions with small recoil energy. Two possibilities can be considered: one can optimize the choice of the gas and the design of the gas cell to thermalize the nuclei in their 1$^+$ (possibly 2$^+$) charge-state and evacuate them as such \cite{Arj1985, Eng2001, Sav2003, Wad2003, Sch2003} or, alternatively, all parameters can be optimized in order to have the nuclei neutralized in their atomic ground state. Figure \ref{Fig:3} shows the generic layout of an In Gas Laser Ionization and Spectroscopy (IGLIS) setup coupled on-line with an in-flight mass separator. Energetic ions from the separator are stopped and neutralized in a well-defined volume inside the gas cell. Argon as a buffer gas is also a better choice for neutralization as its recombination coefficient is one order of magnitude larger  than that for helium \cite{Way1973, Coo1993}. Neutralized radioactive species are transported along with the buffer gas towards the nozzle. In order to have only laser-ionized nuclei injected in the mass separator unwanted ions can be removed from the gas flow with the aid of electric fields applied to collecting electrodes located inside the cell upstream of the nozzle. Since the primary beam and the reaction products create a high plasma density inside the cell, the collection of remaining ions with electrical fields is not a trivial task \cite{Huy2002}. To overcome this problem a dual chamber gas cell \cite{Kud2009} has to be used with the stopping volume out of the direct view of a second adjacent volume, where the remaining ions can be collected safely with electrical fields and subsequent laser ionization can be performed either within the gas cell or in the gas jet. In the setup depicted in Fig. \ref{Fig:3}, the dual chamber effect is accomplished by the displacement of the laser ionization volume relative to the axis of the incoming ion beam.

  In Fig. \ref{Fig:3}, a de Laval nozzle type is used to create an axisymmetric supersonic gas jet. This gas jet is characterized by a low atom density and a low temperature, which result in a strong reduction of the collision- and Doppler broadening mechanisms in comparison with those existing in the gas cell. Since the gas-flow velocity inside the gas cell is low, laser ionization in this region should be avoided, otherwise all atoms would get ionized and collected before they reach the jet. One should thus prevent the two laser beams from reaching inside the gas cell. To achieve such conditions and at the same time  provide high-resolution excitation and efficient ionization in the two-step laser ionization process, a special geometry of the laser beams must be arranged. This can be realized in a crossed-laser-beam configuration with any of the nozzles considered here, or in the case of the spike-nozzle, even using a parallel beam geometry configuration. The former can be realized if the ionization volume in the supersonic atomic beam is defined by the crossing of the first- and second step laser beams. Usually the first excitation step is employed for the laser spectroscopy measurement. In this case the first-step laser beam is directed axially counterpropagating with the supersonic atomic beam. The second step circularly-shaped laser beam is converted by a one-dimensional telescope in a planar beam and is directed perpendicular to the supersonic beam, only in the region where a uniform and cold gas flow is formed, to ionize the excited atoms. Notice that under these temperature conditions the population of those atoms with low-lying metastable states will be kept in the atomic ground state.  Throughout this article only two-step ionization is considered but the same could also be applied to three-step laser ionization schemes. In such a case, the third step laser beam can be directed perpendicular or along the jet axis. The laser-produced ions are then moved by the gas flow in the direction of the S-shaped Radio Frequency Quadrupole (RFQ) ion guide, where they are confined and transported through the gas cell chamber. Segmentation of the RFQ rods allows to use a dc dragging field to improve the transport efficiency. Additionally, a weak electrical field can be applied between the nozzle and the first segment of the RFQ  ion guide to attract the laser ions into the RFQ structure. The bending of the RFQ structure allows to send the full beam of the first-step laser between the rods. In Fig. \ref{Fig:3} an S-shaped RFQ is shown but a 90$^\circ$ bent RFQ can also be used. The laser-produced ions are subsequently transferred to a linear RFQ, acting as a differential pumping region, that guides the ions towards the acceleration region preceding the mass separator. To assure that all atoms get ionized, the laser pulse repetition rate  has to be high enough to irradiate all radioactive atoms in the fast supersonic jet. The following condition needs to be fulfilled to ensure that all radioactive atoms interact with the laser light at least once in their transit through the ionization region
\begin{equation}\label{num:1}
    f_{laser}\geq 1/(L/u)\, ,
\end{equation}					
where $L$ is the length of the ionization zone and $u$ is the jet velocity. For atoms moving in a supersonic argon jet with $u=$ 550 m/s the length of the ionization zone has to be 5.5 cm for a laser pulse repetition rate of 10 kHz. To provide two-step ionization the first-step laser frequency $\nu_1$ has to be red-shifted owing to the Doppler shift undergone by the moving atoms, while the frequency of the second step $\nu_2$ remains unshifted. We find then

\begin{equation}\label{num:2}
\nu_1=\nu_{01}\cdot(1-u/c)\,\,\,\,\,\, \mbox{and}
\end{equation}

\begin{equation}\label{num:3}
\nu_2=\nu_{02}\, ,
\end{equation}
where  $\nu_{01}$ and $\nu_{02}$  are  the atomic transition frequencies of the first and second steps, respectively, and $c$ is the speed of light. The spectral resolution that can be achieved is defined by the temperature and the atom density in the jet.

\section{Characterization of the supersonic jet produced by a de Laval nozzle}

The de Laval nozzle is the most popular type of convergent-divergent nozzle that allows to generate an axisymmetric  supersonic flow of approximately constant temperature and density. Here we discuss only the properties of the gas jet that are important for selective, efficient, and high-resolution laser ionization spectroscopy of radioactive atoms. These atoms can be considered as seeded into the buffer gas and as such they do not have any influence on the properties of the supersonic gas flow. Indeed, for a total fusion cross section $\sim$1 barn in a target of 1 mg/cm$^2$ and a projectile beam current of 1 p$\mu$A the total production rate amounts to about $10^8$ nuclei/s, which is much smaller than the typical  buffer-gas flow of 10$^{21}$ atoms/s. The gas moving from the cell through the converging part of nozzle with a subsonic velocity becomes sonic at the nozzle throat where the cross-sectional area (S$^\ast$) is the smallest. Downstream in the divergent part, the gas expands and the stream velocity becomes progressively more supersonic. Provided that during the isentropic and adiabatic gas expansion the sum of the specific enthalpy and the kinetic energy remains constant, a very low gas temperature can be reached. The expansion is axially symmetric. The shape of the divergent part has to be designed to avoid reflection of secondary expansion waves \cite{Atk1995}. The flow at the exit of the nozzle is uniform with parallel streamlines and it is characterized by the final Mach number ($M$), which is defined as the ratio of the stream velocity $u$ to the local speed of sound $a$, i.e.  $M=u/a$. The speed of sound and the stream velocity are expressed as
\begin{equation}\label{num:4}
    a=\sqrt{\frac{\gamma\, k \, T}{m}}\,\,\,\,\,\,\,\,\,\,\,\, \mbox{and}
\end{equation}

\begin{equation}\label{num:5}
    u=\sqrt{\frac{\gamma \, k \, T_0 \,  M^2}{m \, (1+[(\gamma-1)/2]\, M^2)}}\,\, ,
\end{equation}
where $k$  is the Boltzmann constant, $m$ is the mass of the buffer gas atom, $T$ and $T_0$ are the gas temperatures in the jet and in the gas cell, respectively,  and $\gamma=C_p/C_v$ is the ratio of specific heat capacities, which for monatomic gases like argon and helium equals to 5$/$3.
\begin{figure}
\begin{center}
\includegraphics[scale=0.55]{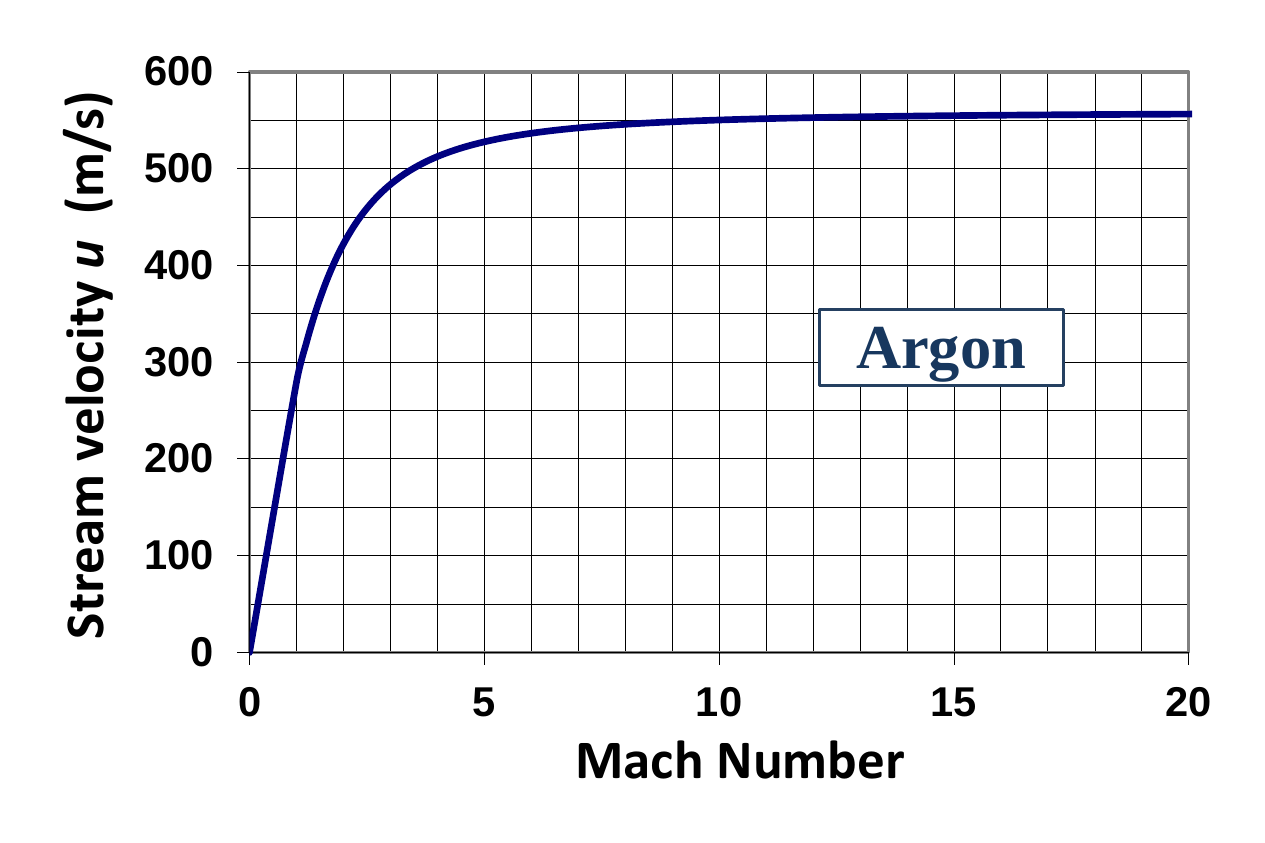} \caption{The argon stream velocity $u$ at the exit of the de Laval nozzle as a function of the Mach number starting with a stagnation temperature $T_0=$ 300 K.}\label{Fig:4}
\end{center}
\end{figure}
The stream velocity for argon as a function of Mach number is shown in Fig. \ref{Fig:4} for a stagnation gas-cell temperature $T_0=$ 300 K. The stream velocity at $M=$ 12 reaches 99\% of its maximum value of 558 m/s. The temperature $T$, the density $\rho$, and the pressure $P$ in the supersonic flow are related to the corresponding values $T_0$, $\rho_0$ and, $P_0$ in the stagnation region of the gas cell via the Mach number as

\begin{equation}\label{num:6}
    \frac{T}{T_0}= [1+\left(\frac{\gamma - 1}{2}\right)\, M^2]^{-1}
\end{equation}

\begin{equation}\label{num:7}
    \frac{\rho}{\rho_0}= [1+\left(\frac{\gamma - 1}{2}\right)\, M^2]^{-\frac{1}{\gamma-1}}
\end{equation}

\begin{equation}\label{num:8}
    \frac{P}{P_0}= [1+\left(\frac{\gamma - 1}{2}\right)\, M^2]^{-\frac{\gamma}{\gamma-1}}\,\, .
\end{equation} 				

\begin{figure}
\begin{center}
\includegraphics[scale=0.55]{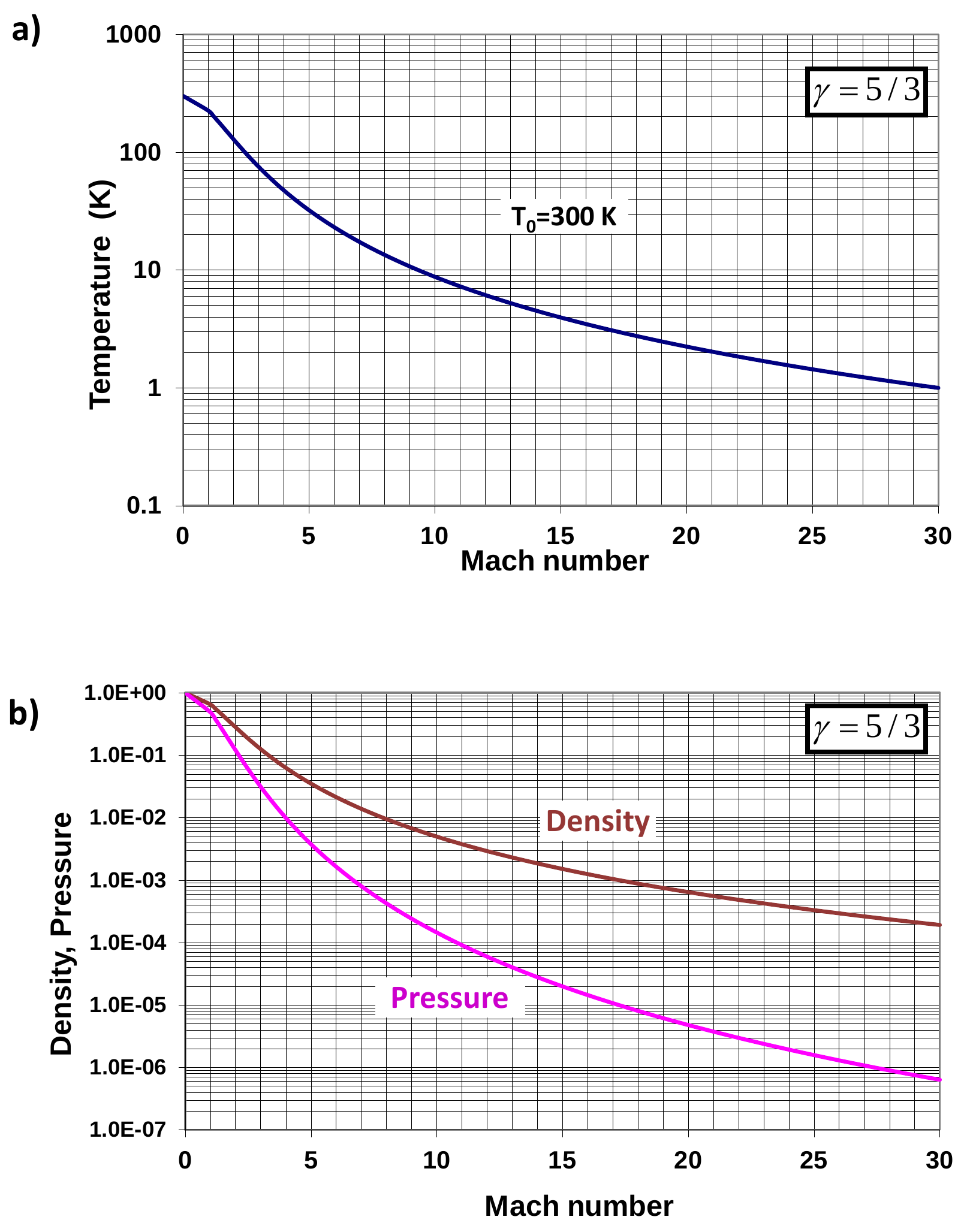} \caption{a) Temperature in the supersonic jet as a function of the Mach number for a monoatomic gas ($\gamma=$ 5/3) at the stagnation temperature $T_0=$ 300 K. b) Density and pressure reduction in the supersonic jet for a monoatomic gas relative to those in the stagnation region as a function of the Mach number.}\label{Fig:5}
\end{center}
\end{figure}
During expansion, the thermal energy of the atoms in the cell is converted into kinetic flow energy. Figures \ref{Fig:5}a,b show the dependence of the temperature, the relative density, and the relative pressure as a function of the Mach number for argon with the stagnation temperature $T_0=$ 300 K.  The gas temperature, density, and pressure drop very fast with increasing Mach number.

\begin{figure}
\begin{center}
\includegraphics[scale=0.55]{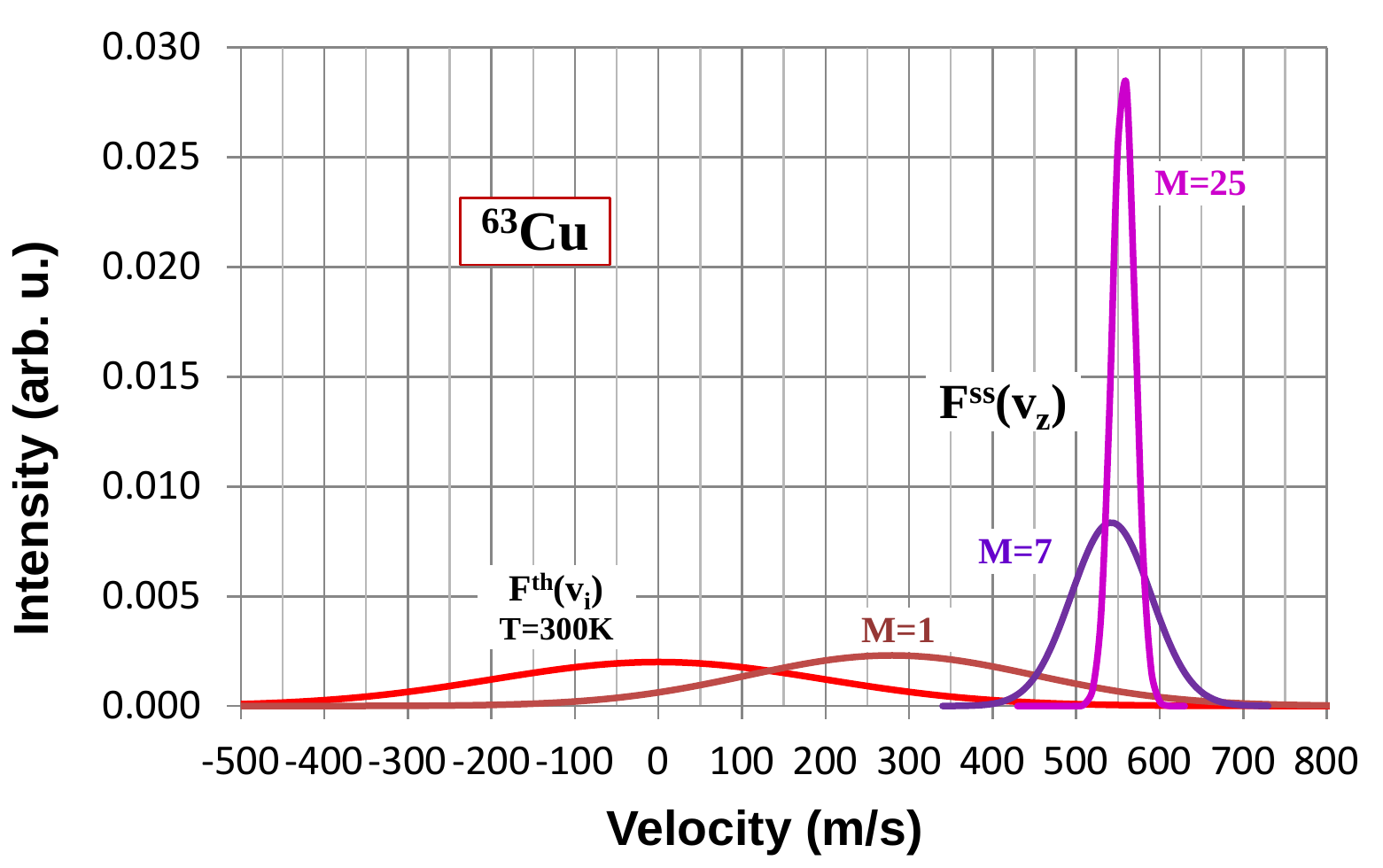} \caption{One-dimensional velocity distribution of $^{63}$Cu atoms in the gas cell (F$^{th}$(v$_i$)) and in the supersonic argon beam (F$^{ss}$(v$_z$)) for different  Mach numbers and a stagnation temperature T$_0=$ 300 K.}\label{Fig:6}
\end{center}
\end{figure}

The radioactive atoms in the cell are in thermal equilibrium with the buffer gas atoms and thus they move with the same flow velocity and have the same temperature.  For laser excitation not the average velocity of the atoms but the velocity distribution components $v_i$ ($i= x, y, z$) are important because they determine the width of the resonance lines. In the gas cell, the one dimensional Maxwell-Boltzmann velocity distribution of atoms $F^{th}(v_i)$  is given by
\begin{equation}\label{num:9}
    F^{th}(v_i)=\sqrt{\frac{m}{2\, \pi\, k\, T_0}}\cdot \exp-\left(\frac{m\, v_i^2}{2\, k\, T_0}\right)\,\, ,
\end{equation}				 		
 with $m$ representing now the mass of the radioactive atom. The three directions $x$, $y$, and $z$ are equivalent. The velocity distribution of the supersonic atomic beam in the direction of the flow ($i= z$) is expressed by the shifted Maxwell-Boltzmann velocity distribution $F^{ss}$ 	
\begin{equation}\label{num:10}
    F^{ss}(v_z)=\sqrt{\frac{m}{2\, \pi\, k\, T}}\cdot \exp-\left(\frac{m\, (v_{z}-u)^2}{2\, k\, T}\right)\,\, ,
\end{equation}
where $u$ is the stream velocity defined by Eqn. (\ref{num:5}). The other two components $v_x$ and $v_y$  of the supersonic velocity distribution are similar to Eqn. (\ref{num:9}) with the temperature $T$ defined by Eqn. (\ref{num:6}). The one-dimensional velocity distributions $F^{ss}(v_z)$  in the supersonic beam for different Mach numbers and for the thermal motion in the gas cell $F^{th}(v_i)$ at $T_0=$ 300 K are shown in Fig. \ref{Fig:6}. The full width at half maximum (FWHM) of these distributions for atoms with mass $m$ depends only on the temperature and can be written as

\begin{equation}\label{num:11}
 \Delta(F)=2\sqrt{\ln 2}\cdot\sqrt{\frac{2\,k\,T}{m}}\,\, .
\end{equation}

The mass flow through the de Laval nozzle is defined by the throat area $S^\ast$. The flow streamlines are all parallel to the nozzle axis and the specified Mach number defines the shape of the nozzle and its exit area $S$. The nozzle can be designed only for the preliminary defined exit temperature $T$, the mass flow and,  the type of gas. The relationship between the exit and the throat diameters can be calculated using the exit to throat areas ratio
\begin{equation}\label{num:12}
   \frac{S}{S^\ast}=\frac{1}{M}\left[\left(\frac{2}{\gamma + 1}\right)\left(1+\frac{\gamma -1}{2}\, M^2\right)\right]^\frac{\gamma+1}{2\,(\gamma-1)}\,\, .
\end{equation}
The nozzle exit diameter for a noble gas and for a throat of 1 mm in diameter results in 4.88 mm and 10.51 mm for Mach numbers 7 and 12, respectively.

In practice, viscous and heat transfer effects from the wall of the nozzle to the gas jet have to be considered, particularly for small-size nozzles. This effects are usually taken into account by splitting the flow into central isentropic core and boundary layer near the wall \cite{Atk1995}.
The flow uniformity outside the nozzle is strongly influenced by the background pressure in the gas cell chamber. Ideally the background pressure has to be equal to the static flow pressure defined by Eqn. (\ref{num:8}). The matching has to be done very accurately since it will define the beam divergence that is crucial for the spectral resolution and for an efficient overlap between the laser beams and the gas jet, as is discussed in the following paragraph.

\section{Laser spectroscopy in a supersonic jet and in a gas cell}

In this section we shall perform a comparison between laser ionization spectroscopy in a gas cell and in a supersonic gas jet in view of the resulting resonance linewidth. The shape of the spectral line is mainly determined by two broadening mechanisms: Doppler broadening and collision broadening. The former is defined by the velocity distribution of atoms and has a Gaussian shape, while the latter follows a Lorentzian distribution. The resulting shape of the spectral line will therefore be a convolution of Lorentzian and Gaussian functions which is known as a Voigt profile or Doppler-broadened Lorentzian.
\begin{figure}
\begin{center}
\includegraphics[scale=0.55]{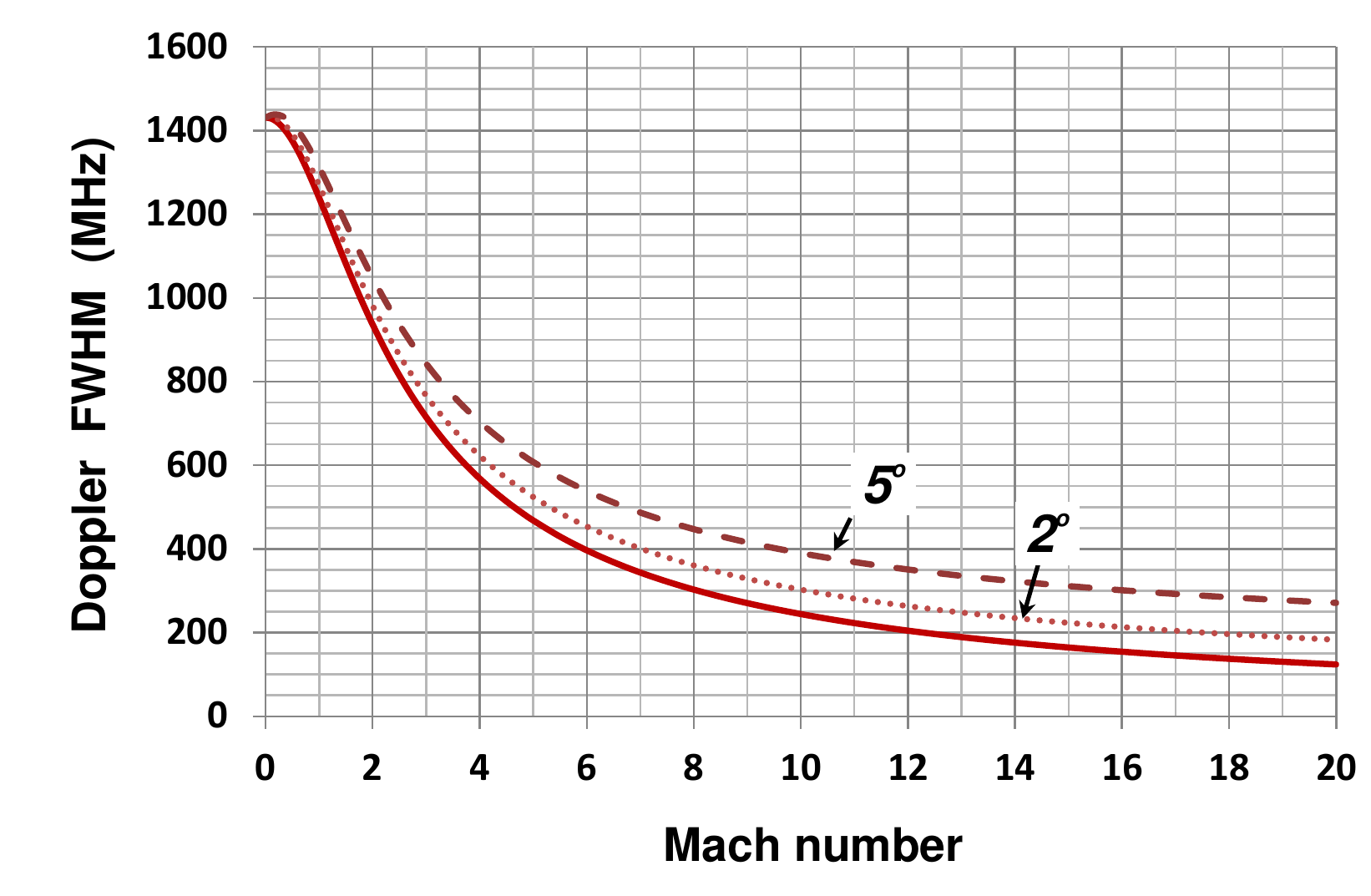} \caption{Doppler broadening (FWHM) of the 327.4 nm 4s$^2$S$_{1/2}$ - 4p$^2$P$_{1/2}$ transition in copper as a function of the Mach number (solid line).  The pointed- and dashed lines show the increase of the linewidth for transverse excitation caused by the jet angular divergence of 2 and 5 degrees, respectively.}\label{Fig:7}
\end{center}
\end{figure}
\subsection{Doppler broadening}
The Gaussian contribution to the line shape associated with the Doppler effect on the Maxwell-Boltzmann velocity distribution of atoms as expressed in Eqn. (\ref{num:9}) is given by
\begin{equation}\label{num:13}
    G(\nu)= G_0 \cdot \exp\left[-\frac{c^2\,(\nu-\nu_{01})^2}{\nu_{01}^2\,\frac{2kT}{m}}\right]\,\, .
\end{equation}
The full width at half maximum of this distribution is
\begin{equation}\label{num:14}
   \Delta\nu_{Doppler}=2\sqrt{\ln 2}\,\frac{\nu_{01}}{c}\,\sqrt{\frac{2\,k\,T}{m}}
\end{equation}
and can be rewritten in the following way
\begin{equation}\label{num:15}
 \Delta\nu_{Doppler}=7.16\cdot10^{-7}\,\nu_{01}\sqrt{T/A}\,\, ,
\end{equation}
where $\nu_{01}$  is the atomic transition frequency in cm$^{-1}$, $T$ is the absolute temperature in K, and $A$ is the atomic mass number. For example, for the 327.4 nm 4s $^2$S$_{1/2}$\,$\rightarrow$\,4p $^2$P$_{1/2}$ atomic transition in copper ($\nu_{01}=$ 30535.3 cm$^{-1}$), the Doppler FWHM at $T=$ 300 K amounts to 0.048 cm$^{-1}$ or 1.43 GHz. In the supersonic beam the Doppler width is diminished owing to the reduced temperature, as given by Eqn. (\ref{num:6}).  The dependence of the Doppler FWHM as a function of the Mach number is illustrated in Fig. \ref{Fig:7} by the solid line. It drops down to 200 MHz for the Mach number $M=$12, which corresponds to a temperature of the atomic beam of 6 K. The effect of the beam divergence is very important for the high Mach number atomic beams in case of transverse direction of the laser beam used for excitation. If the atomic beam is not parallel but has an angle $\theta$ between the stream velocity vector and the beam axis, then an additional broadening
\begin{equation}\label{num:16}
   \Delta\nu_{tr}=\nu_{01}\,u\,\sin\theta/c
\end{equation}
has to be added to the Doppler linewidth given by  Eqn. (\ref{num:14}) owing to the presence of the additional velocity component $u\,\cdot\sin\theta$  in the direction defined by the laser used for excitation. This effect is also shown in Fig. \ref{Fig:6} for the angles $\theta=$2$^\circ$ and 5$^\circ$. The Doppler width is increased from 200 MHz up to 260 MHz or 350 MHz for a beam with Mach number $M=$ 12  with divergence of 2$^\circ$ or 5$^\circ$, respectively. The influence of the beam divergence on the total Doppler width is much smaller if the laser light is directed axially with the atomic beam. In this case only the velocity components along the jet axis $u\cdot\cos\theta$ contribute to the additional broadening. This contribution can be written as
\begin{equation}\label{num:17}
   \Delta\nu_{ax}=\nu_{01}\,u(1-\cos\theta)/c \,\, ,
\end{equation}				 			
which for an atomic beam divergence of 5$^\circ$ results in only 6.4 MHz.

\subsection{Collision and natural broadenings}
Non-resonant collisions of the atoms of interest with the buffer gas atoms cause a shift and a broadening of the spectral lines. The natural broadening associated with spontaneous decay of excited atoms, and the collision-induced shift and broadening are described by a shifted Lorentzian function
\begin{equation}\label{num:18}
    L(\nu-\nu_{01})=\frac{1}{2\pi}\,\frac{\Gamma}{(\nu-\nu_{01}+\Gamma_{sh})^2+(\Gamma/2)^2}\,\, ,
\end{equation}
where $\Gamma=\Gamma_{nat}+\Gamma_{coll}$ represents the FWHM, and with $\Gamma_{sh}$ and $\Gamma_{coll}$ being the line shift- and broadening rates, respectively. The natural linewidth $\Gamma_{nat}$  depends on the atomic transition probability A$_{01}$ between the ground and the excited levels and can be written as
\begin{equation}\label{num:19}
    \Gamma_{nat}=A_{01}/2\pi\,\,\, .
\end{equation}
The shift $\Gamma_{sh}$ and collision $\Gamma_{coll}$ rate parameters are proportional to the density  $\rho$ of the buffer gas in the following way
\begin{equation}\label{num:20}
\Gamma_{coll}=\gamma_{coll}\cdot\rho\,\,\,\,\,\,\,\,\,\,\,\,  \mbox{and}
\end{equation}

\begin{equation}\label{num:21}
\Gamma_{sh}=\gamma_{sh}\cdot\rho\,\, ,
\end{equation}
where the lower case $\gamma_{coll}$ and $\gamma_{sh}$ represent the collision and shift broadening rate coefficients, respectively. They are usually expressed in units of $10^{-20}$ cm$^{-1}$/cm$^{-3}$, which approximately amounts to 8 MHz/mbar at room temperature. For most resonant atomic transitions $\gamma_{sh}$  has a negative sign for argon and a positive sign for helium \cite{All1982}. This means that the resonance is shifted to a smaller frequency (red shift) if argon is used as a buffer gas. For example, the 4s$^2$S$_{1/2}$$\rightarrow$ 4p$^2$P$_{1/2}$ transition at 327.4 nm in copper has a natural linewidth $\Gamma_{nat}=$ 22 MHz   (A$_{01}=$ 1.36$\cdot$10$^8$ s$^{-1}$). The collision broadening rate coefficient $\gamma_{coll}$ in argon for this transition has not been measured but a close look at the literature values for similar transitions seems to indicate a value of about 1.5$\cdot$10$^{-20}$ cm$^{-1}$/cm$^{-3}$.  The Lorentzian contribution to the spectral linewidth $\Gamma$ in the gas cell is mainly defined by the collision broadening and results in 1090 MHz and 5450 MHz for the gas pressure of 100 mbar and 500 mbar, respectively.
\begin{figure}
\begin{center}
\includegraphics[scale=0.55]{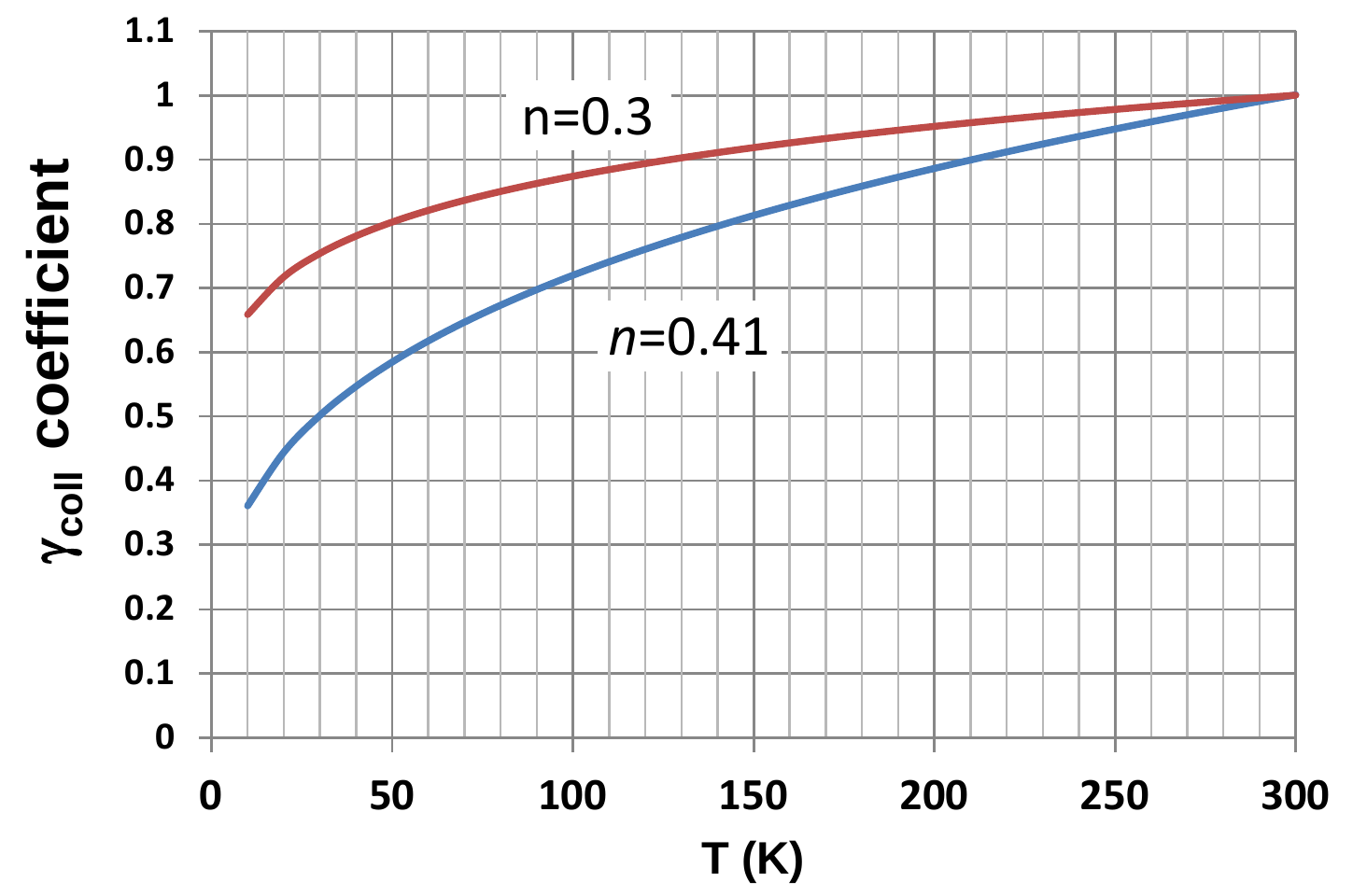} \caption{Normalized collision-broadening coefficient (at T=300K) as a function of the temperature for two temperature dependence indexes of 0.3 and 0.41.}\label{Fig:8}
\end{center}
\end{figure}

In the supersonic jet with Mach number $M=$ 12, the buffer gas density drops down to 0.003 of the stagnation value $\rho_0$ (see Fig. \ref{Fig:5}b) and the collision broadening $\Gamma_{coll}$ for a stagnation gas pressure $P_0=$ 500 mbar equals to 16 MHz. This value is smaller than the natural linewidth  $\Gamma_{nat}=$ 22 MHz, however, estimating the collision effect we assumed that the collision broadening rate coefficient $\gamma_{coll}$  is the same as at T=300 K. In reality, $\gamma_{coll}$  has a power-law dependence on the temperature of the form
\begin{equation}\label{num:22}
\gamma_{coll}\approx T^n\,\,\,\,\,\,\,\,\,\,\,\,  \mbox{with}
\end{equation}

\begin{equation}\label{num:23}
n=\frac{p-3}{2(p-1)}\,\, ,
\end{equation} 	
where $p$ characterizes the type of interaction of the lower and the upper atomic levels with the noble gas \cite{Lwi1977, Hel2001}. For the long-range attractive van der Waals potential ($p=$ 6) the temperature dependence of the pressure broadening coefficient has the form of $\gamma_{coll}\approx T^{0.3}$. In some cases the repulsive part (C$_{12}$R$^{-12}$) of the Lennard-Jons potential ($p=$ 12) correctly describes the temperature dependence of the collision-broadening coefficient as $\gamma_{coll}\approx T^{0.41}$ . This dependence is a more realistic power law for the broadening by light perturbers such as helium, while the $T^{0.3}$ relation is more suitable for heavier perturbers like argon. Experimental values of the temperature dependence index $n$ vary between 0.2 and 0.5 for different spectral lines and can be used as a very crude approximation for the extrapolation of the measured coefficient towards lower temperatures \cite{Bie1989}. Figure \ref{Fig:8} shows the $T^n$ dependence of the collision broadening coefficient for $n=$ 0.3 and $n=$ 0.41. One can safely conclude that by reducing the gas temperature the collision-induced broadening coefficient can only be smaller than that at T$=$ 300 K.

\subsection{Comparison of the spectral resolution in the gas jet and in the gas cell}
\begin{figure}
\begin{center}
\includegraphics[scale=0.55]{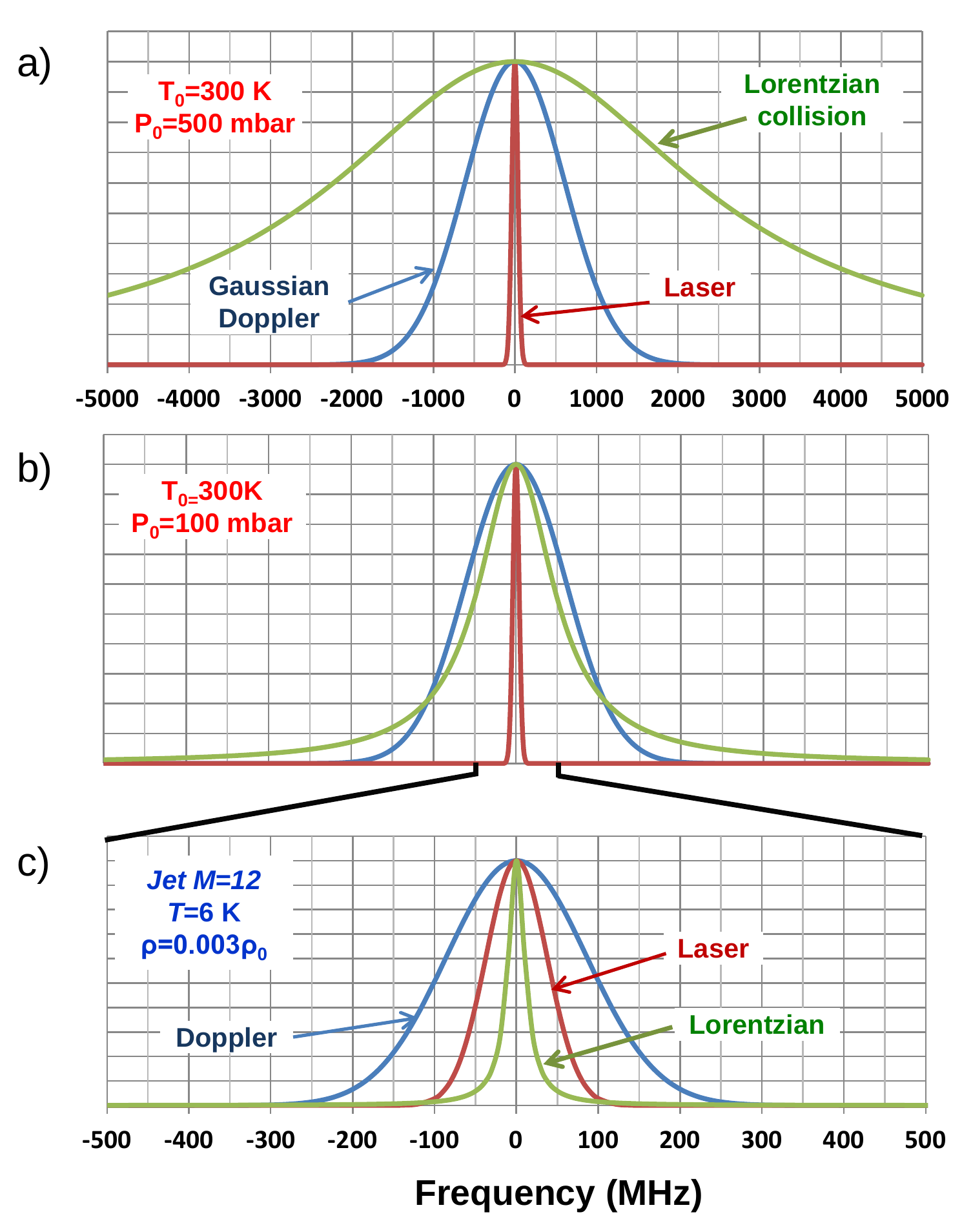} \caption{Gaussian, Lorentzian, and laser contributions to the shape of a spectral line in a) the gas cell $P_0=$ 500 mbar, T$_0=$ 300 K, b) in the gas cell $P_0=$ 100 mbar, T$_0=$ 300 K , and c) in a supersonic gas jet of Mach number M=12, T=6 K, and $\rho=$ 0.003$\rho_0$.}\label{Fig:9}
\end{center}
\end{figure}

The half width of the Voigt profile, which is the convolution of Gaussian and Lorentzian functions, cannot be expressed by an analytical formula. The FWHM of the convoluted shape can be estimated using the following expression \cite{Oli1977}
\begin{equation}\label{num:24}
   \Delta_{Voigt}\approx0.5346\,\Gamma+\sqrt{0.2166\,\Gamma^2+\Delta\nu^2_{Doppler}}\,\, .
\end{equation}
For the 327.4 nm transition in copper the Voigt FWHM in the gas cell at T$_0=$ 300 K results in $\Delta_{Voigt}=$ 2110 MHz and 5830 MHz  for 100 mbar and 500 mbar, respectively. In the supersonic gas jet ($M=$ 12) the Voigt FWHM results in 220 MHz and consists mainly of the Doppler contribution. The different components that contribute to a single spectral line are shown in Figs. \ref{Fig:9}a,b for the gas cell at two different buffer gas pressures and in Fig. \ref{Fig:9}c for the supersonic gas jet with a Mach number $M=$ 12. The spectral linewidth of the laser is also displayed in all figures.

For laser spectroscopy with pulsed laser beams a well suitable way to produce a narrow laser linewidth is by amplification of the light from a continuous wave (CW) single mode laser in a pulsed amplifier. For a Gaussian time profile of the pump laser pulse with a length of $\tau_{pulse}$, the spectral line profile is the Fourier transform and has also a Gaussian shape with a spectral FWHM $\delta_{laser}$ that can be calculated as
\begin{equation}\label{num:25}
   \delta_{laser}=441/\tau_{pulse}\,\, ,
\end{equation}
with $\tau_{pulse}$ given in ns and $\delta_{laser}$ in MHz. A laser pulse of 5 ns length has therefore a spectral bandwidth of 88 MHz, which is smaller than the Doppler width of the $M=$ 12 supersonic beam (Fig. \ref{Fig:9}c). This spectral bandwidth can, however, still provide excitation of the essential part of the atoms in the beam resulting in a high ionization efficiency at resonance.
In our example of the excitation of copper atoms the laser pulse length is shorter than the lifetime of the upper level (7.4 ns),  and in a first approximation the saturation of the atomic transition is defined by the photon fluence (photons/cm$^2$) and can be calculated as $\Phi_{sat}= 1/(2\sigma)$, where $\sigma$ is the atomic transition cross section. However, to ensure the excitation of all atoms in the beam, power broadening has to be involved and the transition has to be oversaturated by a factor of 4.3 \cite{Let1987} to excite all atoms in the inhomogeneous-broadened Doppler line profile. This will cause Loreantzian tails but the resolution will still be defined by the Doppler width (see Sec. \ref{las} for more details). The contribution of different broadening mechanisms to the spectral line broadening in the gas cell and in the supersonic gas jet with Mach numbers 7 and 12 are summarized in Tab. \ref{tab:1}.
\begin{table*}
\centering
\begin{threeparttable}
\begin{tabular}{lccc}
\toprule
\multirow{2}{*}{} & Gas Cell& \multicolumn{2}{c}{Gas Jet} \\
& & Mach 7 & Mach 12 \\ \hline		
Temperature (K)&300&17.3&6 \\
Doppler FWHM (MHz)&1430&344&202\\
Laser spectral width (MHz)&88&88&88\\
Gas density\tnote{a} \,(atoms/cm$^3$)&$2.4\cdot10^{18}$&$3.6\cdot10^{16}$&$7.2\cdot10^{15}$\\
Collision broadening FWHM (MHz)&1086&15&3.3\\
Natural Broadening (MHz)&21.7&21.7&21.7\\
Lorentzian FWHM (MHz)&1108&37&25\\
Voight FWHM (MHz)&2110&364&214\\
\bottomrule		
\end{tabular}
\begin{tablenotes}
\footnotesize
\item[a] Corresponding with a stagnation pressure P$_0=$100 mbar
\end{tablenotes}	
\end{threeparttable}
\caption{P-T parameters and contribution of different broadening mechanisms to the 327.4 nm single spectral line in copper for the gas cell and for the supersonic gas jet.}
	\label{tab:1}
\end{table*}

During the gas expansion of heavy atoms along with the lighter noble gas atoms the translational energy of the first exceeds by tenfold their thermal energy in the gas cell. For example, atoms with atomic mass number $A=$ 100 expanding in a helium jet have an energy of 1.64 eV. Although in collinear photonization spectroscopy much higher energies are used, this acceleration would already affect the line shape \cite{Kau1976, Ant1978, Win1976} and would cause an additional artificial isotope shift \cite{Kud1982}  that has to be taken into consideration while measuring the chain of radioactive isotopes. In contrast to standard collinear spectroscopy where the energy of all isotopes remains constant, in the case of supersonic expansion the velocity of all isotopes is constant. This monokinetization during the gas expansion assures that no additional kinematic isotope shift is produced. A difference between the velocity of the particles under investigation and the buffer-gas velocity, the so-called "velocity slip" effect, has been observed in the supersonic free jet expansion for heavy particles with a big mass difference and seems to be not important for heavy isotopes of the same element \cite{Red1977}.

\subsection{Formation of clusters and chemical reactions}
\begin{figure}
\begin{center}
\includegraphics[scale=0.55]{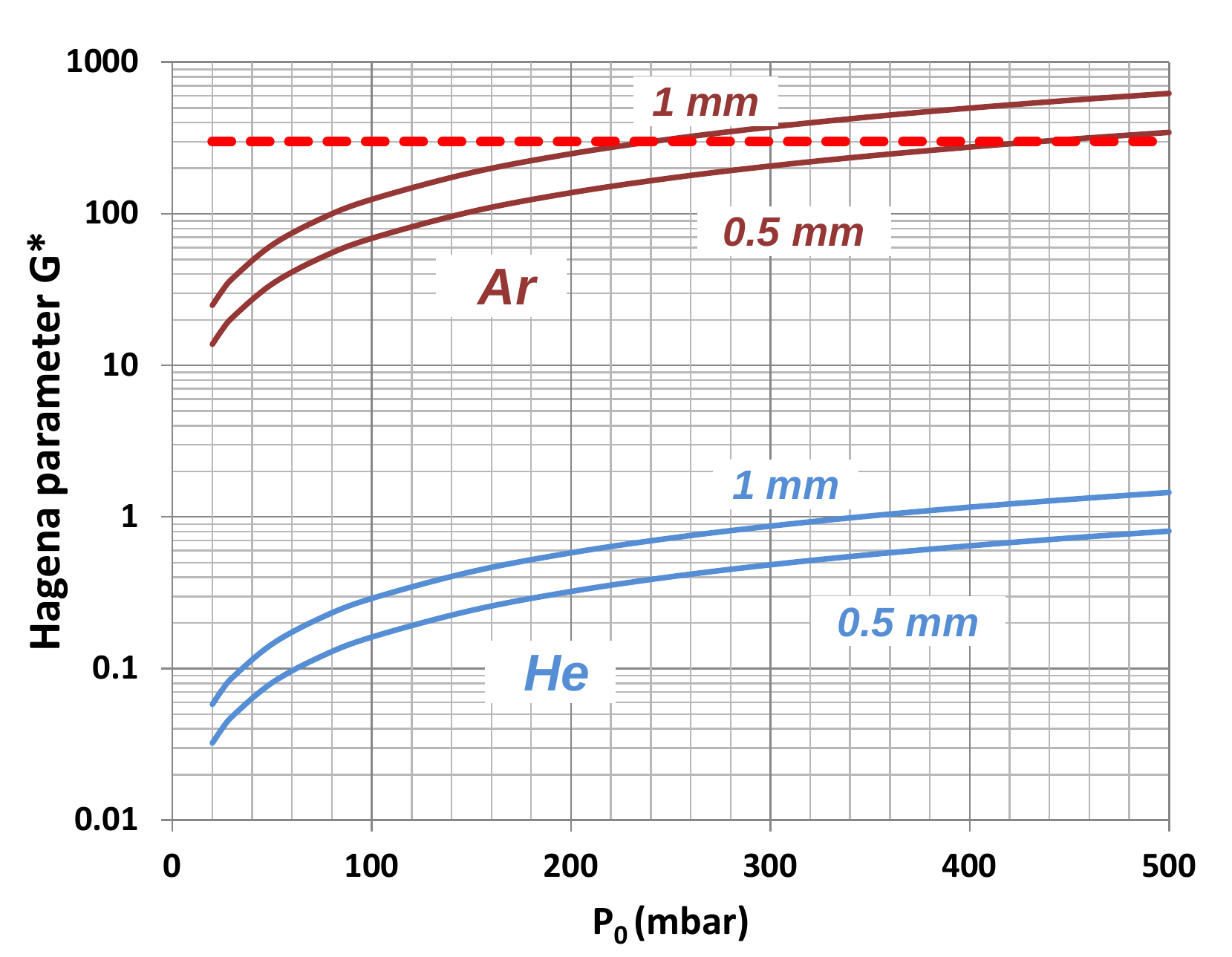} \caption{Hangena parameter $G^\ast$ as a function of the gas-cell pressure $P_0$ for helium and argon at a stagnation temperature $T _0=$ 300 K and a throat diameter of 0.5 mm and 1 mm. The dashed line at $G^\ast=$ 300 indicates the beginning of fast growth of clusters.}\label{Fig:10}
\end{center}
\end{figure}
The expansion of the buffer gas through the nozzle results in a substantial cooling and consequently in small relative atom velocities, hence the interaction between atoms in the beam  can cause formation of dimers, trimers, and even clusters. The dimer formation requires three-body collisions in which the third particle removes the excess of energy of the collision complex. The dimers serve as condensation centers for further growth of clusters. This can happen only in the beginning of the expansion part of the nozzle close to the throat, where the three-body collision probability is high enough.  In principle, formation of multimers consisting only of noble gas atoms does not play an essential role for the laser excitation. However, this process can be important if the concentration of clusters is increased so much that it changes the properties of the gas jet. This nucleation process is related with the van der Waals interaction. The noble gas cluster formation has been studied extensively in many laboratories \cite{Hag1972, Hag1987, Wor1989, Gra2011, Dun2012}. The clustering effect is determined by the temperature $T_{0}$ and the pressure $P_{0}$ in the gas cell, the shape and the size of the nozzle \cite{McD2003}, and the strength of the interatomic bonds. The onset of clustering and the size of created clusters in a free jet can be described by an empirical scaling parameter $G^\ast$ known as the Hagena parameter
\begin{equation}\label{num:26}
   G^\ast=\eta\,\frac{d^{\,0.85}}{T_0^{\,2.29}}\,P_0\,\, ,
\end{equation}
where $d$ is the nozzle diameter in $\mu$m, $T_0$ is in Kelvin, $P_0$ in mbar, and $\eta$ represents the condensation parameter related to the bond formation that results in  3.85 and 1650 for helium and argon, respectively. The dependence of the Hagena parameter as a function of the gas pressure is shown in Fig. \ref{Fig:10} for argon and helium and for a throat diameter of 0.5 mm and 1 mm. The clustering starts when the Hagena parameter $G^\ast>$ 300 \cite{Wor1990}, in such case the average number of atoms per cluster $N_c$ is increased very fast from several atoms at $G^\ast=$ 300 up to 1000 atoms per cluster at $G^\ast=$ 2000. Only for large values of the nozzle diameter a high pressure can cause clustering during the jet formation. The number of atoms per cluster $N_c$ scales as $N_c\sim G^{\ast\,2.0-2.5}$ and is extremely sensitive to the gas temperature in the cell $N_c\sim T_0^{-5}$. Consequently, a small increase of the gas temperature in the case of argon can prevent the formation of big clusters without essentially influencing the Doppler resonance width.

Formation of van der Waals molecules between the atoms of interest and the noble gas atoms causes a reduction of the efficiency. This molecular formation can only happen in the beginning of the expansion, where the three-body collision frequency is high enough. The amount of molecules formed depends on the condensation parameter $\eta$ related to the interatomic bond formation.
The formation of weakly-bound complexes of laser-produced radioactive ions with noble gas- or impurity atoms plays a very important role in the case of laser ionization in the gas cell.  Gas purity at the ppb (part per billion) level is required to minimize the reaction rate of the investigated species in ionic form with the impurity molecules \cite{Kud2001}. In most cases the weakly-bound complexes can be decomposed by an electrical field applied between the gas cell and the RF structure. For in-gas-jet ionization, however, only the loss of the investigated species in atomic form is important. Since the chemical reaction rate for atoms with the impurity molecules is much smaller than that for ions, the requirements on the gas purity can be relaxed. Furthermore, the loss of laser-produced ions due to molecular formation in the jet is smaller than that in the gas cell owing to the low gas density and the short interaction times.

\subsection{Requirements for the pumping system}
\begin{figure}
\begin{center}
\includegraphics[scale=0.55]{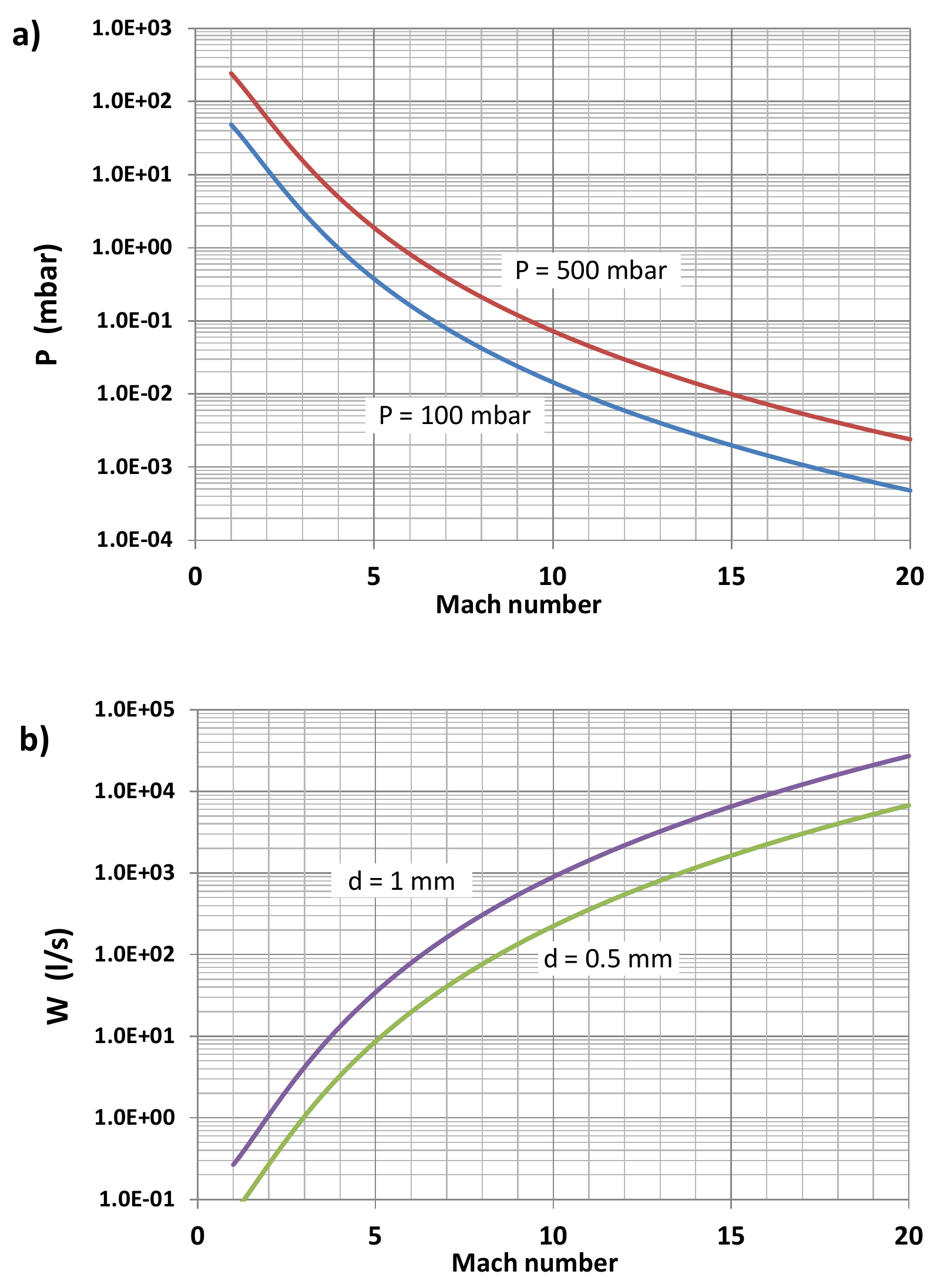} \caption{The required a) pressure $P$ in the gas-cell chamber as a function of the Mach number $M$ for a gas cell $P_0=$ 100 mbar and $P_0=$ 500 mbar at $T_0=$ 300 K and b) pumping speed $W$ as a function of the Mach number $M$ for a nozzle throat diameter of 0.5 mm and 1 mm.}\label{Fig:11}
\end{center}
\end{figure}

To ensure a homogeneous flow of the supersonic gas jet, the pumping system has to be able to provide the required background pressure in the gas cell chamber. The pressure in the gas cell  is defined by the amount of gas needed to stop the energetic reaction products. The gas throughput of the nozzle will then define the transit time of the radioactive atoms in the gas cell and could lead to decay losses for short-lived nuclei. The amount of gas that has to be pumped out depends on the throat area, the pressure, the temperature in the gas cell, and the type of noble gas. As mentioned before, the larger the Mach number one aims for, the smaller the background pressure should be. The volume flow rate of the buffer gas $Q$ (in l/s) with atomic mass number $A$ and stagnation temperature in the gas cell $T_0$ (in K) is given by
\begin{equation}\label{num:27}
   Q=0.052\,d^2\sqrt{\frac{T_0}{A}} \,\, ,
\end{equation}
with the  throat diameter $d$ given in mm. The background pressure in the gas cell chamber $P$ as a function of the required Mach number is obtained using Eqn. (\ref{num:8}), and is shown in Fig. \ref{Fig:11}a for an stagnation argon pressure $P_{0}=$ 100 mbar and 500 mbar. To provide this required pressure, the pumping speed of the vacuum system  $W$ (in l/s) should fulfill that
\begin{equation}\label{num:28}
   W=Q\,\frac{P_0}{P}\,\, .
\end{equation}
The dependence of the pumping speed as a function of the Mach number is shown in Fig. \ref{Fig:11}b for a throat diameter of 0.5 mm and 1 mm. Supersonic beams with a Mach number in the range 5 to 15 require a gas cell chamber background pressure in the range between 2 and 0.001  mbar for a gas cell pressure of 500 mbar, (see Fig. \ref{Fig:11}a). The combination of roots- and big turbo molecular pumps can provide the required pressure range.

To avoid decay losses of the radioactive nuclei, the transit time in the gas cell should be smaller than their half-life. For example, isotopes of $^{100}$Sn produced in the  reaction $^{58}$Ni + $^{46}$Ti $\rightarrow$ $^{100}$Sn + 4n and separated by an in-flight mass separator enter the gas cell with an energy of 88~MeV. They can be stopped in 500 mbar Ar gas at a distance of 12.3 mm  relative to a 4 $\mu$m Mo entrance window with a longitudinal straggling range of 2.8 mm, see Fig. \ref{Fig:3}. The cross section of the $^{100}$Sn beam is about 3 by 4 cm$^2$. The volume of the stopping compartment of the dual chamber gas cell, downstream from the position of the stopped nuclei, together with the laser ionization volume $V_{cell}$  is about 60 cm$^3$. In this case the transit time $T_{tr}$ of nuclei through the cell can be estimated as $T_{tr}=V_{cell}/\,Q$  , where Q is the volume flow rate defined by Eqn. (\ref{num:27}). For an exit orifice diameter of 1 mm, $T_{tr}$ amounts to 0.42 s, which is smaller than the half-life of $^{100}$Sn (T$_{1/2}=$ 0.94 s).

\section{Laser ionization in the jet produced by an axisymmetric spike nozzle}
\begin{figure}
\begin{center}
\includegraphics[scale=0.55]{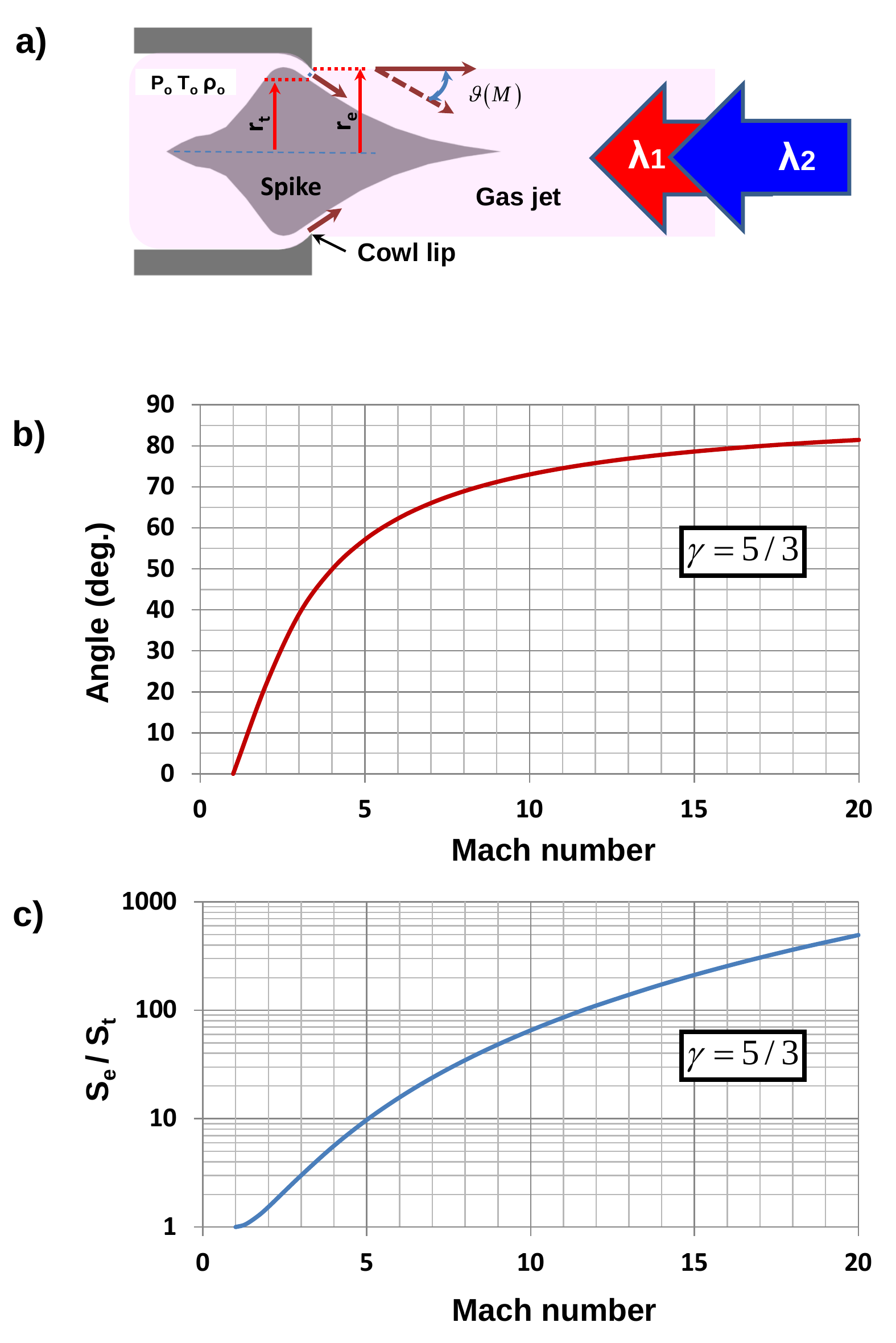} \caption{a) Shape of the spike nozzle and a possible laser beam arrangement for laser ionization and spectroscopy, b) the Prandtl-Meyer expansion fan angle as a function of the Mach number, and c) the exit- to the throat area ratio as a function of the Mach number for monoatomic gases ($\gamma=$ 5/3).}\label{Fig:12}
\end{center}
\end{figure}

The other type of jet that can be used for laser resonance ionization spectroscopy is the one provided by an axisymmetric spike (or plug) nozzle. The geometry of the spike nozzle is shown in Fig. \ref{Fig:12}a. As mentioned earlier usage of this type of nozzle, other than in the field of rocket design, can not be found in the literature. The reason for this might well be in the manufacturing difficulties to provide the tolerances required in smaller scale nozzles to those employed in propulsion engine applications. As an example, the width of the ring slit in a nozzle of 8 mm in diameter giving a throughput equivalent to the 1 mm converging-diverging nozzle would be only 31 $\mu$m. Unlike the de Laval nozzle, the configuration of the spike nozzle allows directing both laser beams for the two-step ionization along the nozzle axis as the laser beams cannot penetrate inside the gas cell. This fact enables excitation and ionization of atoms outside the cell  only, where the gas flow is supersonic. Furthermore, in this beam configuration  one avoids the use of the laser beam expander, which reduces the laser energy density. This can be important if the second step transition is weak and is difficult to saturate  with the available laser power. However, to get a high spectral resolution the ionization should be performed only in the region where a low gas temperature is reached. It should be noted here that this fact could favor the use of a crossed laser beam geometry as well for this kind of nozzles.

In the spike nozzle the gas from the stagnation region is accelerated to sonic speed while moving between two opposite walls, which are coming closer to each other. At the point where one wall ends and the gas expands around its edge, the other wall forms the spike contour. The design approach for the contour of the spike nozzle is based on a Prandtl-Meyer expansion fan around a cowl lip. The optimal contouring of nozzles has been studied by several groups \cite{Kra2007, Rao1961, Gre1961, Ang1964, Lee1963}. In order to get the gas flow with the Mach number $M$ to be parallel to the nozzle axis at its exit, the thruster angle $\theta(M)$ has to fulfill that 	
\begin{equation}\label{num:29}
    \theta(M)=\sqrt{\frac{\gamma+1}{\gamma-1}}\arctan\sqrt{\frac{\gamma-1}{\gamma+1}\,(M^2-1)}-\arctan\sqrt{M^2-1}\,\, .
\end{equation}
This Prandtl-Meyer function is shown in Fig. \ref{Fig:12}b for the ratio of specific heat capacities $\gamma$= 5/3. Similarly as for the de Laval nozzle, the exit- to the throat areas ratio $S_e/S_t$  for the desired Mach number $M$ should be fulfilled (see Eqn. (\ref{num:12})). The required ratio of areas as a function of Mach number is shown in Fig. \ref{Fig:12}c  for $\gamma=$ 5/3.  The throat and the exit areas are defined as
\begin{equation}\label{num:30}
    A_t=\frac{\pi\,(r_e^2-r_t^2)}{cos\,\theta}\,\,\,\,\,\,\,\,\,\,\,\,  \mbox{and}
\end{equation}
\begin{equation}\label{num:31}
    A_e=\pi\,r_e^2\,\,\, ,
\end{equation}
with $r_t$ and $r_e$ representing the radii of the throat and the cowl lip, respectively. Since the Prandtl-Meyer equation is valid only for a planar plug nozzle configuration with a one-dimensional inflow, the method of characteristics is applied for the plug contour definition of  the axisymmetric nozzle \cite{Ang1963, Ono2002}. Despite of the technical limitation found presently to construct a spike nozzle with the suitable dimensions to be used in conventional laser laboratories the authors look forward to the future technological progress that will allow the use of such nozzles in laser spectroscopy experiments.

\section{Laser ionization in a free jet}

\begin{figure}
\begin{center}
\includegraphics[scale=0.55]{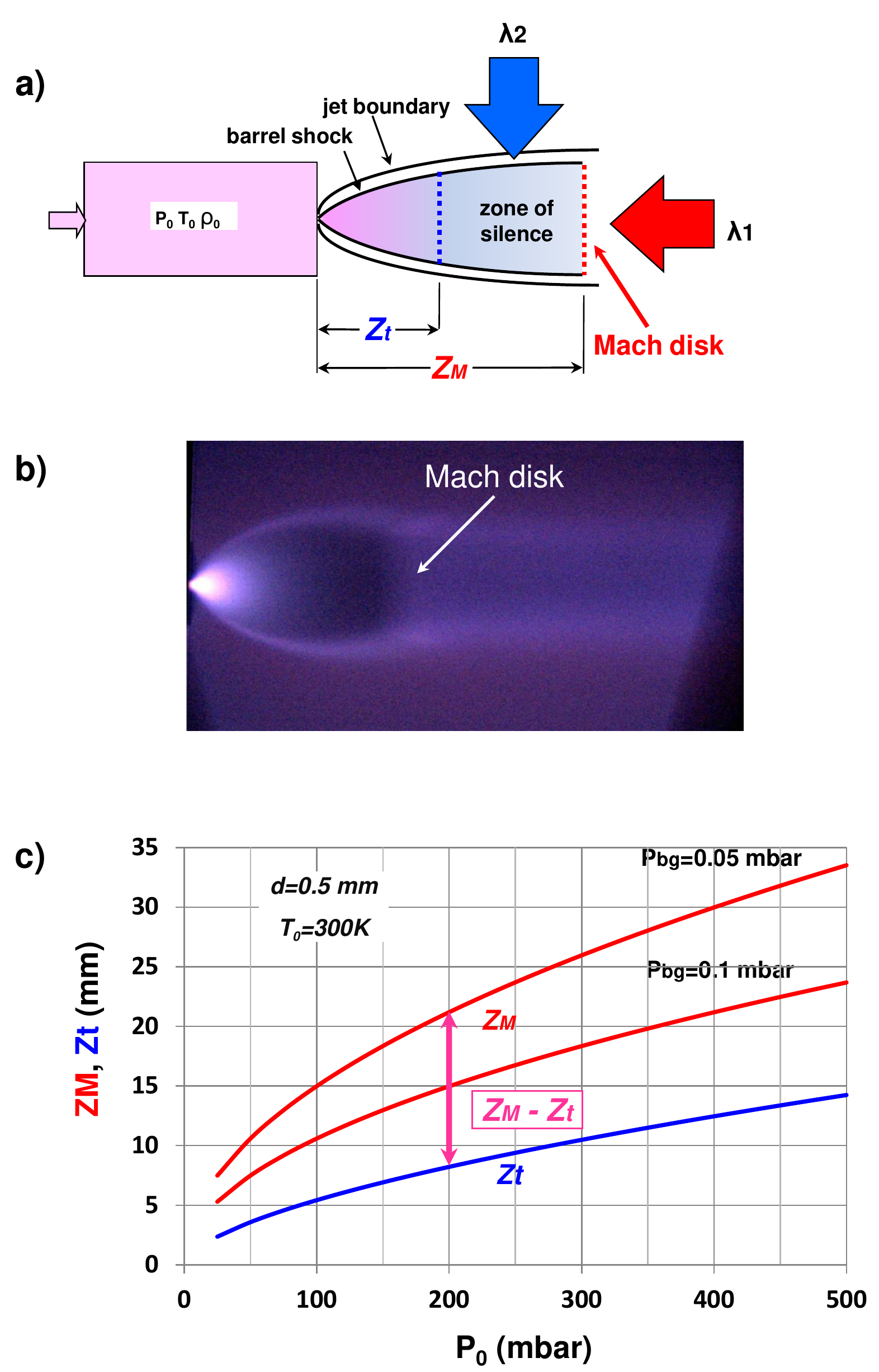} \caption{a) Supersonic free jet expansion in vacuum and a possible arrangement of the laser beams. b) Visualization of an argon jet by the electron beam-induced fluorescence method \cite{Bel2008} using helium as background gas at the pressure ratio of $P_0/P_{bg}=$ 1600 (courtesy of the Department of Aerospace Engineering of the Politecnico di Milano). c) Mach disk position $Z_M$ for a background pressure in the gas cell chamber of 0.1 mbar and 0.05 mbar, and the location $Z_t$ of the terminal Mach number as a function of the argon stagnation pressure in the gas cell for an orifice diameter $d=$ 0.5 mm and  $T_0=$ 300 K.}\label{Fig:13}
\end{center}
\end{figure}

A supersonic free jet can be obtained in the expansion of gas through a round orifice from a high-pressure gas cell into a low-pressure gas cell chamber. The term "free" refers to the absence of external surfaces that restrict the gas expansion, as e.g. in the de Laval or spike nozzles.  The properties of free jets have been investigated in detail \cite{And1974, Com1984, Mil1988}. During the gas expansion two types of shock zones are developed, see Fig. \ref{Fig:13}a. A barrel shock is formed around the center line of expansion starting from the exit orifice. This expansion terminates at the second shock zone, referred to as the Mach disk, which is perpendicular to the centerline of the beam. Currently there exist many techniques based on electron beam-induced fluorescence \cite{But1974, Bel2008} and light-induced scattering- and fluorescence \cite{Cle2002, Moh2007, Nar2011} techniques  that allow visualization of the free-jet shock structures.
Figure \ref{Fig:13}b illustrates the visualization of an argon jet in a helium background with a pressure ratio of $P_0/P_{bg}=$ 1600 and a terminal Mach number $M=$ 32. The analysis of images obtained through electron beam-induced fluorescence permits accurate density measurements that are important for detailed studies of the barrel shock and Mach disk morphology \cite{Bel2010}. The location of the Mach disk depends only on the ratio between the stagnation gas cell pressure $P_0$ and the background pressure $P_{bg}$ in the gas cell chamber.  The Mach disk distance expressed relative to the orifice diameter $d$ can be written as
\begin{equation}\label{num:32}
    \frac{Z_M}{d}=\,0.67\sqrt{\frac{P_0}{P_{bg}}}\,\, .
\end{equation}
Notice that $Z_M$ is not sensitive to the ratio of specific heats $\gamma$. The thickness of the Mach disk is of the order of the local mean free path and depends on the background pressure. The diameter of the Mach disk is more difficult to correlate since it depends on both $P_0/P_{bg}$ and $\gamma$. For an argon jet it is of the order of 0.45$Z_M$ \cite{Ash1966}. The position of the Mach disk for an orifice diameter of 0.5 mm and a stagnation temperature $T_0=$ 300 K is shown in Fig. \ref{Fig:13}c as a function of the argon stagnation pressure $P_0$ for the background pressure 0.05 mbar  and 0.1 mbar. The core of the expansion, limited by the barrel- and the Mach disk shocks, is isentropic and its properties do not depend on $P_{bg}$. The expanding gas can be considered as ideal and heat conduction and viscous effects can be neglected. In the beginning of the expansion, where the flow is continuous, the gas temperature, pressure, and density are described by the same equations (Eqns. (\ref{num:6}-\ref{num:8})) as for the Laval nozzle.
\begin{figure}
\begin{center}
\includegraphics[scale=0.55]{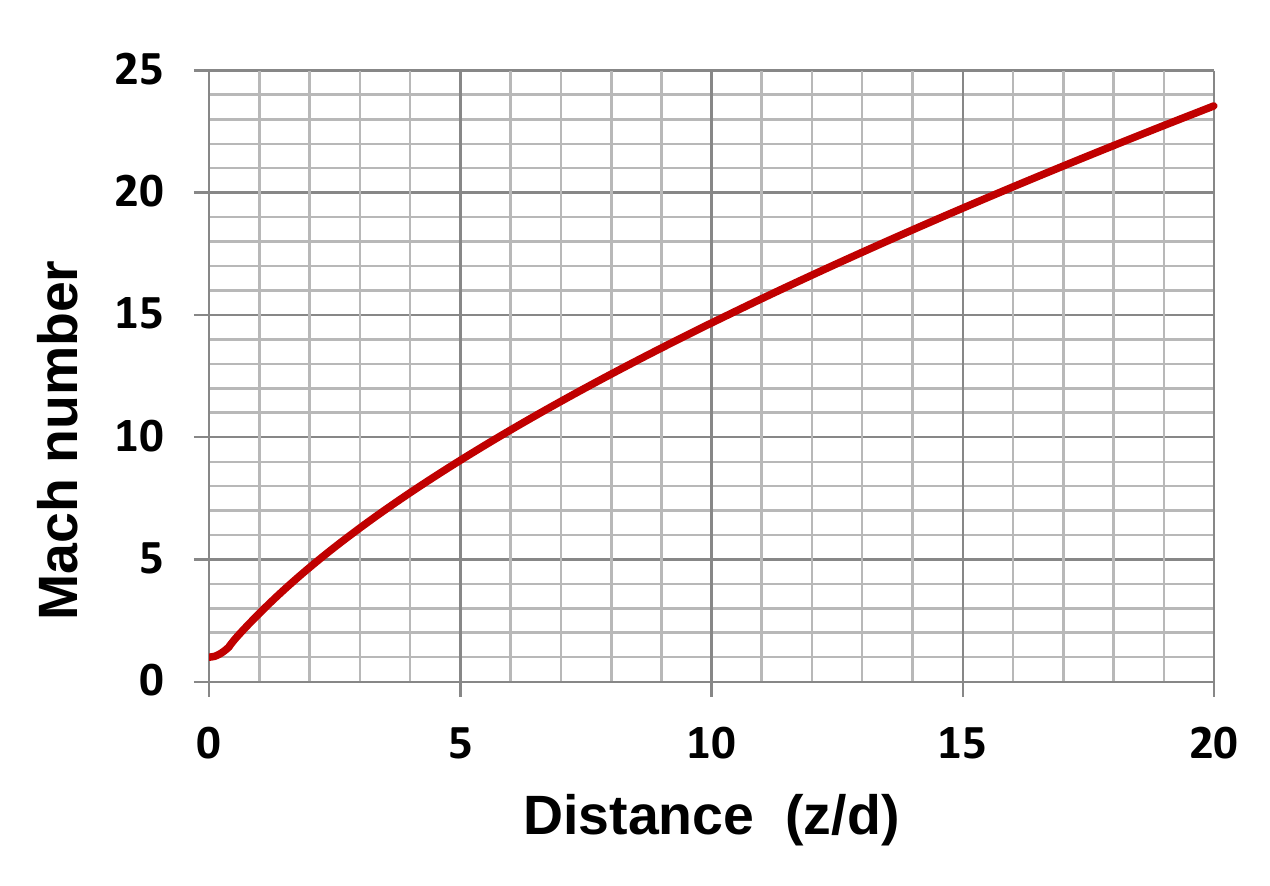} \caption{Mach number in the free-jet expansion for a monoatomic gas ($\gamma=$ 5/3) as a function of the distance from the exit orifice measured in units of the exit orifice diameter.}\label{Fig:14}
\end{center}
\end{figure}
The gas undergoes isentropic wall-free expansion and the collision rate, responsible for the cooling, falls rapidly with increasing distance from the jet orifice. At some point in the expansion the collision rate is too low to provide continuum flow and the transition to a free-molecular expansion begins. At this point the axial velocity distribution and the Mach number is getting frozen. This terminal Mach number $M_t$ is defined by the total number of collisions that atoms undergo during the continuum expansion  and can be calculated for argon as \cite{And1966, Lub1982}
\begin{equation}\label{num:33}
    M_t=\,3.32\,(P_0\,d)^{0.4}\,\, ,
\end{equation}
where $P_0$ is the stagnation gas cell pressure in mbar and $d$ is the orifice diameter in mm. The distance in the orifice diameters at which this terminal  Mach number is reached can be calculated in first approximation  as
\begin{equation}\label{num:34}
    \frac{Z_t}{d}=\left(\frac{M_t}{3.26}\right)^{1.5}\,\, .
\end{equation}
This distance is also shown in Fig. \ref{Fig:13}c as a function of the argon pressure $P_0$ for $d=$ 0.5 mm and $T_0=$ 300 K. Since the distance $Z_t$ does not depend on the background pressure it is possible to increase the difference $Z_M-Z_t$ to allow laser ionization in the so-called 'zone of silence' \cite{Mil1988}. In this case a minimum in the Doppler broadening can be obtained. In the example of Fig. \ref{Fig:13}c  the terminal Mach number $M_t$ amounts to 21 at the stagnation pressure of 200 mbar and $Z_M-Z_t=$ 13 mm. The temperature corresponding to $M=$ 21 is equal to 2 K and the Doppler broadening at this temperature is much smaller than that associated with the divergence of the supersonic beam. To estimate the influence of the divergence on the spectral linewidth one should know the flow-field properties of the free jet.

The variation of the centerline Mach number $M$ as a function of the distance from the exit orifice $z/d$ for $z/d>$ 2.5 is given by the following formula \cite{Ash1966, And1972}

\begin{equation}\label{num:35}
    M=\,B\left(\frac{z-z_0}{d}\right)^{\gamma-1}-\frac{\frac{1}{2}\left(\frac{\gamma+1}{\gamma-1}\right)}{B\,\left(\frac{z-z_0}{d}\right)^{\gamma-1}}\,\, .
\end{equation}
This equation describes the expansion as spherical with streamlines starting as a point source located at $z_0/d$. A better fit of the central line Mach number at smaller distances can be performed using the following formulas \cite{Mur1984, Mil1988}
\begin{equation}\label{num:36}
    M=\left(\frac{z}{d}\right)^{\gamma-1}\left[C_1+\frac{C_2}{(\frac{z}{d})}+\frac{C_3}{(\frac{z}{d})^2}+\frac{C_4}{(\frac{z}{d})^3}\right] \,\,\,\,\, \mbox{for}\,\,\, \frac{z}{d}>0.5\,\,\,\,\, \mbox{and}
\end{equation}

\begin{equation}\label{num:37}
    M=\,1.0+D\,\left(\frac{z}{d}\right)^{-2}+E\,\left(\frac{z}{d}\right)^3\,\,\,\,\, \mbox{for}\,\,\, 0<\frac{z}{d}<1.0\,\,\,.
\end{equation}
Figure \ref{Fig:14} shows the dependence of the central line Mach number as a function of the distance from the exit orifice as given by Eqns. (\ref{num:36}) and (\ref{num:37}). Already at the distance of 6 exit orifice diameters a Mach number greater than 10 is reached. The parameters used in Eqns. (\ref{num:35}-\ref{num:37}) are given in Tab. \ref{tab:2}.

\begin{table*}
\centering
\begin{tabular}{ccccccccc}
\toprule
$\gamma$&$Z_0/d$&B&$C_1$&$C_2$&$C_3$&$C_4$&$D$&$E$\\ \hline
5/3&0.075&3.26&3.232&-0.7563&0.3937&-0.0729&3.337&-1.541\\
\bottomrule		
\end{tabular}
\caption{Parameters employed in the calculations of the centerline Mach number for the axisymmetric free jet flow in Eqns. (\ref{num:35}) through (\ref{num:37})}
	\label{tab:2}
\end{table*}

\begin{figure}
\begin{center}
\includegraphics[scale=0.55]{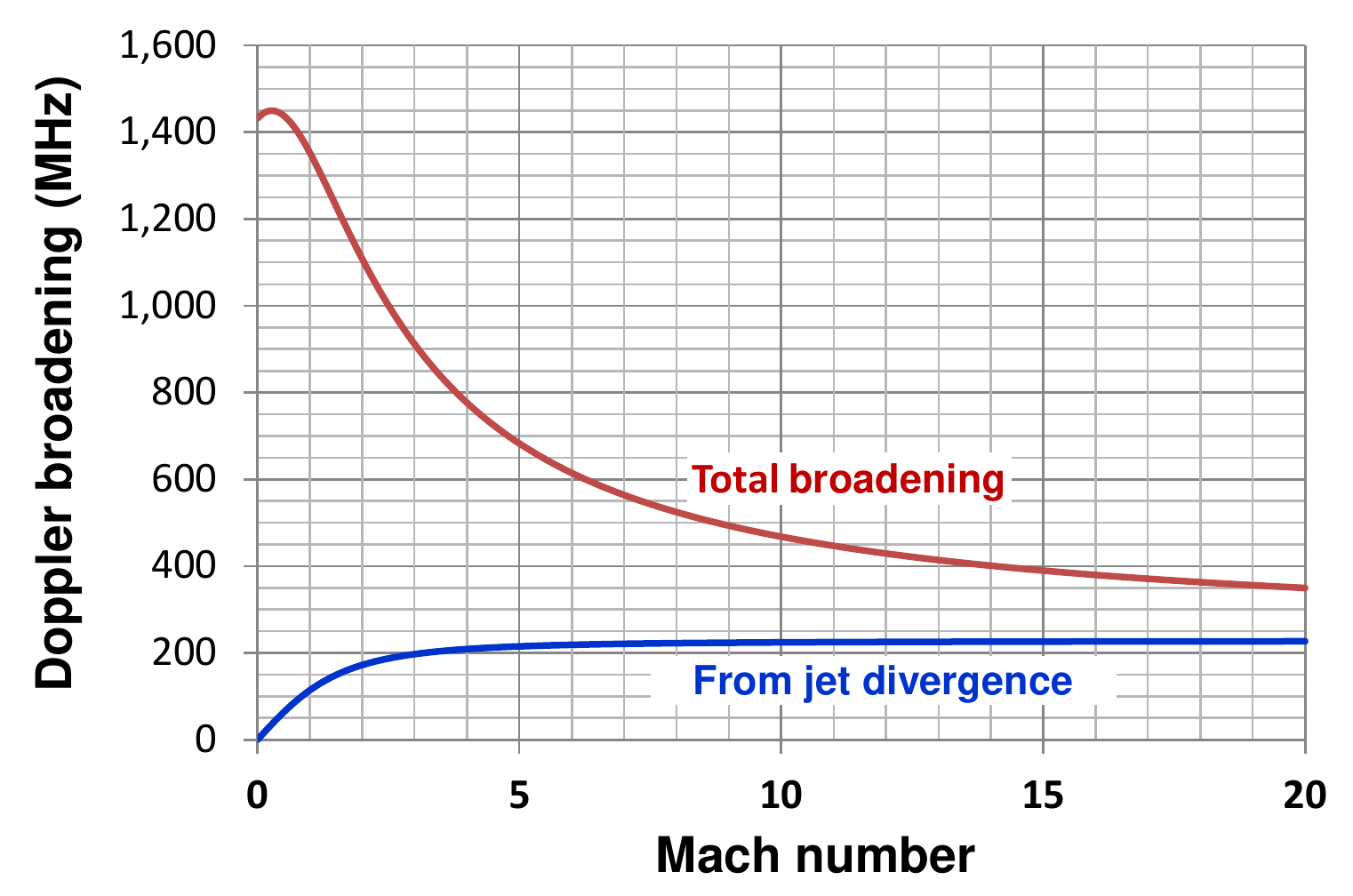} \caption{Total Doppler broadening in the supersonic free jet of the 327.4 nm 4s$^2$S$_{1/2}$$\rightarrow$4p$^2$P$_{1/2}$ transition in copper. The contribution to the total broadening caused by the divergence of the jet is also shown.}\label{Fig:16}
\end{center}
\end{figure}

The atom density distribution for directions perpendicular to the jet axis is given by

\begin{equation}\label{num:38}
    \frac{\rho(y,z)}{\rho(0,z)}=\cos^2\theta\cdot\cos^2\left(\frac{\pi\,\theta}{2\,\phi}\right)\,\,\,\mbox{and}\,\,\,\frac{\rho(R,\theta)}{\rho(R,0)}=cos^2\left(\frac{\pi\,\theta}{2\,\phi}\right)\,\, ,
\end{equation}
where $\tan\theta=\frac{\mbox{y}}{z}$ and $R^2=z^2+y^2$.
\begin{figure}
\begin{center}
\includegraphics[scale=0.55]{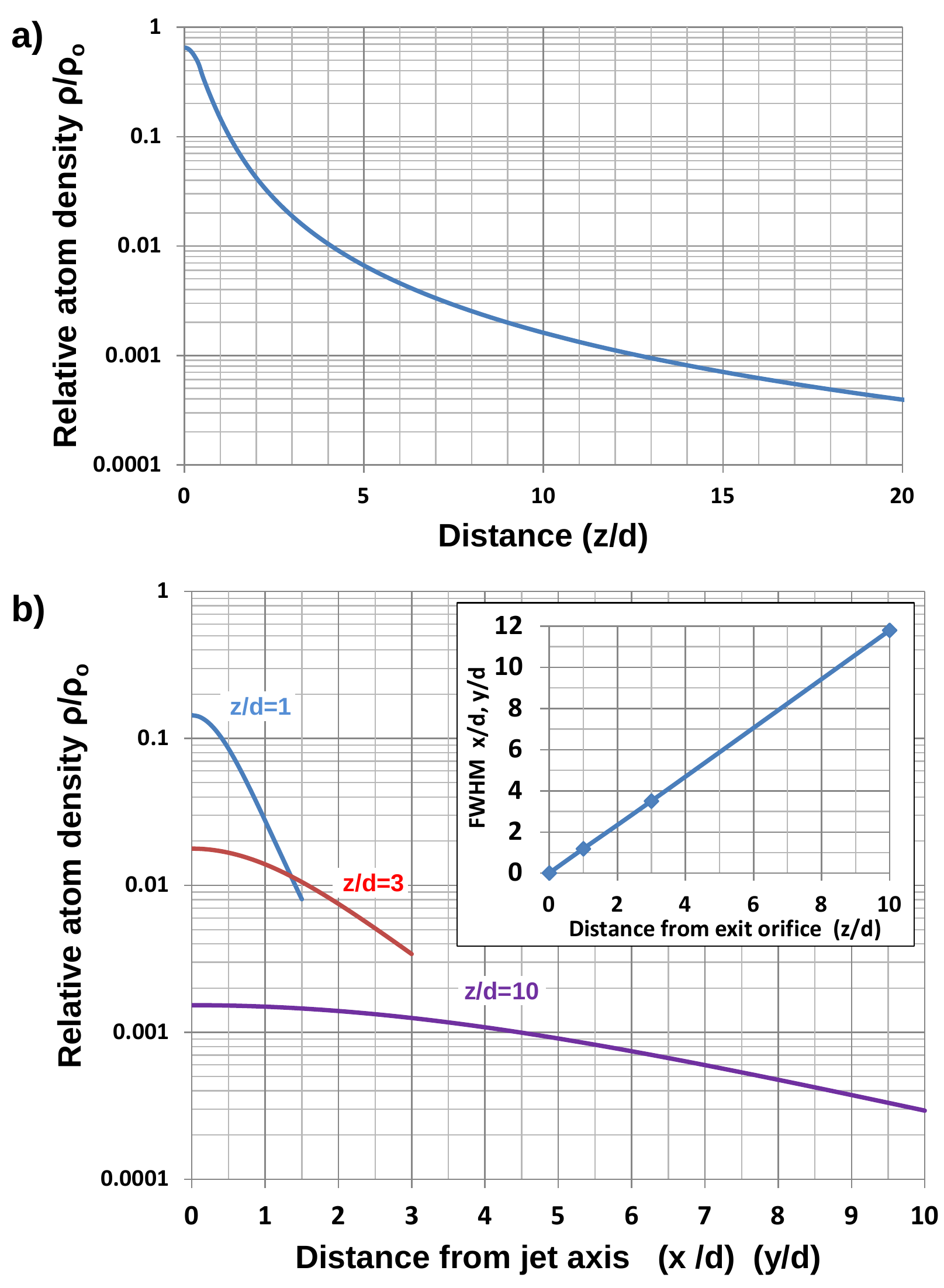} \caption{a) The centerline atom density in the supersonic free jet normalized to the density in the stagnation region is shown as a function of the distance from the exit orifice. b) Atom density in the direction perpendicular to the jet axis normalized to the density in the stagnation region as a function of the distance from the jet axis, the inset shows the FWHM of the atom-density distribution as a function of the distance from the exit orifice. All distances are give relative to the exit orifice diameter and the value of the parameter $\phi$ in Eqn. (\ref{num:38}) amounts to 1.365  for $\gamma=$ 5/3.}\label{Fig:15}
\end{center}
\end{figure}
The dependence of the centerline atom density as a function of the distance from the exit orifice is shown in Fig. \ref{Fig:15}a and the off-axis atom density distribution at different distances from the orifice in Fig. \ref{Fig:15}b. The off-axis distance is also defined in units of the orifice diameter. The inset in Fig. \ref{Fig:15}b displays the full width at half maximum of the atom density distribution as a function of the distance from the orifice. From this linear dependence one can determine the angle of the jet relative to the central beam axis to be equal to 30.5$^\circ$. This angle is obviously much larger than that obtained using the de Laval nozzle. Nevertheless, by using the axial direction for the first step laser beam one can obtain a reasonable spectral resolution as given by Eqn. (\ref{num:17}). In Fig. \ref{Fig:16} the contributions of the jet divergence ($\theta=$ 30.5$^\circ$) to the resonance width and of the total Doppler broadening for the 327.4 nm line in copper are shown as a function of the Mach number. For $M=$ 11 the contribution due to the divergence results in about 225 MHz, while the total Doppler broadening amounts to 440 MHz.

\section{Requirements for the laser system}
\label{las}
To ensure an efficient excitation and subsequent efficient ionization of radioactive atoms the laser system should provide temporal and spatial overlap of the pulsed laser beams with the radioactive atoms in the supersonic jet and, in addition, the laser energy density should be high enough to saturate both transitions.  These conditions can be fulfilled more easily for the de Laval and the spike nozzles because they can produce long and well-collimated jets. A full  temporal overlap is guaranteed if the laser pulse repetition rate is high enough to irradiate all atoms in the continuous atomic jet; for this to happen the condition given by  Eqn. (\ref{num:1}) should be fulfilled. For the spatial overlap the dimensions of the laser beams should not be smaller than the size of the gas jet. For two-step ionization processes via a short-lived auotioinizig state or through the continuum, the duration of the laser pulses of both steps and the time delay between them should be shorter than the relaxation time of the population of the intermediate level. At the same time, to get ionized the maximum fraction of atoms, the energy fluence (photons/cm$^2$) of the  pulse $\Phi_1$, $\Phi_2$ must satisfy respectively the conditions

\begin{equation}\label{num:39}
   \Phi_1\geq\Phi_1^{sat}=(2\sigma_{exc})^{-1}\,\,\,\,\,\,\,\,\,\,\,\,  \mbox{and}
\end{equation}

\begin{equation}\label{num:40}
   \Phi_2\geq\Phi_2^{sat}=\sigma_{ion}^{-1}\,\, ,
\end{equation}
with $\Phi_1^{sat}$, $\sigma_{exc}$ and $\Phi_2^{sat}$, $\sigma_{ion}$ representing the saturation energy fluence and cross sections for the excitation and ionization steps, respectively \cite{Let1987}. The saturation energy fluence for the resonant transition is very low and can be easily obtained. For the excitation cross section in the range of 10$^{-10}$-10$^{-12}$ cm$^{-2}$ it is found to be between 3-300 nJ/cm$^2$ ($\lambda_1=$ 330 nm). Equation (\ref{num:39})
is valid for a laser bandwidth  $\delta_{laser}$ of the first step smaller than the homogeneous-broadened spectral line $\Gamma_{nat}$ (Eqn. (\ref{num:19})). If the laser bandwidth given by Eqn. \ref{num:25} is bigger than $\Gamma_{nat}$, the saturation energy fluence should be increased by a factor  $\delta_{laser}/\Gamma_{nat}$. The width of the resonance in a supersonic jet is mainly determined by the Doppler broadening (Eqns. (\ref{num:14},\ref{num:15})). The ideal condition for the first excitation step would be the equality of the laser and the Doppler width. For the Doppler linewidth shown in Fig. \ref{Fig:9}c this condition can only be satisfied for Mach numbers greater than 20, see Fig. \ref{Fig:7}.  The only way to excite all atoms in an inhomogeneous-broadened Doppler line for smaller Mach numbers is to oversaturate the transition by increasing the energy fluence. The disadvantage of this is the presence of Lorentzian tails in a power-broadened spectral line. The energy fluence condition for the second step (Eqn. \ref{num:39}) is more easily  fulfilled for the ionization via an autoionization level owing to the higher cross section in comparison to non-resonant ionization into the continuum. The laser bandwidth should be smaller than the linewidth of the autoionizing transition, which is usually bigger than 1 cm$^{-1}$. For a second-step transition cross section of $1\cdot10^{-15}$ cm$^{-2}$, the saturation energy fluence results in 0.4 mJ/cm$^2$  ($\lambda_2=$ 500 nm).

\section{Experimental proof-of-principle of laser spectroscopy in a free jet.}
The proof-of-principle for the in-gas-jet laser ion source using a free jet has been demonstrated at the Leuven Isotope Separator On-Line (LISOL)  in a series of off-line experiments. In order to perform these tests, the front end of the mass separator had to be modified with the incorporation of a 90$^\circ$  bent RFQ  to allow for the possibility of using a cross laser beam geometry with one of the laser beams counterpropagating to the atomic jet, see Fig. \ref{Fig:17}a. In previous experiments \cite{Son2009, Fer2012} either the laser beams passed first through the gas cell or were both sent transversally to the jet, thus limiting the performance of the technique.



\subsection{Experimental setup}
\begin{figure}
\begin{center}
\includegraphics[scale=0.55]{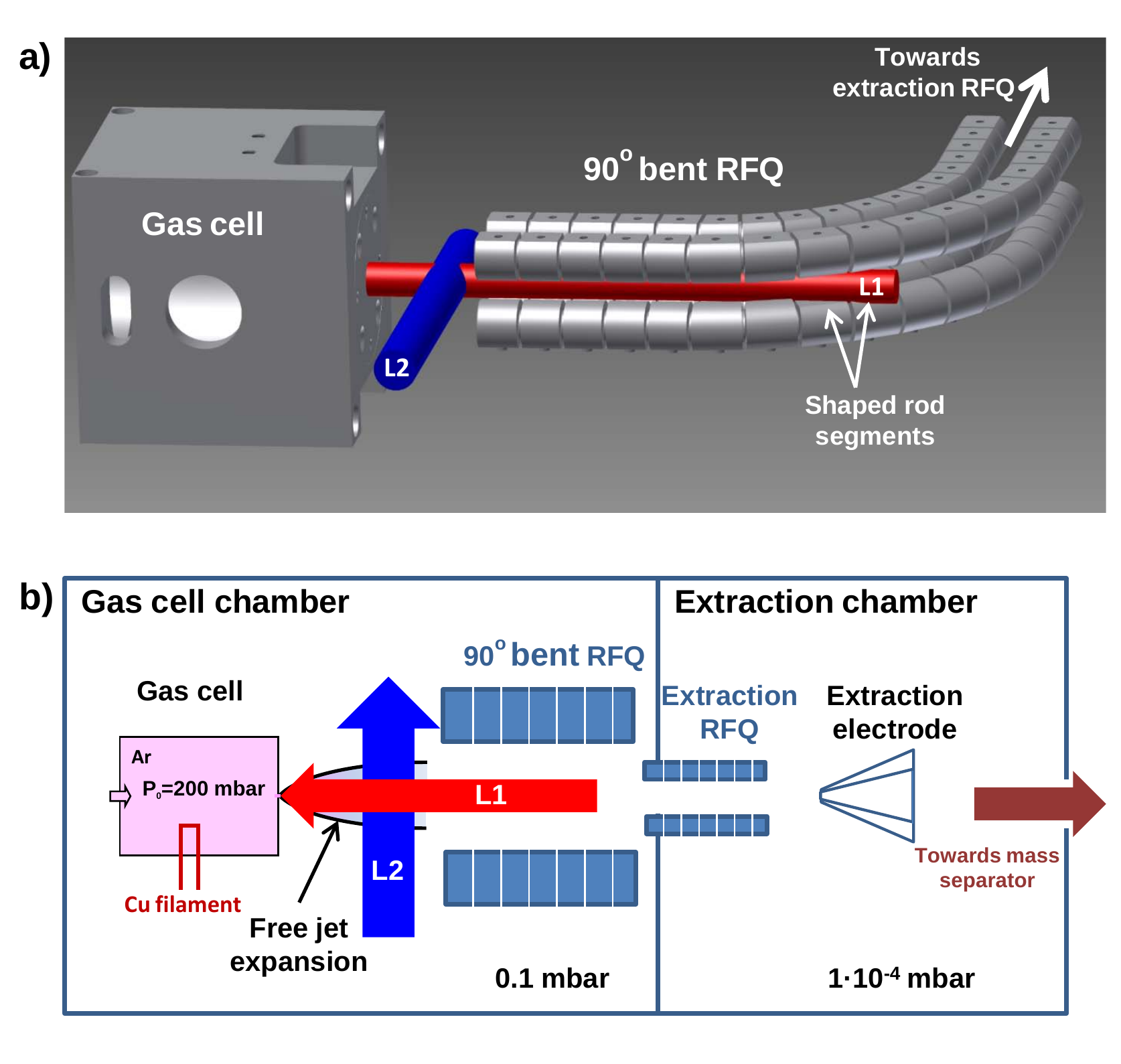} \caption{a) 3D view of the laser beams defining the ionization region and of the 90$^\circ$ bent RFQ ion guide. b) Layout of the experimental setup employed at LISOL for the high-resolution laser resonance ionization spectroscopy measurements performed in the supersonic free jet with a crossed laser beam (L1 and L2) geometry. }\label{Fig:17}
\end{center}
\end{figure}

In these tests the standard LISOL gas cell for fusion-evaporation reactions \cite{Fac2004} was placed in the gas cell chamber, see layout in Fig. \ref{Fig:17}b. Argon as buffer gas was supplied into the cell after additional purification to sub-ppb level in a getter-based purifier. Copper atoms from a resistively-heated filament were seeded into the argon gas. Argon expanded from the gas cell into the gas-cell chamber through the nozzle, which consisted of a sharp-edged orifice of 1 mm in diameter with a flat surface at the outer side of the nozzle flange and a spherical surface of radius $r=$ 5 mm at the gas cell side.

The segmented 90$^\circ$  bent RFQ ion guide has an entrance- and exit linear parts, and a curved part with a central radius of 60 mm and an inter-rod spacing (inscribed diameter) of 12 mm. The rod segments are 9 mm long, 12 mm in diameter and are separated by 1 mm gaps in the axial direction. The distance between the exit orifice and the first segment of the entrance part of the RFQ is 8 mm. The design of the RFQ ion guide allows to send the first step laser beam between the rods counterpropagating with the supersonic gas jet.  Four of the segments in the curved part are especially shaped to let the full laser beam interact with the jet. A laser beam diameter of up to 8 mm can be inserted through these segments towards the ionization region, see Fig. \ref{Fig:17}a. The second step laser beam was sent perpendicular to the jet axis. The crossing of the two laser beams and the gas jet defined the zone of selective laser ionization. After the 90$^\circ$ bending the ions were transferred into a smaller RFQ structure (inter-rod spacing of 4 mm,  length of 4 mm, and diameter of the segments of 4 mm) acting as a pumping barrier that transported the ions into the extraction chamber Fig. \ref{Fig:17}b. The combination of the LISOL pumping system \cite{Kud2003} with the differential pumping element provided by the extraction RFQ resulted in a pressure suppression factor between the gas cell chamber and the extraction chamber of three orders of magnitude. Finally, the extracted ions were accelerated to an energy of 40 keV and transported to a dipole magnet, where those ions with an A/Q$=$ 63 were mass selected and subsequently detected by a Faraday cup or by a Secondary Electron Multiplier (SEM). The background pressure in the gas cell chamber, which defines the size of the jet, could be precisely adjusted to a value of 0.1 mbar by changing the pumping capacity of the roots pump system by means of a movable shutter.

The copper atoms in the supersonic free gas jet were ionized in a two-step process according to Fig. \ref{Fig:18}a. The ground state 3d$^{10}$4s $^2$S$_{1/2}$  atoms, excited  by the 327.4 nm first step laser beam to the intermediate  3d$^{10}$4p $^2$P$_{1/2}$ level, were further excited by the 287.9 nm second step laser beam to the 3d$^9$4s5s $^2$D$_{3/2}$  autoionizing state leading to ionization. The laser setup employed is similar to that described in \cite{Kud1996} with the only exception that for these tests a narrow- band laser  was used for the first excitation step. To accomplish this, a single mode tunable laser beam of 654.8 nm delivered by a continuous wave (CW) diode laser (Ta-pro, Toptica Photonics)  was amplified in a two-stage pulsed dye amplifier. To get the required radiation at 327.4 nm, the amplified light was frequency-doubled in a second-harmonic generation unit. The pulse length of the 327.4 nm radiation was 5 ns, which resulted in a spectral bandwidth of 88 MHz (Eqn. (\ref{num:25})). The first-step laser beam directed to the jet was additionally attenuated to avoid power broadening of the atomic transition.
The second-step laser light at 287.9 nm was produced by frequency doubling of the 575.8 nm (0.15 cm$^{-1}$ bandwidth) radiation from a dye laser (Scanmate, Lambda Physik). The  amplifier and the dye laser were pumped by two time-synchronized XeCl excimer lasers (LPX 240i, Lambda Physik) with a pulse repetition rate of 50 Hz. Both laser beams were transported a distance of 15 m to the front end of the mass separator. The first- and second-step laser beams had a diameter of about 3 mm in the jet region and the center of the second step beam crossed the gas jet 6.5 mm away from the exit orifice, thus copper atoms in the region between 5 and 8 mm were ionized. The jet at this distance was about 7.5 mm in diameter (see inset in Fig. \ref{Fig:15}b).

In the reference cell, located in the laser hut, a collimated atomic beam of copper atoms with a natural abundance ($^{63}$Cu$=$ 69\%, $^{65}$Cu$=$ 31\%) was produced by resistive heating of a graphite crucible at a temperature of 1250 K. The residual pressure in the reference cell was 1$\cdot$10$^{-6}$ mbar. Atoms from the crucible entered the laser ionization zone through a collimating orifice of 3 mm in diameter. The collimation ratio of the atomic beam  in the setup amounts to 1/12. About 5\% of the  laser power was directed towards the reference cell. The first and the second step laser beams were parallel to each other and crossed the atomic beam at 90$^\circ$. The laser-produced ions were pushed out the ionization zone by an electrical field and detected by a SEM. No mass separation was available in the reference cell. The ion signals from the gas jet and from the reference cell were recorded simultaneously as a function of the first-step laser wavelength. The wavelength was measured by a lambda meter LM 007 (ATOS).

\subsection{Results}
\begin{figure}
\begin{center}
\includegraphics[scale=0.6]{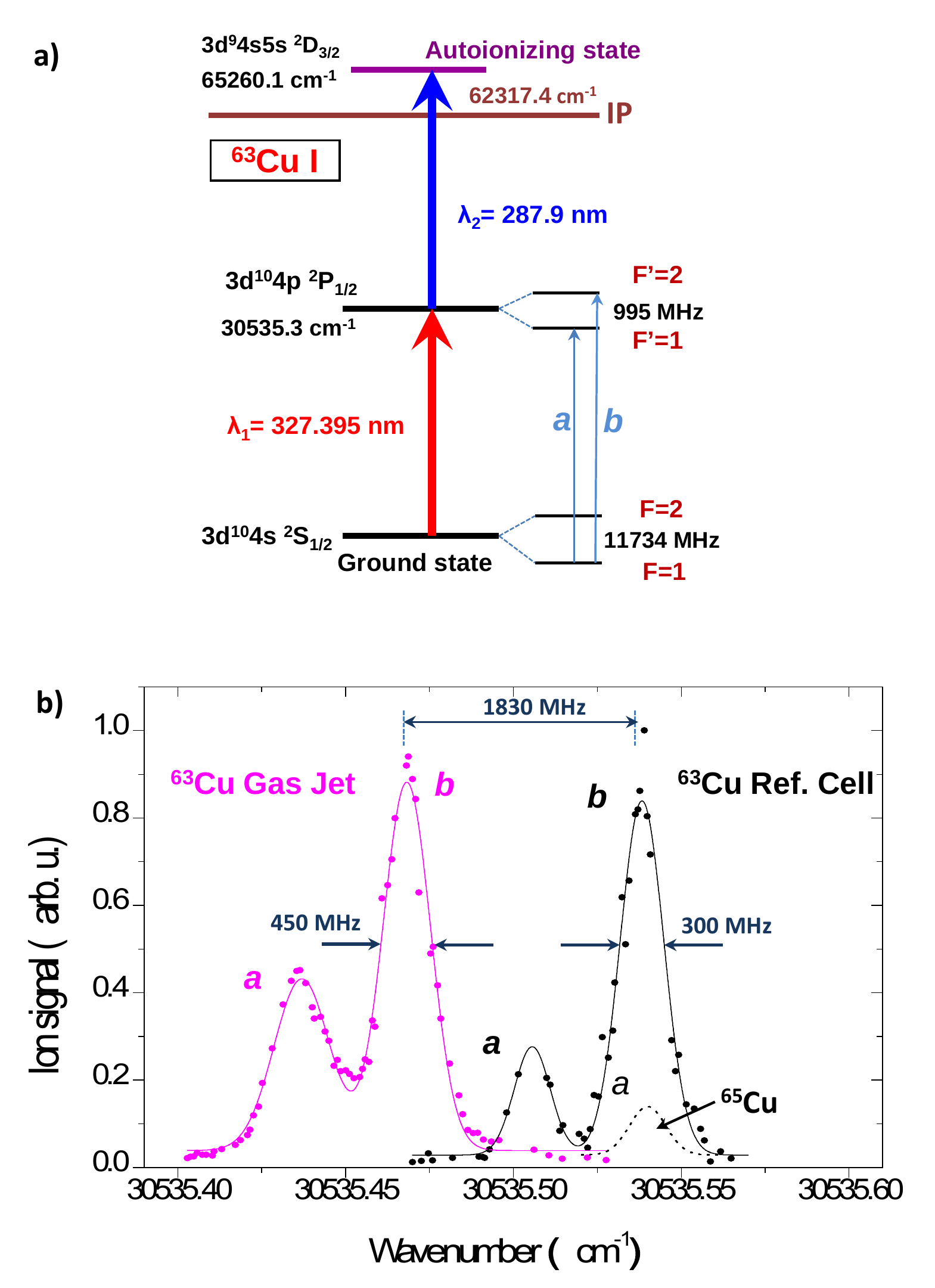} \caption{a) The two-step ionization scheme of copper atoms (not to scale) used in these experiments, b) ion signal from  $^{63}$Cu in the gas jet (purple) and from the copper sample of natural abundance in the reference cell (black). Points represent experimental  values, while the solid lines are the best Gaussian fits to the experimental data in order to determine the total FWHM (see table 3). The dashed line is the contribution of the line $a$ of $^{65}$Cu to the line $b$ of $^{63}$Cu in the reference cell.}\label{Fig:18}
\end{center}
\end{figure}

The results presented here were taken at a stagnation pressure $P_0=$ 200 mbar (pressure in the gas cell) and background pressure $P_{bg}=$ 0.1 mbar (pressure in the gas cell chamber), hence resulting in a pressure ratio of 2000. The stagnation temperature was estimated to be $T_0$= 300 K. The measured ion signal as a function of the frequency of the first step laser, given in  wavenumbers, for the $^{63}$Cu isotopes in the gas jet and for the natural copper in the reference cell are shown in Fig. \ref{Fig:18}. The displayed frequency range covers two resonances corresponding to transitions from the ground state, with a total angular momentum F=1, to the first excited state, with F'=1 and F'=2, denoted in Fig. \ref{Fig:18} by $a$ and $b$, respectively. In the reference cell, the $b$ line of $^{63}$Cu is mixed with a small contribution from the $a$ line of the lower-abundant copper $^{65}$Cu isotope, illustrated by a dashed line. The centroid of line $a$ of $^{65}$Cu is 60 MHz away from the center of line $b$ of $^{63}$Cu towards higher wavenumbers  and the contribution of this line to the width and the position of the $^{63}$Cu $b$ line is not more than 10 MHz.

The measured hyperfine splitting (the distance between the lines $a$ and $b$) of 995(30) MHz both in the gas jet and in the reference cell is in agreement with the literature values of 1013.2(20) MHz \cite{Ber1989} and 960(30) MHz \cite{Fis1961}. The resonances in the jet are shifted to lower wave numbers relative to those in the reference cell owing to the counterpropagating direction of the first step laser with respect to the atomic beam. This Doppler shift amounts to 1830(30) MHz and results, using Eqn. (\ref{num:2}), in a stream velocity in the ionization region of $u=$ 599(10) m/s. Applying Eqn. (\ref{num:33}), the terminal Mach number $M_t$ is found to be 28. However, this value is only reached at a distance of 25 mm from the orifice. At the position of the second-step laser beam of 6.5 mm, the Mach number is equal to 11, see Fig. \ref{Fig:14}. At this Mach number, the stream velocity reaches almost the maximum value (see Fig. \ref{Fig:4}) and it can be used to estimate the gas temperature in the cell (see Eqn.(\ref{num:5})). If the temperature in the cell is 300 K, the stream velocity should be 552 m/s, which should correspond to a Doppler shift of $\Delta\nu_{Doppler}=$ 1683 MHz. The measured shift points to higher gas cell temperature of 355 K and can be explained by the additional heating of the gas caused by the glowing filament. Collisions in the free jet shift the spectral line position to smaller wave numbers. The shift-rate coefficient $\gamma_{sh}$  is usually smaller than the collision-broadening coefficient $\gamma_{coll}$ \cite{All1982}. The shift rate $\Gamma_{sh}$  (Eqn. (\ref{num:21})) is estimated to be not more than 8 MHz, which gives the maximum error on the  jet velocity of 2.6 m/s and on the gas cell temperature of about 3 K.

\begin{table*}
\centering
\begin{threeparttable}
\begin{tabular}{lcc}
\toprule
&Free Jet&Reference Cell\\ \hline
Stagnation temperature T$_0$ (K)&355&\\
Stagnation pressure P$_0$ (mbar)&200&\\
Mach number M&11&\\
Jet temperature T (K)&8.6&\\
Stream velocity (m/s)&599&\\
Doppler FWHM (MHz)&242&\\
Divergence broadening (MHz)&150&\\
Laser bandwidth (MHz)&88&\\
Gaussian  FWHM (MHz)&402&\\
Atom density (cm$^{-3}$)&1.8$\cdot10^{16}$&\\
Collision broadening FWHM (MHz)&8&\\
Natural broadening FWHM (MHz)&22&\\
Lorentzian FWHM (MHz)&30&\\
Voigt FWHM (MHz)&420&\\
Experimental FWHM (MHz)&450&\\
Temperature of the crucible (K)&&1250\\
Doppler FWHM\tnote{a} (MHz)&&2920\\
Collimation ratio&&1$/$12\\
Doppler FWHM (MHz)&&249\\
Total width,FWHM (MHz)&&260\\
Experimental FWHM (MHz)&&300\\
\bottomrule		
\end{tabular}
\begin{tablenotes}
\footnotesize
\item[a] For a temperature T$=1250$ K
\end{tablenotes}	
\end{threeparttable}
\caption{Conditions in the stagnation region (gas cell) and contributions of different broadening mechanisms to the 327.4 nm spectral lines in copper (3d$^{10}$4s $^2$S$_{1/2}$\,$\rightarrow$\,3d$^{10}$4p $^2$P$_{1/2}$) in the ionization region of the supersonic free gas jet and in the reference cell.}
	\label{tab:3}
\end{table*}
A key point in these studies was to find the answer to the question of what spectral resolution can be obtained by using a combination of axial excitation and  transverse ionization in the supersonic free jet. The measured width (FWHM) of the resonance was found to be 450 MHz for both lines. Calculated contributions of different broadening mechanisms to the 327.4 nm spectral lines (F=1, F'=1,2) in copper in the ionization region of the supersonic free gas jet and in the reference cell are given in Tab. 3. In the jet ionization region, the main contribution to the total Voigt line profile with a FWHM of 420 MHz is due to the Gaussian contribution with FWHM of 402 MHz. The Gaussian width is defined by the residual Doppler broadening with FWHM of 242 MHz at the jet temperature of 8.6 K, the additional broadening due to the divergence of the jet is estimated to be 150 MHz, and the laser bandwidth amounts to 88 MHz (5 ns laser pulse length). A  Lorentzian FWHM of 30 MHz is mainly given by the natural linewidth of 22 MHz, resulting in a much smaller contribution due to the collision broadening. The experimentally found FWHM (450 MHz) is very close to that estimated (420 MHz). A smaller Doppler width can be obtained in the case of performing the ionization farther from the exit orifice, where larger Mach numbers can be reached. The essential contribution to the linewidth comes from the jet divergence. This contribution was reduced in the experiments to 150 MHz by restricting the ionization volume. If all atoms in the jet would undergo laser ionization the additional broadening would be 225 MHz and the total FWHM close to 500 MHz. Hence the gas jet divergence is the limiting factor for high-resolution spectroscopy in the free jet.

In the reference cell the Doppler FWHM of the spectral line $\Delta\nu^\ast_{Doppler}$  was found to be 249 MHz. It was calculated using the expression $\Delta\nu^\ast_{Doppler}= \Delta\nu_{Doppler}\cdot\sin\epsilon$, where $\Delta\nu_{Doppler}$ is the Doppler FWHM at the crucible temperature of 1250 K (15) and $\sin\epsilon$  is the collimation ratio \cite{Dem2002}. Taking into account the laser width of 88 MHz, the total FWHM of 260 MHz is close to the experimentally obtained value of 300 MHz.

\section{Conclusions and outlook}

Different approaches for selective two-step laser resonance ionization that can be used for high-resolution spectroscopy of radioactive isotopes produced in nuclear reactions have been considered in various types of supersonic gas jets. Although the efficiency of the total ionization process was not determined in this study, it will be low due to the low repetition rate of the lasers (50 Hz) and to the limited overlap of the laser beams with the gas jet. An improvement of the spatial overlap will be possible when the de Laval or spike nozzles are used. A better collimation of the gas jet by these two types of nozzle will allows to obtain a much more efficient overlap with the laser beams over a much longer distance.  The cylindrical shape of such jets can be made easier to match with the shape of the laser beams. Since the radial size of the de Laval and the spike nozzle beams can be kept smaller than the radial size of the free-jet beam, a higher laser-energy density will be available to saturate the atomic transitions. In addition, optimum temporal overlap will be provided by high-repetition rate pulsed lasers of typically 10 kHz, that will enable irradiation of all isotopes in the well collimated fast-moving gas jet.

The crossed first- and second-step laser beams interacting with the supersonic jet are critical to select the ionization region with cold atoms when a high spectral resolution is needed. Here, it has been shown experimentally that a spectral resolution $\delta\nu/\nu=$ 4.9$\cdot$10$^{-7}$ can be obtained using laser ionization in a supersonic free-gas jet. In this jet the resolution is limited by the intrinsic divergence of the atomic beam. The spectral  resolution can be further improved up to 2.3$\cdot$10$^{-7}$ by using better collimated supersonic jets produced by the de Laval- or by the spike nozzles.

The IGLIS technique combining laser ionization in a gas cell or in a gas jet, is especially adapted for production and spectroscopy of rare radioactive isotopes (see Fig. \ref{Fig:3}). Broad resonance structures can be first investigated using in-gas cell laser ionization and the obtained Z-selective beams can be mass separated and sent to the typical experimental setups developed at ISOL facilities. When maximum selectivity and resolution is needed, repelling fields can be used to eliminate all contaminating ions and laser ionization might be performed in the gas jet. This will allow high-resolution and high-sensitivity laser spectroscopy studies in different exotic regions of the nuclear chart and also the production of purified-isomeric beams.

The gas-jet properties that are important to achieve an optimum spectral resolution were described without taking into account the influence of mechanical structures (such as the RFQ) on the gas flow field. For the final design of the set-up, 3D fluid dynamics flow calculations will be needed. Efforts in this direction are underway and prototypes will be validated. Moreover, optimal conditions to perform high-resolution spectroscopy with collimated supersonic beams and with high repetition rate lasers will be investigated in a new off-line laboratory that is being commissioned at KU Leuven. As a next step for the on-line studies, high-resolution spectroscopy of $^{58}$Cu will be performed at LISOL.

An optimized IGLIS setup to perform laser ionization spectroscopy in the gas cell and in the gas jet, including high repetition lasers, will be installed at the Super Separator Spectrometer (S3), which will be coupled to the superconducting linear accelerator of the SPIRAL2 facility at GANIL \cite{GAN2012}.

\section*{Acknowledgments}
This work was supported by FWO-Vlaanderen (Belgium), by GOA/2010/010 (BOF KU Leuven), by the IAP Belgian Science Policy (BriX network P6/23), by the European Commission within the Seventh Framework Programme through I3-ENSAR (contract No. RII3-CT-2010-262010), and by a Grant from the European Research Council (ERC-2011-AdG-291561-HELIOS). We are grateful to E. Mogilevskiy for fruitful discussions during the preparation of the manuscript.



\begin{thebibliography}{119}

\expandafter\ifx\csname natexlab\endcsname\relax\def\natexlab#1{#1}\fi
\providecommand{\bibinfo}[2]{#2}
\ifx\xfnm\relax \def\xfnm[#1]{\unskip,\space#1}\fi
\bibitem[{Lethokov(1987)}]{Let1987}
\bibinfo{author}{V.~S. Lethokov}, \bibinfo{title}{Laser Photoionization
  Spectroscopy}, \bibinfo{publisher}{Academic Press, Orlando},
  \bibinfo{year}{1987}.
\bibitem[{Hurst and Payne(1989)}]{Hur1988}
\bibinfo{author}{G.~S. Hurst}, \bibinfo{author}{M.~G. Payne},
  \bibinfo{title}{Principles and Applications of Resonance Ionization
  Spectroscopy}, \bibinfo{publisher}{Hilger, London}, \bibinfo{year}{1989}.
\bibitem[{Alkhazov et~al.(1989)Alkhazov, Berlovich, and Panteleyev}]{Alk1989}
\bibinfo{author}{G.~D. Alkhazov}, \bibinfo{author}{E.~{\relax Ye}. Berlovich},
  \bibinfo{author}{V.~N. Panteleyev}, \bibinfo{journal}{Nucl. Instr. Methods A}
  \bibinfo{volume}{288} (\bibinfo{year}{1989}) \bibinfo{pages}{141}.
\bibitem[{Barzakh et~al.(2012)Barzakh, Fedorov, Ivanov, Molkanov, Panteleev,
  and Volkov}]{Bar2012}
\bibinfo{author}{A.~E. Barzakh}, \bibinfo{author}{D.~V. Fedorov},
  \bibinfo{author}{V.~S. Ivanov}, \bibinfo{author}{P.~L. Molkanov},
  \bibinfo{author}{V.~N. Panteleev}, \bibinfo{author}{{\relax Yu}.~M. Volkov},
  \bibinfo{journal}{Rev. Sci. Instrum.} \bibinfo{volume}{83}
  (\bibinfo{year}{2012}) \bibinfo{pages}{02B306}.
\bibitem[{Vermeeren et~al.(1994)Vermeeren, Bijnens, Huyse, Kudryavtsev, {V}an
  Duppen, Wauters, Qamhieh, Silverans, Thoen, and Vandeweert}]{Ver1994}
\bibinfo{author}{L.~Vermeeren}, \bibinfo{author}{N.~Bijnens},
  \bibinfo{author}{M.~Huyse}, \bibinfo{author}{{\relax Yu}.~A. Kudryavtsev},
  \bibinfo{author}{P.~{V}an Duppen}, \bibinfo{author}{J.~Wauters},
  \bibinfo{author}{Z.~N. Qamhieh}, \bibinfo{author}{R.~E. Silverans},
  \bibinfo{author}{P.~Thoen}, \bibinfo{author}{E.~Vandeweert},
  \bibinfo{journal}{Phys. Rev. Lett.} \bibinfo{volume}{73}
  (\bibinfo{year}{1994}) \bibinfo{pages}{1935}.
\bibitem[{Kudryavtsev et~al.(1996)Kudryavtsev, Andrzejewski, Bijnens, Franchoo,
  Gentens, Huyse, Piechaczek, Reusen, Szerypo, {V}an~den Bergh, {V}an Duppen,
  Vermeeren, Wauters, and W\"{o}hr}]{Kud1996}
\bibinfo{author}{{\relax Yu}.~A. Kudryavtsev},
  \bibinfo{author}{J.~Andrzejewski}, \bibinfo{author}{N.~Bijnens},
  \bibinfo{author}{S.~Franchoo}, \bibinfo{author}{J.~Gentens},
  \bibinfo{author}{M.~Huyse}, \bibinfo{author}{A.~Piechaczek},
  \bibinfo{author}{I.~Reusen}, \bibinfo{author}{J.~Szerypo},
  \bibinfo{author}{P.~{V}an~den Bergh}, \bibinfo{author}{P.~{V}an Duppen},
  \bibinfo{author}{L.~Vermeeren}, \bibinfo{author}{J.~Wauters},
  \bibinfo{author}{A.~W\"{o}hr}, \bibinfo{journal}{Nucl. Instr. Methods Phys.
  Res.} \bibinfo{volume}{114} (\bibinfo{year}{1996}) \bibinfo{pages}{350}.
\bibitem[{Cocolios et~al.(2009)Cocolios, Andreyev, Bastin, Bree, Buscher,
  Elseviers, Gentens, Huyse, Kudryavtsev, Pauwels, Sonoda, {V}an~den Bergh, and
  {V}an Duppen}]{Coc2009}
\bibinfo{author}{T.~E. Cocolios}, \bibinfo{author}{A.~N. Andreyev},
  \bibinfo{author}{B.~Bastin}, \bibinfo{author}{N.~Bree},
  \bibinfo{author}{J.~Buscher}, \bibinfo{author}{J.~Elseviers},
  \bibinfo{author}{J.~Gentens}, \bibinfo{author}{M.~Huyse},
  \bibinfo{author}{{\relax Yu}.~Kudryavtsev}, \bibinfo{author}{D.~Pauwels},
  \bibinfo{author}{T.~Sonoda}, \bibinfo{author}{P.~{V}an~den Bergh},
  \bibinfo{author}{P.~{V}an Duppen}, \bibinfo{journal}{Phys. Rev. Lett.}
  \bibinfo{volume}{103} (\bibinfo{year}{2009}) \bibinfo{pages}{102501}.
\bibitem[{Mishin et~al.(1993)Mishin, Fedoseyev, Kluge, Letokhov, Ravn,
  Scheerer, Shirakabe, Sundell, and Tengblad}]{Mis1993}
\bibinfo{author}{V.~I. Mishin}, \bibinfo{author}{V.~N. Fedoseyev},
  \bibinfo{author}{H.-J. Kluge}, \bibinfo{author}{V.~S. Letokhov},
  \bibinfo{author}{H.~L. Ravn}, \bibinfo{author}{F.~Scheerer},
  \bibinfo{author}{Y.~Shirakabe}, \bibinfo{author}{S.~Sundell},
  \bibinfo{author}{O.~Tengblad}, \bibinfo{journal}{Nucl. Instr. Methods B}
  \bibinfo{volume}{73} (\bibinfo{year}{1993}) \bibinfo{pages}{550}.
\bibitem[{Fedoseyev et~al.(2000)Fedoseyev, Huber, K\"{o}ster, Lettry, Mishin,
  Ravn, and Sebastian}]{Fed2000}
\bibinfo{author}{V.~N. Fedoseyev}, \bibinfo{author}{G.~Huber},
  \bibinfo{author}{U.~K\"{o}ster}, \bibinfo{author}{J.~Lettry},
  \bibinfo{author}{V.~I. Mishin}, \bibinfo{author}{H.~Ravn},
  \bibinfo{author}{V.~Sebastian}, \bibinfo{journal}{Hyperfine Interact.}
  \bibinfo{volume}{127} (\bibinfo{year}{2000}) \bibinfo{pages}{409}.
\bibitem[{Backe et~al.(1998)Backe, Hies, Kunz, Lauth, Curtze, Schwamb, Sewtz,
  Theobald, Zahn, Eberhardt, Trautmann, Habs, Repnow, and Fricke}]{Bac1998}
\bibinfo{author}{H.~Backe}, \bibinfo{author}{M.~Hies},
  \bibinfo{author}{H.~Kunz}, \bibinfo{author}{W.~Lauth},
  \bibinfo{author}{O.~Curtze}, \bibinfo{author}{P.~Schwamb},
  \bibinfo{author}{M.~Sewtz}, \bibinfo{author}{W.~Theobald},
  \bibinfo{author}{R.~Zahn}, \bibinfo{author}{K.~Eberhardt},
  \bibinfo{author}{N.~Trautmann}, \bibinfo{author}{D.~Habs},
  \bibinfo{author}{R.~Repnow}, \bibinfo{author}{B.~Fricke},
  \bibinfo{journal}{Phys. Rev. Lett.} \bibinfo{volume}{80}
  (\bibinfo{year}{1998}) \bibinfo{pages}{920}.
\bibitem[{Backe et~al.(2000)Backe, Dretzke, Hies, Kube, Kunz, Lauth, Sewtz,
  Trautmann, and R.~Repnow}]{Bac2000}
\bibinfo{author}{H.~Backe}, \bibinfo{author}{A.~Dretzke},
  \bibinfo{author}{M.~Hies}, \bibinfo{author}{G.~Kube},
  \bibinfo{author}{H.~Kunz}, \bibinfo{author}{W.~Lauth},
  \bibinfo{author}{M.~Sewtz}, \bibinfo{author}{N.~Trautmann},
  \bibinfo{author}{H.~J.~M. R.~Repnow}, \bibinfo{journal}{Hyperfine Interact.}
  \bibinfo{volume}{127} (\bibinfo{year}{2000}) \bibinfo{pages}{35}.
\bibitem[{Lassen et~al.(2005)Lassen, Bricault, Dombsky, Lavoie, Geppert, and
  Wendt}]{Las2005}
\bibinfo{author}{J.~Lassen}, \bibinfo{author}{P.~Bricault},
  \bibinfo{author}{M.~Dombsky}, \bibinfo{author}{J.~P. Lavoie},
  \bibinfo{author}{C.~Geppert}, \bibinfo{author}{K.~Wendt},
  \bibinfo{journal}{Hyperfine Interact.} \bibinfo{volume}{162}
  (\bibinfo{year}{2005}) \bibinfo{pages}{69}.
\bibitem[{Moore et~al.(2005)Moore, Nieminen, Billowes, Campbell, Geppert,
  Jokinen, Kessler, Marsh, Penttil\"{a}, Rinta-Antila, Tordoff, Wendt, and
  \"{A}yst\"{o}}]{Moo2005}
\bibinfo{author}{I.~D. Moore}, \bibinfo{author}{A.~Nieminen},
  \bibinfo{author}{J.~Billowes}, \bibinfo{author}{P.~Campbell},
  \bibinfo{author}{C.~Geppert}, \bibinfo{author}{A.~Jokinen},
  \bibinfo{author}{T.~Kessler}, \bibinfo{author}{B.~Marsh},
  \bibinfo{author}{H.~Penttil\"{a}}, \bibinfo{author}{S.~Rinta-Antila},
  \bibinfo{author}{B.~Tordoff}, \bibinfo{author}{K.~D.~A. Wendt},
  \bibinfo{author}{J.~\"{A}yst\"{o}}, \bibinfo{journal}{J. Phys. G}
  \bibinfo{volume}{31} (\bibinfo{year}{2005}) \bibinfo{pages}{S1499}.
\bibitem[{Moore et~al.(2010)Moore, Kessler, Sonoda, Kudryavstev,
  Per\"{a}j\"{a}rvi, Popov, Wendt, and \"{A}yst\"{o}}]{Moo2010}
\bibinfo{author}{I.~D. Moore}, \bibinfo{author}{T.~Kessler},
  \bibinfo{author}{T.~Sonoda}, \bibinfo{author}{{\relax Yu}.~Kudryavstev},
  \bibinfo{author}{K.~Per\"{a}j\"{a}rvi}, \bibinfo{author}{A.~Popov},
  \bibinfo{author}{K.~D.~A. Wendt}, \bibinfo{author}{J.~\"{A}yst\"{o}},
  \bibinfo{journal}{Nucl. Instr. Methods B} \bibinfo{volume}{268}
  (\bibinfo{year}{2010}) \bibinfo{pages}{657}.
\bibitem[{Liu et~al.(2006)Liu, Baktash, Beene, Bilheux, Havener, Krause,
  Schultz, Stracener, Vane, Br\"{u}ck, Geppert, Kessler, and Wendt}]{Liu2006}
\bibinfo{author}{Y.~Liu}, \bibinfo{author}{C.~Baktash}, \bibinfo{author}{J.~R.
  Beene}, \bibinfo{author}{H.~Z. Bilheux}, \bibinfo{author}{C.~C. Havener},
  \bibinfo{author}{H.~F. Krause}, \bibinfo{author}{D.~R. Schultz},
  \bibinfo{author}{D.~W. Stracener}, \bibinfo{author}{C.~R. Vane},
  \bibinfo{author}{K.~Br\"{u}ck}, \bibinfo{author}{C.~Geppert},
  \bibinfo{author}{T.~Kessler}, \bibinfo{author}{K.~Wendt},
  \bibinfo{journal}{Nucl. Instr. Methods B} \bibinfo{volume}{243}
  (\bibinfo{year}{2006}) \bibinfo{pages}{442}.
\bibitem[{Jeong et~al.(2010)Jeong, Imai, Ishiyama, Hirayama, Miyatake, and
  Watanabe}]{Jeo2010}
\bibinfo{author}{S.~C. Jeong}, \bibinfo{author}{N.~Imai},
  \bibinfo{author}{H.~Ishiyama}, \bibinfo{author}{Y.~Hirayama},
  \bibinfo{author}{H.~Miyatake}, \bibinfo{author}{Y.~X. Watanabe},
  \bibinfo{journal}{KEK Report} \bibinfo{volume}{2010-2}
  (\bibinfo{year}{2010}).
\bibitem[{Sonoda et~al.(2009)Sonoda, Wada, Takamine, Okada, Schury, Yoshida,
  Kubo, Matsuo, Furukawa, Wakui, Shinozuka, Iimura, Yamazaki, Katayama, Ohtani,
  Wollnik, Schuessler, Kudryavtsev, {V}an Duppen, and Huyse}]{Son2009a}
\bibinfo{author}{T.~Sonoda}, \bibinfo{author}{M.~Wada},
  \bibinfo{author}{A.~Takamine}, \bibinfo{author}{K.~Okada},
  \bibinfo{author}{P.~Schury}, \bibinfo{author}{A.~Yoshida},
  \bibinfo{author}{T.~Kubo}, \bibinfo{author}{Y.~Matsuo},
  \bibinfo{author}{T.~Furukawa}, \bibinfo{author}{T.~Wakui},
  \bibinfo{author}{T.~Shinozuka}, \bibinfo{author}{H.~Iimura},
  \bibinfo{author}{Y.~Yamazaki}, \bibinfo{author}{I.~Katayama},
  \bibinfo{author}{S.~Ohtani}, \bibinfo{author}{H.~Wollnik},
  \bibinfo{author}{H.~A. Schuessler}, \bibinfo{author}{{\relax
  Yu}.~Kudryavtsev}, \bibinfo{author}{P.~{V}an Duppen},
  \bibinfo{author}{M.~Huyse}, \bibinfo{journal}{AIP Conf. Proc.}
  \bibinfo{volume}{1104} (\bibinfo{year}{2009}) \bibinfo{pages}{132}.
\bibitem[{Lecesne et~al.(2010)Lecesne, Alv\`{e}s-Cond\'{e}, Coterreau,
  Oliveira, Dubois, Flambard, Franberg, Gottwald, Jardin, Lassen, {L}e Blanc,
  Leroy, Mattolat, Olivier, Pacquet, Pichard, Rothe, Saint-Laurent, and
  Wendt}]{Lec2010}
\bibinfo{author}{N.~Lecesne}, \bibinfo{author}{R.~Alv\`{e}s-Cond\'{e}},
  \bibinfo{author}{E.~Coterreau}, \bibinfo{author}{F.~D. Oliveira},
  \bibinfo{author}{M.~Dubois}, \bibinfo{author}{J.~L. Flambard},
  \bibinfo{author}{H.~Franberg}, \bibinfo{author}{T.~Gottwald},
  \bibinfo{author}{P.~Jardin}, \bibinfo{author}{J.~Lassen},
  \bibinfo{author}{F.~{L}e Blanc}, \bibinfo{author}{R.~Leroy},
  \bibinfo{author}{C.~Mattolat}, \bibinfo{author}{A.~Olivier},
  \bibinfo{author}{J.~Y. Pacquet}, \bibinfo{author}{A.~Pichard},
  \bibinfo{author}{S.~Rothe}, \bibinfo{author}{M.~G. Saint-Laurent},
  \bibinfo{author}{K.~Wendt}, \bibinfo{journal}{Rev. Sci. Instrum.}
  \bibinfo{volume}{81} (\bibinfo{year}{2010}) \bibinfo{pages}{02A910}.
\bibitem[{Otten(1989)}]{Ott1989}
\bibinfo{author}{E.~W. Otten}, \bibinfo{title}{Nuclear radii and moments of
  unstable isotopes. In \emph{Treatise on heavy-ion science}},
  \bibinfo{publisher}{Plenum Publishing Corporation}, pp.
  \bibinfo{pages}{517--638}, \bibinfo{year}{1989}.
\bibitem[{Billowes and Campbell(1995)}]{Bil1995}
\bibinfo{author}{J.~Billowes}, \bibinfo{author}{P.~Campbell},
  \bibinfo{journal}{J. Phys. G} \bibinfo{volume}{21} (\bibinfo{year}{1995})
  \bibinfo{pages}{701}.
\bibitem[{Neugart(2002)}]{Neu2002}
\bibinfo{author}{R.~Neugart}, \bibinfo{journal}{Eur. Phys. J. A}
  \bibinfo{volume}{15} (\bibinfo{year}{2002}) \bibinfo{pages}{35}.
\bibitem[{Kluge and N\"{o}rtersh\"{a}user(2003)}]{Klu2003}
\bibinfo{author}{H.-J. Kluge}, \bibinfo{author}{W.~N\"{o}rtersh\"{a}user},
  \bibinfo{journal}{Spectrochim. Acta B} \bibinfo{volume}{58}
  (\bibinfo{year}{2003}) \bibinfo{pages}{1031}.
\bibitem[{Neugart and Neyens(2006)}]{Neu2006}
\bibinfo{author}{R.~Neugart}, \bibinfo{author}{G.~Neyens},
  \bibinfo{title}{Nuclear Moments. In \emph{The Euroschool Lectures on Physics
  with Exotic Beams, Vol. II}}, \bibinfo{publisher}{Springer, Berlin
  Heidelberg}, p. \bibinfo{pages}{135}, \bibinfo{year}{2006}.
\bibitem[{Cheal and Flanagan(2010)}]{Che2010}
\bibinfo{author}{B.~Cheal}, \bibinfo{author}{K.~T. Flanagan},
  \bibinfo{journal}{J. Phys. G} \bibinfo{volume}{37} (\bibinfo{year}{2010})
  \bibinfo{pages}{113101}.
\bibitem[{Kluge(2010)}]{Klu2010}
\bibinfo{author}{H.-J. Kluge}, \bibinfo{journal}{Hyperfine Interact.}
  \bibinfo{volume}{196} (\bibinfo{year}{2010}) \bibinfo{pages}{295}.
\bibitem[{Fedosseev et~al.(2012)Fedosseev, Kudryavtsev, and Mishin}]{Fed2012}
\bibinfo{author}{V.~N. Fedosseev}, \bibinfo{author}{{\relax Yu}.~Kudryavtsev},
  \bibinfo{author}{V.~I. Mishin}, \bibinfo{journal}{Phys. Scr.}
  \bibinfo{volume}{85} (\bibinfo{year}{2012}) \bibinfo{pages}{058104}.
\bibitem[{Kudriavtsev and Letokhov(1982)}]{Kud1982}
\bibinfo{author}{Y.~A. Kudriavtsev}, \bibinfo{author}{V.~Letokhov},
  \bibinfo{journal}{Appl. Phys. B} \bibinfo{volume}{29} (\bibinfo{year}{1982})
  \bibinfo{pages}{219}.
\bibitem[{Schulz et~al.(1991)Schulz, Arnold, Borchers, Neu, Neugart, Neuroth,
  Otten, Scherf, Wendt, Lievens, Kudryavtsev, Letokhov, Mishin, and
  Petrunin}]{Sch1991}
\bibinfo{author}{C.~Schulz}, \bibinfo{author}{E.~Arnold},
  \bibinfo{author}{W.~Borchers}, \bibinfo{author}{W.~Neu},
  \bibinfo{author}{R.~Neugart}, \bibinfo{author}{M.~Neuroth},
  \bibinfo{author}{E.~W. Otten}, \bibinfo{author}{M.~Scherf},
  \bibinfo{author}{K.~Wendt}, \bibinfo{author}{P.~Lievens},
  \bibinfo{author}{{\relax Yu}.~A. Kudryavtsev}, \bibinfo{author}{V.~S.
  Letokhov}, \bibinfo{author}{V.~I. Mishin}, \bibinfo{author}{V.~V. Petrunin},
  \bibinfo{journal}{J. Phys. B} \bibinfo{volume}{24} (\bibinfo{year}{1991})
  \bibinfo{pages}{4831}.
\bibitem[{Wendt et~al.(2000)Wendt, Trautmann, and Bushaw}]{Wen2000}
\bibinfo{author}{K.~Wendt}, \bibinfo{author}{N.~Trautmann},
  \bibinfo{author}{B.~A. Bushaw}, \bibinfo{journal}{Nucl. Instr. Methos B}
  \bibinfo{volume}{172} (\bibinfo{year}{2000}) \bibinfo{pages}{162}.
\bibitem[{Flanagan(2008)}]{Fla2008}
\bibinfo{author}{K.~T. Flanagan}, \bibinfo{journal}{CERN-INTC}
  \bibinfo{volume}{2008-010} (\bibinfo{year}{2008}) \bibinfo{pages}{240}.
\bibitem[{Procter et~al.(2012)Procter, Aghaei-Khozani, Billowes, Bissell, {L}e
  Blanc, Cheal, Cocolios, Flanagan, Hori, Kobayashi, Lunney, Lynch, Marsh,
  Neyens, Papuga, Rajabali, Rothe, Simpson, Smith, Stroke, Vanderheijden, and
  Wendt}]{Pro2012}
\bibinfo{author}{T.~J. Procter}, \bibinfo{author}{H.~Aghaei-Khozani},
  \bibinfo{author}{J.~Billowes}, \bibinfo{author}{M.~L. Bissell},
  \bibinfo{author}{F.~{L}e Blanc}, \bibinfo{author}{B.~Cheal},
  \bibinfo{author}{T.~E. Cocolios}, \bibinfo{author}{K.~T. Flanagan},
  \bibinfo{author}{H.~Hori}, \bibinfo{author}{T.~Kobayashi},
  \bibinfo{author}{D.~Lunney}, \bibinfo{author}{K.~M. Lynch},
  \bibinfo{author}{B.~A. Marsh}, \bibinfo{author}{G.~Neyens},
  \bibinfo{author}{J.~Papuga}, \bibinfo{author}{M.~M. Rajabali},
  \bibinfo{author}{S.~Rothe}, \bibinfo{author}{G.~Simpson},
  \bibinfo{author}{A.~J. Smith}, \bibinfo{author}{H.~H. Stroke},
  \bibinfo{author}{W.~Vanderheijden}, \bibinfo{author}{K.~Wendt},
  \bibinfo{journal}{J. Phys. Conf. Series} \bibinfo{volume}{381}
  (\bibinfo{year}{2012}) \bibinfo{pages}{012070}.
\bibitem[{Kaufman(1976)}]{Kau1976}
\bibinfo{author}{S.~Kaufman}, \bibinfo{journal}{Opt. Commun.}
  \bibinfo{volume}{17} (\bibinfo{year}{1976}) \bibinfo{pages}{309}.
\bibitem[{Kudryavtsev(1992)}]{Kud1992}
\bibinfo{author}{{\relax Yu}.~A. Kudryavtsev}, \bibinfo{journal}{Hyperfine
  Interact.} \bibinfo{volume}{74} (\bibinfo{year}{1992}) \bibinfo{pages}{171}.
\bibitem[{Monz et~al.(1993)Monz, Hohmann, Kluge, Kunze, Lantzsch, Otten,
  Passler, Senne, Stenner, Stratmann, Wendt, Zimmer, Herrmann, Trautmann, and
  Walter}]{Mon1993}
\bibinfo{author}{L.~Monz}, \bibinfo{author}{R.~Hohmann}, \bibinfo{author}{H.-J.
  Kluge}, \bibinfo{author}{S.~Kunze}, \bibinfo{author}{J.~Lantzsch},
  \bibinfo{author}{E.~W. Otten}, \bibinfo{author}{G.~Passler},
  \bibinfo{author}{P.~Senne}, \bibinfo{author}{J.~Stenner},
  \bibinfo{author}{K.~Stratmann}, \bibinfo{author}{K.~Wendt},
  \bibinfo{author}{K.~Zimmer}, \bibinfo{author}{G.~Herrmann},
  \bibinfo{author}{N.~Trautmann}, \bibinfo{author}{K.~Walter},
  \bibinfo{journal}{Spectrochim. Acta. B} \bibinfo{volume}{48}
  (\bibinfo{year}{1993}) \bibinfo{pages}{1655}.
\bibitem[{Wendt et~al.(1997)Wendt, Bhowmick, Herrmann, Kratz, Lantzsch,
  M\"{u}ller, N\"{o}rtersh\"{a}user, Otten, Schwalbach, Seibert, Trautmann, and
  Waldek}]{Wen1997}
\bibinfo{author}{K.~Wendt}, \bibinfo{author}{G.~K. Bhowmick},
  \bibinfo{author}{G.~Herrmann}, \bibinfo{author}{J.~V. Kratz},
  \bibinfo{author}{J.~Lantzsch}, \bibinfo{author}{P.~M\"{u}ller},
  \bibinfo{author}{W.~N\"{o}rtersh\"{a}user}, \bibinfo{author}{E.~W. Otten},
  \bibinfo{author}{R.~Schwalbach}, \bibinfo{author}{U.~A. Seibert},
  \bibinfo{author}{N.~Trautmann}, \bibinfo{author}{A.~Waldek},
  \bibinfo{journal}{Radiochim. Acta} \bibinfo{volume}{79}
  (\bibinfo{year}{1997}) \bibinfo{pages}{183}.
\bibitem[{Alkhazov et~al.(1992)Alkhazov, Barzakh, Denisov, Mezilev, Novikov,
  Panteleyev, Popov, Sudentas, Letokhov, Mishin, Fedoseyev, Andreyev,
  Vedeneyev, and Zyuzikov}]{Alk1992}
\bibinfo{author}{G.~D. Alkhazov}, \bibinfo{author}{A.~E. Barzakh},
  \bibinfo{author}{V.~P. Denisov}, \bibinfo{author}{K.~A. Mezilev},
  \bibinfo{author}{{\relax Yu}.~N. Novikov}, \bibinfo{author}{V.~N.
  Panteleyev}, \bibinfo{author}{A.~N. Popov}, \bibinfo{author}{E.~P. Sudentas},
  \bibinfo{author}{V.~S. Letokhov}, \bibinfo{author}{V.~I. Mishin},
  \bibinfo{author}{V.~N. Fedoseyev}, \bibinfo{author}{S.~.~V. Andreyev},
  \bibinfo{author}{D.~S. Vedeneyev}, \bibinfo{author}{A.~D. Zyuzikov},
  \bibinfo{journal}{Nucl. Instr. Methods B} \bibinfo{volume}{69}
  (\bibinfo{year}{1992}) \bibinfo{pages}{517}.
\bibitem[{Fedosseev et~al.(2003)Fedosseev, Fedorov, Horn, Huber, K\"{o}ster,
  Lassen, Mishin, Seliverstov, Weissman, and Wendt}]{Fed2003}
\bibinfo{author}{V.~N. Fedosseev}, \bibinfo{author}{D.~V. Fedorov},
  \bibinfo{author}{R.~Horn}, \bibinfo{author}{G.~Huber},
  \bibinfo{author}{U.~K\"{o}ster}, \bibinfo{author}{J.~Lassen},
  \bibinfo{author}{V.~I. Mishin}, \bibinfo{author}{M.~D. Seliverstov},
  \bibinfo{author}{L.~Weissman}, \bibinfo{author}{K.~Wendt},
  \bibinfo{journal}{Nucl. Instr. Methods B} \bibinfo{volume}{204}
  (\bibinfo{year}{2003}) \bibinfo{pages}{353}.
\bibitem[{K\"{o}ster et~al.(2003)K\"{o}ster, Fedoseyev, and Mishin}]{Kos2003}
\bibinfo{author}{U.~K\"{o}ster}, \bibinfo{author}{V.~N. Fedoseyev},
  \bibinfo{author}{V.~I. Mishin}, \bibinfo{journal}{Spectrochim. Acta B}
  \bibinfo{volume}{58} (\bibinfo{year}{2003}) \bibinfo{pages}{1047}.
\bibitem[{Facina et~al.(2004)Facina, Bruyneel, Dean, Gentens, Huyse,
  Kudryavtsev, {V}an~den Bergh, and {V}an Duppen}]{Fac2004}
\bibinfo{author}{M.~Facina}, \bibinfo{author}{B.~Bruyneel},
  \bibinfo{author}{S.~Dean}, \bibinfo{author}{J.~Gentens},
  \bibinfo{author}{M.~Huyse}, \bibinfo{author}{{\relax Yu}.~Kudryavtsev},
  \bibinfo{author}{P.~{V}an~den Bergh}, \bibinfo{author}{P.~{V}an Duppen},
  \bibinfo{journal}{Nucl. Instr. Methods B} \bibinfo{volume}{226}
  (\bibinfo{year}{2004}) \bibinfo{pages}{401}.
\bibitem[{{D}e Witte et~al.(2007){D}e Witte, Andreyev, Barre, Bender, Cocolios,
  Dean, Fedorov, Fedoseyev, Fraile, Franchoo, Hellemans, Heenen, Heyde, Huber,
  Huyse, Jeppessen, K\"{o}ster, Kunz, Lesher, Marsh, Mukha, Roussiere, Sauvage,
  Seliverstov, Stefanescu, Tengborn, {V}an~de Vel, {V}an~de Walle, {V}an
  Duppen, and Volkov}]{DeW2007}
\bibinfo{author}{H.~{D}e Witte}, \bibinfo{author}{A.~N. Andreyev},
  \bibinfo{author}{N.~Barre}, \bibinfo{author}{M.~Bender},
  \bibinfo{author}{T.~E. Cocolios}, \bibinfo{author}{S.~Dean},
  \bibinfo{author}{D.~Fedorov}, \bibinfo{author}{V.~N. Fedoseyev},
  \bibinfo{author}{L.~M. Fraile}, \bibinfo{author}{S.~Franchoo},
  \bibinfo{author}{V.~Hellemans}, \bibinfo{author}{P.~H. Heenen},
  \bibinfo{author}{K.~Heyde}, \bibinfo{author}{G.~Huber},
  \bibinfo{author}{M.~Huyse}, \bibinfo{author}{H.~Jeppessen},
  \bibinfo{author}{U.~K\"{o}ster}, \bibinfo{author}{P.~Kunz},
  \bibinfo{author}{S.~R. Lesher}, \bibinfo{author}{B.~A. Marsh},
  \bibinfo{author}{I.~Mukha}, \bibinfo{author}{B.~Roussiere},
  \bibinfo{author}{J.~Sauvage}, \bibinfo{author}{M.~Seliverstov},
  \bibinfo{author}{I.~Stefanescu}, \bibinfo{author}{E.~Tengborn},
  \bibinfo{author}{K.~{V}an~de Vel}, \bibinfo{author}{J.~{V}an~de Walle},
  \bibinfo{author}{P.~{V}an Duppen}, \bibinfo{author}{{\relax Yu}.~Volkov},
  \bibinfo{journal}{Phys. Rev. Lett} \bibinfo{volume}{98}
  (\bibinfo{year}{2007}) \bibinfo{pages}{112502}.
\bibitem[{Cocolios et~al.(2011)Cocolios, Dexters, Seliverstov, Andreyev,
  Antalic, Barzakh, Bastin, B\"{u}scher, Darby, Fedorov, Fedosseyev, Flanagan,
  Franchoo, Fritzsche, Huber, Huyse, Keupers, K\"{o}ster, Kudryavtsev,
  Man\'{e}, Marsh, Molkanov, Page, Sjoedin, Stefan, {V}an~de Walle, {V}an
  Duppen, Venhart, Zemlyanoy, Bender, and Heenen}]{Coc2011}

\bibinfo{author}{T.~E. Cocolios}, \bibinfo{author}{W.~Dexters},
  \bibinfo{author}{M.~D. Seliverstov}, \bibinfo{author}{A.~N. Andreyev},
  \bibinfo{author}{S.~Antalic}, \bibinfo{author}{A.~E. Barzakh},
  \bibinfo{author}{B.~Bastin}, \bibinfo{author}{J.~B\"{u}scher},
  \bibinfo{author}{I.~G. Darby}, \bibinfo{author}{D.~Fedorov},
  \bibinfo{author}{V.~N. Fedosseyev}, \bibinfo{author}{K.~T. Flanagan},
  \bibinfo{author}{S.~Franchoo}, \bibinfo{author}{S.~Fritzsche},
  \bibinfo{author}{G.~Huber}, \bibinfo{author}{M.~Huyse},
  \bibinfo{author}{M.~Keupers}, \bibinfo{author}{U.~K\"{o}ster},
  \bibinfo{author}{{\relax Yu}.~Kudryavtsev}, \bibinfo{author}{E.~Man\'{e}},
  \bibinfo{author}{B.~A. Marsh}, \bibinfo{author}{P.~L. Molkanov},
  \bibinfo{author}{R.~D. Page}, \bibinfo{author}{A.~M. Sjoedin},
  \bibinfo{author}{I.~Stefan}, \bibinfo{author}{J.~{V}an~de Walle},
  \bibinfo{author}{P.~{V}an Duppen}, \bibinfo{author}{M.~Venhart},
  \bibinfo{author}{S.~G. Zemlyanoy}, \bibinfo{author}{M.~Bender},
  \bibinfo{author}{P.-H. Heenen}, \bibinfo{journal}{Phys. Rev. Lett}
  \bibinfo{volume}{106} (\bibinfo{year}{2011}) \bibinfo{pages}{052503}.
\bibitem[{Lauth et~al.(1992)Lauth, Backe, Dahlinger, Klaft, Schwamb,
  Schwickert, Trautmann, and Othmer}]{Lau1992}
\bibinfo{author}{W.~Lauth}, \bibinfo{author}{H.~Backe},
  \bibinfo{author}{M.~Dahlinger}, \bibinfo{author}{I.~Klaft},
  \bibinfo{author}{P.~Schwamb}, \bibinfo{author}{G.~Schwickert},
  \bibinfo{author}{N.~Trautmann}, \bibinfo{author}{U.~Othmer},
  \bibinfo{journal}{Rev. Lett.} \bibinfo{volume}{68} (\bibinfo{year}{1992})
  \bibinfo{pages}{1675}.
\bibitem[{Cocolios et~al.(2010)Cocolios, Andreyev, Bastin, Bree, Buscher,
  Elseviers, Gentens, Huyse, Kudryavtsev, Pauwels, Sonoda, {V}an~den Bergh, and
  {V}an Duppen}]{Coc2010}
\bibinfo{author}{T.~E. Cocolios}, \bibinfo{author}{A.~N. Andreyev},
  \bibinfo{author}{B.~Bastin}, \bibinfo{author}{N.~Bree},
  \bibinfo{author}{J.~Buscher}, \bibinfo{author}{J.~Elseviers},
  \bibinfo{author}{J.~Gentens}, \bibinfo{author}{M.~Huyse},
  \bibinfo{author}{{\relax Yu}.~Kudryavtsev}, \bibinfo{author}{D.~Pauwels},
  \bibinfo{author}{T.~Sonoda}, \bibinfo{author}{P.~{V}an~den Bergh},
  \bibinfo{author}{P.~{V}an Duppen}, \bibinfo{journal}{Phy. Rev. C}
  \bibinfo{volume}{81} (\bibinfo{year}{2010}) \bibinfo{pages}{014314}.
\bibitem[{Stone et~al.(2008)Stone, K\"{o}ster, Stone, Fedorov, Fedoseyev,
  Flanagan, Hass, and Lakshmi}]{Sto2008}
\bibinfo{author}{N.~J. Stone}, \bibinfo{author}{U.~K\"{o}ster},
  \bibinfo{author}{J.~R. Stone}, \bibinfo{author}{D.~V. Fedorov},
  \bibinfo{author}{V.~N. Fedoseyev}, \bibinfo{author}{K.~T. Flanagan},
  \bibinfo{author}{M.~Hass}, \bibinfo{author}{S.~Lakshmi},
  \bibinfo{journal}{Phys. Rev. C} \bibinfo{volume}{77} (\bibinfo{year}{2008})
  \bibinfo{pages}{067302}.
\bibitem[{Qamhieh et~al.(1992)Qamhieh, Vandeweert, Silverans, {V}an Duppen,
  Huyse, and Vermeeren}]{Qam1992}
\bibinfo{author}{Z.~N. Qamhieh}, \bibinfo{author}{E.~Vandeweert},
  \bibinfo{author}{R.~E. Silverans}, \bibinfo{author}{P.~{V}an Duppen},
  \bibinfo{author}{M.~Huyse}, \bibinfo{author}{L.~Vermeeren},
  \bibinfo{journal}{Nucl. Instr. Methods B} \bibinfo{volume}{70}
  (\bibinfo{year}{1992}) \bibinfo{pages}{131}.
\bibitem[{{V}an Duppen et~al.(1992){V}an Duppen, Dendooven, Huyse, Vermeeren,
  Qamhien, Silverans, and Vandeweert}]{VDu1992}
\bibinfo{author}{P.~{V}an Duppen}, \bibinfo{author}{P.~Dendooven},
  \bibinfo{author}{M.~Huyse}, \bibinfo{author}{L.~Vermeeren},
  \bibinfo{author}{Z.~N. Qamhien}, \bibinfo{author}{R.~E. Silverans},
  \bibinfo{author}{E.~Vandeweert}, \bibinfo{journal}{Hyperfine Interact.}
  \bibinfo{volume}{74} (\bibinfo{year}{1992}) \bibinfo{pages}{193}.
\bibitem[{Kudryavtsev et~al.(2003)Kudryavtsev, Facina, Huyse, Gentens,
  {V}an~den Bergh, and {V}an Duppen}]{Kud2003}
\bibinfo{author}{{\relax Yu}.~Kudryavtsev}, \bibinfo{author}{M.~Facina},
  \bibinfo{author}{M.~Huyse}, \bibinfo{author}{J.~Gentens},
  \bibinfo{author}{P.~{V}an~den Bergh}, \bibinfo{author}{P.~{V}an Duppen},
  \bibinfo{journal}{Nucl. Instr. Methods B} \bibinfo{volume}{204}
  (\bibinfo{year}{2003}) \bibinfo{pages}{336}.
\bibitem[{{V}an~den Bergh et~al.(1997){V}an~den Bergh, Franchoo, Gentens,
  Huyse, Kudryavtsev, Piechaczek, Raabe, Reusen, {V}an Duppen, Vermeeren, and
  W\"{o}hr}]{VdB1997}
\bibinfo{author}{P.~{V}an~den Bergh}, \bibinfo{author}{S.~Franchoo},
  \bibinfo{author}{J.~Gentens}, \bibinfo{author}{M.~Huyse},
  \bibinfo{author}{{\relax Yu}.~A. Kudryavtsev},
  \bibinfo{author}{A.~Piechaczek}, \bibinfo{author}{R.~Raabe},
  \bibinfo{author}{I.~Reusen}, \bibinfo{author}{P.~{V}an Duppen},
  \bibinfo{author}{L.~Vermeeren}, \bibinfo{author}{A.~W\"{o}hr},
  \bibinfo{journal}{Nucl. Instr. Methos B} \bibinfo{volume}{126}
  (\bibinfo{year}{1997}) \bibinfo{pages}{194}.
\bibitem[{Blaum et~al.(2003)Blaum, Geppert, Kluge, Mukherjee, Schwarz, and
  Wendt}]{Bla2003}
\bibinfo{author}{K.~Blaum}, \bibinfo{author}{C.~Geppert},
  \bibinfo{author}{H.-J. Kluge}, \bibinfo{author}{M.~Mukherjee},
  \bibinfo{author}{S.~Schwarz}, \bibinfo{author}{K.~Wendt},
  \bibinfo{journal}{Nucl. Instr. Methods B} \bibinfo{volume}{204}
  (\bibinfo{year}{2003}) \bibinfo{pages}{331}.
\bibitem[{Fink and {R}ichter {\relax et al.}(2012)}]{Fin2012}
\bibinfo{author}{D.~Fink}, \bibinfo{author}{S.~{R}ichter {\relax et al.}},
  \bibinfo{year}{(2012)}. \bibinfo{note}{In preparation}.
\bibitem[{Moore et~al.(2006)Moore, Billowes, Campbell, Eronen, Geppert,
  Jokinen, Karvonen, Kessler, Marsh, Nieminen, Penttil\"{a}, Rinta-Antila,
  Sonoda, Tordoff, Wendt, and \"{A}ysto}]{Moo2006}
\bibinfo{author}{I.~D. Moore}, \bibinfo{author}{J.~Billowes},
  \bibinfo{author}{P.~Campbell}, \bibinfo{author}{T.~Eronen},
  \bibinfo{author}{C.~Geppert}, \bibinfo{author}{A.~Jokinen},
  \bibinfo{author}{P.~Karvonen}, \bibinfo{author}{T.~Kessler},
  \bibinfo{author}{B.~Marsh}, \bibinfo{author}{A.~Nieminen},
  \bibinfo{author}{H.~Penttil\"{a}}, \bibinfo{author}{S.~Rinta-Antila},
  \bibinfo{author}{T.~Sonoda}, \bibinfo{author}{B.~Tordoff},
  \bibinfo{author}{K.~Wendt}, \bibinfo{author}{J.~\"{A}ysto},
  \bibinfo{journal}{AIP Conf. Proc.} \bibinfo{volume}{831}
  (\bibinfo{year}{2006}) \bibinfo{pages}{511}.
\bibitem[{Sonoda et~al.(2009)Sonoda, Cocolios, Gentens, Huyse, Ivanov,
  Kudryavtsev, Pauwels, {V}an~den Bergh, and {V}an Duppen}]{Son2009}
\bibinfo{author}{T.~Sonoda}, \bibinfo{author}{T.~Cocolios},
  \bibinfo{author}{J.~Gentens}, \bibinfo{author}{M.~Huyse},
  \bibinfo{author}{O.~Ivanov}, \bibinfo{author}{{\relax Yu}.~Kudryavtsev},
  \bibinfo{author}{D.~Pauwels}, \bibinfo{author}{P.~{V}an~den Bergh},
  \bibinfo{author}{P.~{V}an Duppen}, \bibinfo{journal}{Nucl. Instr. Methods B}
  \bibinfo{volume}{267} (\bibinfo{year}{2009}) \bibinfo{pages}{2918}.
\bibitem[{Ferrer et~al.(2012)Ferrer, Sonnenschein, Bastin, Franchoo, Huyse,
  Kudryavtsev, Kron, Lecesne, Moore, Osmond, Pauwels, Radulov, Reader, Rens,
  Reponen, Ro{\ss}nagel, Savajols, Sonoda, Thomas, {V}an~den Bergh, {V}an
  Duppen, Wendt, and Zemlyanoy}]{Fer2012}
\bibinfo{author}{R.~Ferrer}, \bibinfo{author}{V.~T. Sonnenschein},
  \bibinfo{author}{B.~Bastin}, \bibinfo{author}{S.~Franchoo},
  \bibinfo{author}{M.~Huyse}, \bibinfo{author}{{\relax Yu}.~Kudryavtsev},
  \bibinfo{author}{T.~Kron}, \bibinfo{author}{N.~Lecesne},
  \bibinfo{author}{I.~D. Moore}, \bibinfo{author}{B.~Osmond},
  \bibinfo{author}{D.~Pauwels}, \bibinfo{author}{D.~Radulov},
  \bibinfo{author}{S.~Reader}, \bibinfo{author}{L.~Rens},
  \bibinfo{author}{M.~Reponen}, \bibinfo{author}{J.~Ro{\ss}nagel},
  \bibinfo{author}{H.~Savajols}, \bibinfo{author}{T.~Sonoda},
  \bibinfo{author}{J.~C. Thomas}, \bibinfo{author}{P.~{V}an~den Bergh},
  \bibinfo{author}{P.~{V}an Duppen}, \bibinfo{author}{K.~Wendt},
  \bibinfo{author}{S.~Zemlyanoy}, \bibinfo{journal}{Nucl. Instr. Methods B}
  \bibinfo{volume}{In press} (\bibinfo{year}{2012}).
\bibitem[{Kantrowitz and Grey(1951)}]{Kan1951}
\bibinfo{author}{A.~Kantrowitz}, \bibinfo{author}{J.~Grey},
  \bibinfo{journal}{Rev. Sci. Instrum} \bibinfo{volume}{22}
  (\bibinfo{year}{1951}) \bibinfo{pages}{328}.
\bibitem[{Smalley et~al.(1977)Smalley, Wharton, and Levy}]{Sma1977}
\bibinfo{author}{R.~E. Smalley}, \bibinfo{author}{L.~Wharton},
  \bibinfo{author}{D.~H. Levy}, \bibinfo{journal}{Acc. Chem. Res.}
  \bibinfo{volume}{10} (\bibinfo{year}{1977}) \bibinfo{pages}{139}.
\bibitem[{Levy(1980)}]{Lev1980}
\bibinfo{author}{D.~H. Levy}, \bibinfo{journal}{Annu. Rev. Phys. Chem.}
  \bibinfo{volume}{31} (\bibinfo{year}{1980}) \bibinfo{pages}{197}.
\bibitem[{Koperski and Fry(2006)}]{Kop2006}
\bibinfo{author}{J.~Koperski}, \bibinfo{author}{E.~S. Fry},
  \bibinfo{journal}{J. Phys. B} \bibinfo{volume}{39} (\bibinfo{year}{2006})
  \bibinfo{pages}{S1125}.
\bibitem[{Reponen et~al.(2011)Reponen, Moore, Pohjalainen, Kessler, , Karvonen,
  Kurpeta, Marsh, Piszczek, Sonnenschein, and \"{A}yst\"{o}}]{Rep2011}
\bibinfo{author}{M.~Reponen}, \bibinfo{author}{I.~D. Moore},
  \bibinfo{author}{I.~Pohjalainen}, \bibinfo{author}{T.~Kessler}, ,
  \bibinfo{author}{P.~Karvonen}, \bibinfo{author}{J.~Kurpeta},
  \bibinfo{author}{B.~Marsh}, \bibinfo{author}{S.~Piszczek},
  \bibinfo{author}{V.~Sonnenschein}, \bibinfo{author}{J.~\"{A}yst\"{o}},
  \bibinfo{journal}{Nucl. Instr. Methods A} \bibinfo{volume}{635}
  (\bibinfo{year}{2011}) \bibinfo{pages}{24}.
\bibitem[{Reponen et~al.(2012)Reponen, Moore, Kessler, Pohjalainen, Rothe, and
  Sonnenschein}]{Rep2012}
\bibinfo{author}{M.~Reponen}, \bibinfo{author}{I.~Moore},
  \bibinfo{author}{T.~Kessler}, \bibinfo{author}{I.~Pohjalainen},
  \bibinfo{author}{S.~Rothe}, \bibinfo{author}{V.~Sonnenschein},
  \bibinfo{journal}{Eur. Phys. J. A} \bibinfo{volume}{48}
  (\bibinfo{year}{2012}) \bibinfo{pages}{45}.
\bibitem[{Rowe et~al.(1984)Rowe, Dupeyrat, Marquette, and Gaucherel}]{Row1984}
\bibinfo{author}{B.~R. Rowe}, \bibinfo{author}{G.~Dupeyrat},
  \bibinfo{author}{J.~B. Marquette}, \bibinfo{author}{P.~Gaucherel},
  \bibinfo{journal}{J. Chem. Phys.} \bibinfo{volume}{80} (\bibinfo{year}{1984})
  \bibinfo{pages}{4915}.
\bibitem[{Rowe and Marquette(1987)}]{Row1987}
\bibinfo{author}{R.~Rowe}, \bibinfo{author}{J.~B. Marquette},
  \bibinfo{journal}{Int. J. Mass Spectr.} \bibinfo{volume}{80}
  (\bibinfo{year}{1987}) \bibinfo{pages}{239}.
\bibitem[{Sims et~al.(1994)Sims, Queffelec, Defrance, Rebrion-Rowe, Travers,
  Bocherel, Rowe, and Smith}]{Sim1994}
\bibinfo{author}{I.~R. Sims}, \bibinfo{author}{J.-L. Queffelec},
  \bibinfo{author}{A.~Defrance}, \bibinfo{author}{C.~Rebrion-Rowe},
  \bibinfo{author}{D.~Travers}, \bibinfo{author}{P.~Bocherel},
  \bibinfo{author}{B.~R. Rowe}, \bibinfo{author}{I.~W.~M. Smith},
  \bibinfo{journal}{J. Chem. Phys.} \bibinfo{volume}{100}
  (\bibinfo{year}{1994}) \bibinfo{pages}{4229}.
\bibitem[{Atkinson and Smith(1995)}]{Atk1995}
\bibinfo{author}{D.~B. Atkinson}, \bibinfo{author}{M.~A. Smith},
  \bibinfo{journal}{Rev. Sci. Instrum.} \bibinfo{volume}{66}
  (\bibinfo{year}{1995}) \bibinfo{pages}{4434}.
\bibitem[{Lee et~al.(2000)Lee, Hoobler, and Leone}]{Lee2000}
\bibinfo{author}{S.~Lee}, \bibinfo{author}{R.~J. Hoobler},
  \bibinfo{author}{S.~R. Leone}, \bibinfo{journal}{Rev. Scien. Instrum.}
  \bibinfo{volume}{71} (\bibinfo{year}{2000}) \bibinfo{pages}{1816}.
\bibitem[{Dupeyrat et~al.(1985)Dupeyrat, Marquette, and Rowe}]{Dup1985}
\bibinfo{author}{G.~Dupeyrat}, \bibinfo{author}{J.~B. Marquette},
  \bibinfo{author}{B.~R. Rowe}, \bibinfo{journal}{Phys. Fluids}
  \bibinfo{volume}{28} (\bibinfo{year}{1985}) \bibinfo{pages}{1273}.
\bibitem[{Benidar et~al.(2000)Benidar, Georges, {L}e Doucen, Boissoles, Hamon,
  Canosa, and Rowe}]{Ben2000}
\bibinfo{author}{A.~Benidar}, \bibinfo{author}{R.~Georges},
  \bibinfo{author}{R.~{L}e Doucen}, \bibinfo{author}{J.~Boissoles},
  \bibinfo{author}{S.~Hamon}, \bibinfo{author}{A.~Canosa},
  \bibinfo{author}{B.~R. Rowe}, \bibinfo{journal}{J. Mol. Spectrosc.}
  \bibinfo{volume}{199} (\bibinfo{year}{2000}) \bibinfo{pages}{92}.
\bibitem[{Tordella et~al.(2011)Tordella, Belan, Massaglia, {D}e Ponte, Mignone,
  Bodenschatz, and Ferrari}]{Tor2011}
\bibinfo{author}{D.~Tordella}, \bibinfo{author}{M.~Belan},
  \bibinfo{author}{S.~Massaglia}, \bibinfo{author}{S.~{D}e Ponte},
  \bibinfo{author}{A.~Mignone}, \bibinfo{author}{E.~Bodenschatz},
  \bibinfo{author}{A.~Ferrari}, \bibinfo{journal}{New Journal of Physics}
  \bibinfo{volume}{13} (\bibinfo{year}{2011}) \bibinfo{pages}{043011}.
\bibitem[{Lemieux(2009)}]{Lem2009}
\bibinfo{author}{P.~Lemieux}, in: \bibinfo{booktitle}{39th AIAA Fluid Dynamics
  Conference}.
\bibitem[{Besnard et~al.(2002)Besnard, Chen, and Mueller}]{Bes2002}
\bibinfo{author}{E.~Besnard}, \bibinfo{author}{H.~H. Chen},
  \bibinfo{author}{T.~Mueller}, \bibinfo{journal}{AIAA,} \bibinfo{volume}{02}
  (\bibinfo{year}{2002}) \bibinfo{pages}{4038}.
\bibitem[{\"{A}rje et~al.(1985)\"{A}rje, \"{A}yst\"{o}, Hyv\"{o}nen, Taskinen,
  Koponen, Honkanen, Hautoj\"{a}rvi, and Vierinen}]{Arj1985}
\bibinfo{author}{J.~\"{A}rje}, \bibinfo{author}{J.~\"{A}yst\"{o}},
  \bibinfo{author}{H.~Hyv\"{o}nen}, \bibinfo{author}{P.~Taskinen},
  \bibinfo{author}{V.~Koponen}, \bibinfo{author}{J.~Honkanen},
  \bibinfo{author}{A.~Hautoj\"{a}rvi}, \bibinfo{author}{K.~Vierinen},
  \bibinfo{journal}{Phys. Rev. Lett.} \bibinfo{volume}{54}
  (\bibinfo{year}{1985}) \bibinfo{pages}{99}.
\bibitem[{Engels et~al.(2001)Engels, Beck, Bollen, Habs, Marx, Neumayr,
  Schramm, Schwarz, Thirolf, and Varentsov}]{Eng2001}
\bibinfo{author}{O.~Engels}, \bibinfo{author}{L.~Beck},
  \bibinfo{author}{G.~Bollen}, \bibinfo{author}{D.~Habs},
  \bibinfo{author}{G.~Marx}, \bibinfo{author}{J.~Neumayr},
  \bibinfo{author}{U.~Schramm}, \bibinfo{author}{S.~Schwarz},
  \bibinfo{author}{P.~Thirolf}, \bibinfo{author}{V.~Varentsov},
  \bibinfo{journal}{Hyperfine Interact.} \bibinfo{volume}{132}
  (\bibinfo{year}{2001}) \bibinfo{pages}{505}.
\bibitem[{Savard et~al.(2003)Savard, Clark, Boudreau, Buchinger, Crawford,
  Geisse, Greene, Gulick, Heinz, Lee, Levand, Maier, M\"{u}nzenberg,
  Scheidenberger, and Seweryniak}]{Sav2003}
\bibinfo{author}{G.~Savard}, \bibinfo{author}{J.~Clark},
  \bibinfo{author}{C.~Boudreau}, \bibinfo{author}{F.~Buchinger},
  \bibinfo{author}{J.~E. Crawford}, \bibinfo{author}{H.~Geisse},
  \bibinfo{author}{J.~P. Greene}, \bibinfo{author}{S.~Gulick},
  \bibinfo{author}{A.~Heinz}, \bibinfo{author}{J.~K.~P. Lee},
  \bibinfo{author}{A.~Levand}, \bibinfo{author}{M.~Maier},
  \bibinfo{author}{G.~M\"{u}nzenberg}, \bibinfo{author}{C.~Scheidenberger},
  \bibinfo{author}{D.~Seweryniak}, \bibinfo{journal}{Nucl. Instr. Methods B}
  \bibinfo{volume}{204} (\bibinfo{year}{2003}) \bibinfo{pages}{582}.
\bibitem[{Wada et~al.(2003)Wada, Ishida, Nakamura, Yamazaki, Kambara, Ohyama,
  Kanai, Kojima, Nakai, Ohshima, Yoshida, Kubo, Matsuo, Fukuyama, Okada,
  Sonoda, Ohtani, Noda, Kawakami, and Katayama}]{Wad2003}
\bibinfo{author}{M.~Wada}, \bibinfo{author}{Y.~Ishida},
  \bibinfo{author}{T.~Nakamura}, \bibinfo{author}{Y.~Yamazaki},
  \bibinfo{author}{T.~Kambara}, \bibinfo{author}{H.~Ohyama},
  \bibinfo{author}{Y.~Kanai}, \bibinfo{author}{T.~M. Kojima},
  \bibinfo{author}{Y.~Nakai}, \bibinfo{author}{N.~Ohshima},
  \bibinfo{author}{A.~Yoshida}, \bibinfo{author}{T.~Kubo},
  \bibinfo{author}{Y.~Matsuo}, \bibinfo{author}{Y.~Fukuyama},
  \bibinfo{author}{K.~Okada}, \bibinfo{author}{T.~Sonoda},
  \bibinfo{author}{S.~Ohtani}, \bibinfo{author}{K.~Noda},
  \bibinfo{author}{H.~Kawakami}, \bibinfo{author}{I.~Katayama},
  \bibinfo{journal}{Nucl. Instr. Methos B} \bibinfo{volume}{204}
  (\bibinfo{year}{2003}) \bibinfo{pages}{570}.
\bibitem[{Schwarz et~al.(2003)Schwarz, Bollen, Lawton, Lofy, Morrissey,
  Ottarson, Ringle, Schury, Sun, Varentsov, and Weissman}]{Sch2003}
\bibinfo{author}{S.~Schwarz}, \bibinfo{author}{G.~Bollen},
  \bibinfo{author}{D.~Lawton}, \bibinfo{author}{P.~Lofy},
  \bibinfo{author}{D.~J. Morrissey}, \bibinfo{author}{J.~Ottarson},
  \bibinfo{author}{R.~Ringle}, \bibinfo{author}{P.~Schury},
  \bibinfo{author}{T.~Sun}, \bibinfo{author}{V.~Varentsov},
  \bibinfo{author}{L.~Weissman}, \bibinfo{journal}{Nucl. Instr. Methods B}
  \bibinfo{volume}{204} (\bibinfo{year}{2003}) \bibinfo{pages}{507}.
\bibitem[{{W}ayne Johnson and Gerardo(1973)}]{Way1973}
\bibinfo{author}{A.~{W}ayne Johnson}, \bibinfo{author}{J.~B. Gerardo},
  \bibinfo{journal}{Phys. Rev. A} \bibinfo{volume}{7} (\bibinfo{year}{1973})
  \bibinfo{pages}{1339}.
\bibitem[{Cooper et~al.(1993)Cooper, {V}an Sonsbeek, and Bhave}]{Coo1993}
\bibinfo{author}{R.~Cooper}, \bibinfo{author}{R.~J. {V}an Sonsbeek},
  \bibinfo{author}{R.~N. Bhave}, \bibinfo{journal}{J. Chem. Phys.}
  \bibinfo{volume}{98} (\bibinfo{year}{1993}) \bibinfo{pages}{383}.
\bibitem[{Huyse et~al.(2002)Huyse, Facina, Kudryavtsev, and {V}an
  Duppen}]{Huy2002}
\bibinfo{author}{M.~Huyse}, \bibinfo{author}{M.~Facina},
  \bibinfo{author}{{\relax Yu}.~Kudryavtsev}, \bibinfo{author}{P.~{V}an
  Duppen}, \bibinfo{journal}{Nucl. Instr. Methods B} \bibinfo{volume}{187}
  (\bibinfo{year}{2002}) \bibinfo{pages}{535}.
\bibitem[{Kudryavtsev et~al.(2009)Kudryavtsev, Cocolios, Gentens, Huyse,
  Ivanov, Pauwels, Sonoda, {V}an~den Bergh, and {V}an Duppen}]{Kud2009}
\bibinfo{author}{{\relax Yu}.~Kudryavtsev}, \bibinfo{author}{T.~Cocolios},
  \bibinfo{author}{J.~Gentens}, \bibinfo{author}{M.~Huyse},
  \bibinfo{author}{O.~Ivanov}, \bibinfo{author}{D.~Pauwels},
  \bibinfo{author}{T.~Sonoda}, \bibinfo{author}{P.~{V}an~den Bergh},
  \bibinfo{author}{P.~{V}an Duppen}, \bibinfo{journal}{Nucl. Instr. Methods B}
  \bibinfo{volume}{267} (\bibinfo{year}{2009}) \bibinfo{pages}{2908}.
\bibitem[{Allard(1982)}]{All1982}
\bibinfo{author}{N.~Allard}, \bibinfo{journal}{Rev. Mod. Phys.}
  \bibinfo{volume}{54} (\bibinfo{year}{1982}) \bibinfo{pages}{1103}.
\bibitem[{Lwin et~al.(1977)Lwin, {M}e Cartan, and Levis}]{Lwi1977}
\bibinfo{author}{A.~Lwin}, \bibinfo{author}{D.~D. {M}e Cartan},
  \bibinfo{author}{E.~L. Levis}, \bibinfo{journal}{J. Astrophys.}
  \bibinfo{volume}{213} (\bibinfo{year}{1977}) \bibinfo{pages}{599}.
\bibitem[{Helmi and Roston(2001)}]{Hel2001}
\bibinfo{author}{M.~S. Helmi}, \bibinfo{author}{G.~D. Roston},
  \bibinfo{journal}{AIP Conf. Proc.} \bibinfo{volume}{559}
  (\bibinfo{year}{2001}) \bibinfo{pages}{347}.
\bibitem[{Bielski et~al.(1989)Bielski, Bobkowski, and Szudy}]{Bie1989}
\bibinfo{author}{A.~Bielski}, \bibinfo{author}{R.~Bobkowski},
  \bibinfo{author}{J.~Szudy}, \bibinfo{journal}{Astrom. Astrophys.}
  \bibinfo{volume}{208} (\bibinfo{year}{1989}) \bibinfo{pages}{357}.
\bibitem[{Olivero and Longbothum(1977)}]{Oli1977}
\bibinfo{author}{J.~J. Olivero}, \bibinfo{author}{R.~L. Longbothum},
  \bibinfo{journal}{Quant. Spectrosc. Radiat. Transfer} \bibinfo{volume}{17}
  (\bibinfo{year}{1977}) \bibinfo{pages}{233}.
\bibitem[{Anton et~al.(1978)Anton, Kaufmann, Klempt, Moruzzi, Neugart, Otten,
  and Schinzler}]{Ant1978}
\bibinfo{author}{K.~R. Anton}, \bibinfo{author}{S.~L. Kaufmann},
  \bibinfo{author}{W.~Klempt}, \bibinfo{author}{G.~Moruzzi},
  \bibinfo{author}{R.~Neugart}, \bibinfo{author}{E.~W. Otten},
  \bibinfo{author}{B.~Schinzler}, \bibinfo{journal}{Phys. Rev. Lett.}
  \bibinfo{volume}{40} (\bibinfo{year}{1978}) \bibinfo{pages}{642}.
\bibitem[{Wing et~al.(1976)Wing, Ruff, Lamb, and Spezeski}]{Win1976}
\bibinfo{author}{W.~H. Wing}, \bibinfo{author}{G.~A. Ruff},
  \bibinfo{author}{W.~E. Lamb}, \bibinfo{author}{J.~J. Spezeski},
  \bibinfo{journal}{Phys. Rev. Lett.} \bibinfo{volume}{36}
  (\bibinfo{year}{1976}) \bibinfo{pages}{1488}.
\bibitem[{Redpath and Menzingerc(1977)}]{Red1977}
\bibinfo{author}{A.~E. Redpath}, \bibinfo{author}{M.~Menzingerc},
  \bibinfo{journal}{J. Chem.} \bibinfo{volume}{55} (\bibinfo{year}{1977})
  \bibinfo{pages}{1055}.
\bibitem[{Hagena and Obert(1972)}]{Hag1972}
\bibinfo{author}{O.~F. Hagena}, \bibinfo{author}{W.~Obert},
  \bibinfo{journal}{J. Chem. Phys.} \bibinfo{volume}{56} (\bibinfo{year}{1972})
  \bibinfo{pages}{1793}.
\bibitem[{Hagena(1987)}]{Hag1987}
\bibinfo{author}{O.~F. Hagena}, \bibinfo{journal}{Z. Phys. D}
  \bibinfo{volume}{4} (\bibinfo{year}{1987}) \bibinfo{pages}{291}.
\bibitem[{W\"{o}rmer et~al.(1989)W\"{o}rmer, Guzielski, Stapelfeldt, and
  M\"{o}ller}]{Wor1989}
\bibinfo{author}{J.~W\"{o}rmer}, \bibinfo{author}{V.~Guzielski},
  \bibinfo{author}{J.~Stapelfeldt}, \bibinfo{author}{T.~M\"{o}ller},
  \bibinfo{journal}{Chem. Phys. Lett.} \bibinfo{volume}{159}
  (\bibinfo{year}{1989}) \bibinfo{pages}{321}.
\bibitem[{Grandinetti(2011)}]{Gra2011}
\bibinfo{author}{F.~Grandinetti}, \bibinfo{journal}{Eur. J. Mass. Spectrom.}
  \bibinfo{volume}{17} (\bibinfo{year}{2011}) \bibinfo{pages}{423}.
\bibitem[{Duncan(2012)}]{Dun2012}
\bibinfo{author}{M.~A. Duncan}, \bibinfo{journal}{Rev. Sci. Instrum.}
  \bibinfo{volume}{83} (\bibinfo{year}{2012}) \bibinfo{pages}{041101}.
\bibitem[{McDaniels et~al.(2003)McDaniels, Continetti, and Miller}]{McD2003}
\bibinfo{author}{J.~T. McDaniels}, \bibinfo{author}{R.~E. Continetti},
  \bibinfo{author}{D.~R. Miller}, \bibinfo{journal}{AIP Conf. Proc.}
  \bibinfo{volume}{663} (\bibinfo{year}{2003}) \bibinfo{pages}{670}.
\bibitem[{W\"{o}rmer et~al.(1990)W\"{o}rmer, Guzielski, Stapelfeldt, Zimmerer,
  and M\"{o}ller}]{Wor1990}
\bibinfo{author}{J.~W\"{o}rmer}, \bibinfo{author}{V.~Guzielski},
  \bibinfo{author}{J.~Stapelfeldt}, \bibinfo{author}{G.~Zimmerer},
  \bibinfo{author}{T.~M\"{o}ller}, \bibinfo{journal}{Phys. Scr.}
  \bibinfo{volume}{40} (\bibinfo{year}{1990}) \bibinfo{pages}{490}.
\bibitem[{Kudryavtsev et~al.(2001)Kudryavtsev, Bruyneel, Huyse, Gentens,
  {V}an~den Bergh, {V}an Duppen, and Vermeeren}]{Kud2001}
\bibinfo{author}{{\relax Yu}.~Kudryavtsev}, \bibinfo{author}{B.~Bruyneel},
  \bibinfo{author}{M.~Huyse}, \bibinfo{author}{J.~Gentens},
  \bibinfo{author}{P.~{V}an~den Bergh}, \bibinfo{author}{P.~{V}an Duppen},
  \bibinfo{author}{L.~Vermeeren}, \bibinfo{journal}{Nucl. Instr. Methods B}
  \bibinfo{volume}{179} (\bibinfo{year}{2001}) \bibinfo{pages}{412}.
\bibitem[{Kraiko and Tillyayeva(2007)}]{Kra2007}
\bibinfo{author}{A.~N. Kraiko}, \bibinfo{author}{N.~I. Tillyayeva},
  \bibinfo{journal}{Fluid Dynam.} \bibinfo{volume}{42} (\bibinfo{year}{2007})
  \bibinfo{pages}{321}.
\bibitem[{Rao(1961)}]{Rao1961}
\bibinfo{author}{G.~V.~R. Rao}, \bibinfo{journal}{Planet. Space Sci.}
  \bibinfo{volume}{4} (\bibinfo{year}{1961}) \bibinfo{pages}{92}.
\bibitem[{Greer(1961)}]{Gre1961}
\bibinfo{author}{H.~Greer}, \bibinfo{journal}{ARS J} \bibinfo{volume}{31}
  (\bibinfo{year}{1961}) \bibinfo{pages}{560}.
\bibitem[{Angelino(1964)}]{Ang1964}
\bibinfo{author}{G.~Angelino}, \bibinfo{journal}{AIAA} \bibinfo{volume}{J. 2}
  (\bibinfo{year}{1964}) \bibinfo{pages}{1834}.
\bibitem[{Lee(1963)}]{Lee1963}
\bibinfo{author}{C.~C. Lee}, \bibinfo{title}{Fortran programs for plug nozzle
  design}, \bibinfo{type}{Technical Note} \bibinfo{number}{R-41}, Scientific
  research laboratories {B}rown engineering company, INC.,
  \bibinfo{year}{1963}.
\bibitem[{Angelino(1963)}]{Ang1963}
\bibinfo{author}{G.~Angelino}, \bibinfo{title}{Theoretical and Experimental
  Investigation of the Design and Performance of a Plug-Type Nozzle},
  \bibinfo{type}{Technical Note} \bibinfo{number}{12}, Training Center Exptl.
  Aerodyn., \bibinfo{year}{1963}.
\bibitem[{Onofri(2002)}]{Ono2002}
\bibinfo{author}{M.~Onofri}, \bibinfo{journal}{AIAA}  (\bibinfo{year}{2002})
  \bibinfo{pages}{0584}.
\bibitem[{Anderson(1974)}]{And1974}
\bibinfo{author}{J.~B. Anderson}, \bibinfo{title}{Molecular Beams and low
  density gas dynamics}, \bibinfo{publisher}{Dekker, New York}, pp.
  \bibinfo{pages}{1--91}, \bibinfo{year}{1974}.
\bibitem[{Campargue(1984)}]{Com1984}
\bibinfo{author}{R.~Campargue}, \bibinfo{journal}{J. Phys. Chem.}
  \bibinfo{volume}{88} (\bibinfo{year}{1984}) \bibinfo{pages}{4466}.
\bibitem[{Miller(1988)}]{Mil1988}
\bibinfo{author}{D.~R. Miller}, \bibinfo{title}{Free jet sources. In
  \emph{Atommic and molecular beam methods}}, \bibinfo{publisher}{Oxford
  University Press}, \bibinfo{year}{1988}.
\bibitem[{B\"{u}terfish and Vennemann(1974)}]{But1974}
\bibinfo{author}{K.~A. B\"{u}terfish}, \bibinfo{author}{D.~Vennemann},
  \bibinfo{journal}{Prog. Airosp. Sci.} \bibinfo{volume}{15}
  (\bibinfo{year}{1974}) \bibinfo{pages}{217}.
\bibitem[{Belan et~al.(2008)Belan, {D}e Ponte, and Tordella}]{Bel2008}
\bibinfo{author}{C.~M. Belan}, \bibinfo{author}{S.~{D}e Ponte},
  \bibinfo{author}{D.~Tordella}, \bibinfo{journal}{Exp. Fluids}
  \bibinfo{volume}{45} (\bibinfo{year}{2008}) \bibinfo{pages}{501}.
\bibitem[{Clemens(2002)}]{Cle2002}
\bibinfo{author}{N.~T. Clemens}, \bibinfo{title}{Flow Imaging. In
  \emph{Encyclopedia of Imaging Science and Technology}},
  \bibinfo{publisher}{John Wiley and Sons}, \bibinfo{year}{2002}.
\bibitem[{Mohamed and Bonnet(2007)}]{Moh2007}
\bibinfo{author}{A.~K. Mohamed}, \bibinfo{author}{J.~Bonnet},
  \bibinfo{title}{Advanced Concept for Air Data System using EBF and Lidar. In
  \emph{Flight Experiments for Hypersonic Vehicle Development}},
  \bibinfo{publisher}{Paper 16. Neuilly-sur-Seine}, pp. \bibinfo{pages}{1--32},
  \bibinfo{year}{2007}.
\bibitem[{Narayanaswamy et~al.(2011)Narayanaswamy, Burns, and
  Clemens}]{Nar2011}
\bibinfo{author}{V.~Narayanaswamy}, \bibinfo{author}{R.~Burns},
  \bibinfo{author}{N.~T. Clemens}, \bibinfo{journal}{Opt. Lett.}
  \bibinfo{volume}{36} (\bibinfo{year}{2011}) \bibinfo{pages}{4185}.
\bibitem[{Belan et~al.(2010)Belan, {D}e Ponte, and Tordella}]{Bel2010}
\bibinfo{author}{M.~Belan}, \bibinfo{author}{S.~{D}e Ponte},
  \bibinfo{author}{D.~Tordella}, \bibinfo{journal}{Phys. Rev. E}
  \bibinfo{volume}{82} (\bibinfo{year}{2010}) \bibinfo{pages}{026303}.
\bibitem[{Ashkenas and Sherman(1966)}]{Ash1966}
\bibinfo{author}{H.~Ashkenas}, \bibinfo{author}{F.~S. Sherman},
  \bibinfo{title}{The structure and utilization of supersonic free jets in low
  density wind tunnels. In \emph{Rarefied Gas Dynamics}},
  \bibinfo{publisher}{Academic Press, Ney York}, p.~\bibinfo{pages}{94},
  \bibinfo{year}{1966}.
\bibitem[{Anderson et~al.(1966)Anderson, Andres, and Fenn}]{And1966}
\bibinfo{author}{J.~B. Anderson}, \bibinfo{author}{R.~P. Andres},
  \bibinfo{author}{J.~B. Fenn}, \bibinfo{journal}{Adv. Chem. Phys.}
  \bibinfo{volume}{10} (\bibinfo{year}{1966}) \bibinfo{pages}{275}.
\bibitem[{Lubman et~al.(1982)Lubman, Rettner, and Zare}]{Lub1982}
\bibinfo{author}{D.~M. Lubman}, \bibinfo{author}{C.~N. Rettner},
  \bibinfo{author}{R.~N. Zare}, \bibinfo{journal}{J. Phys. Chem.}
  \bibinfo{volume}{86} (\bibinfo{year}{1982}) \bibinfo{pages}{1129}.
\bibitem[{Anderson(1972)}]{And1972}
\bibinfo{author}{J.~B. Anderson}, \bibinfo{journal}{AIAA} \bibinfo{volume}{10}
  (\bibinfo{year}{1972}) \bibinfo{pages}{112}.
\bibitem[{Murphy(1984)}]{Mur1984}
\bibinfo{author}{H.~Murphy}, \bibinfo{title}{The effects of source geometry on
  free jet expansions}, Ph.D. thesis, University of California, San Diego,
  \bibinfo{year}{1984}.
\bibitem[{Bergstr\"{o}m et~al.(1989)Bergstr\"{o}m, Peng, and Persson}]{Ber1989}
\bibinfo{author}{H.~Bergstr\"{o}m}, \bibinfo{author}{W.~X. Peng},
  \bibinfo{author}{A.~Persson}, \bibinfo{journal}{Z. Phys. D}
  \bibinfo{volume}{13} (\bibinfo{year}{1989}) \bibinfo{pages}{203}.
\bibitem[{Fischer(1961)}]{Fis1961}
\bibinfo{author}{W.~Fischer}, \bibinfo{journal}{Z. Phys.} \bibinfo{volume}{161}
  (\bibinfo{year}{1961}) \bibinfo{pages}{89}.
\bibitem[{Demtr\"{o}der(2002)}]{Dem2002}
\bibinfo{author}{W.~Demtr\"{o}der}, \bibinfo{title}{Laser Spectroscopy. Basic
  Concepts and Instrumentation.}, \bibinfo{publisher}{Springer},
  \bibinfo{edition}{third edition} edition, \bibinfo{year}{2002}.
\bibitem[{GAN(2012)}]{GAN2012}
\bibinfo{title}{{SPIRAL}2 website},
  \bibinfo{howpublished}{http://pro.ganil-spiral2.eu/spiral2/instrumentation/s%
3}, \bibinfo{year}{October 2012}.


\end{thebibliography}



\end{document}